\documentclass[useAMS,usenatbib]{mn2e}
\usepackage{ifpdf}
\ifpdf
\end{document}
\fi

\usepackage{amsmath,fleqn,graphicx,amssymb,xcolor,ifpdf}
\usepackage{multirow}
\renewcommand{\[}{\begin{equation}}
\renewcommand{\]}{\end{equation}}
\def\p{\partial}

\def\ex#1{\left\langle#1\right\rangle}
\def\agama{{\sc agama}}

\def\rd{}
\def\H{\kpc\kms}
\let\boldgrk=\gkvecten
\let\boldgrksc=\gkvecseven

\def\gkthing#1{{\mathchoice%
	{\hbox{{\boldgrk\char#1}}}
	{\hbox{{\boldgrk\char#1}}}
	{\hbox{{\boldgrksc\char#1}}}
	{\hbox{{\boldgrksc\char#1}}}}}

\def\vtheta{\gkthing{18}}

{\newif\ifnotend
\notendtrue
\def\veclist{ABCDEFGHIJKLMNOPQRSTUVWXYZabcdefghijklmnopqrstuvwxyz.}
\def\top#1#2.{#1}
\def\tail#1#2.{#2.}
\loop\expandafter\xdef\csname v\expandafter\top\veclist\endcsname%
{{\noexpand\bf\expandafter\top\veclist}}
\edef\veclist{\expandafter\tail\veclist}
\if\veclist.\notendfalse\fi\ifnotend\repeat}
{\newif\ifnotend
\notendtrue
\def\veclist{ABCDEFGHIJKLMNOPQRSTUVWXYZ.}
\def\top#1#2.{#1}
\def\tail#1#2.{#2.}
\loop\expandafter\xdef\csname c\expandafter\top\veclist\endcsname%
{{\noexpand\cal\expandafter\top\veclist}}
\edef\veclist{\expandafter\tail\veclist}
\if\veclist.\notendfalse\fi\ifnotend\repeat}
\def\d{{\rm d}}

\def\Myr{\,\mathrm{Myr}}
\def\kpc{\,\mathrm{kpc}}

\def\kms{\,\mathrm{km\,s}^{-1}}

\def\dex{\,{\rm dex}}

\def\msun{\,{\rm M}_\odot}

\def\pc{\,\mathrm{pc}}
\def\e{\mathrm{e}}

\def\fracj#1#2{{\textstyle{#1\over#2}}}

\title[Chemodynamical models of our Galaxy]
{Chemodynamical models of our Galaxy}

\author[James Binney \& Eugene Vasiliev]{
  James Binney$^1$\thanks{E-mail: binney@physics.ox.ac.uk} and Eugene
  Vasiliev$^{2}$\\  
  $^1$Rudolf Peierls Centre for Theoretical Physics, Clarendon Laboratory,
  Oxford, OX1 3PU, UK\\
  $^2$Institute of Astronomy, Madingley Road, Cambridge CB3 0HA
}

\begin{document}
\maketitle

\begin{abstract}
A chemodynamical model of our galaxy is fitted to data from DR17 of the
APOGEE survey supplemented with data from the StarHorse catalogue and Gaia
DR3. Dynamically, the model is defined by action-based distribution functions
for dark matter and six stellar components plus a gas disc.  The
gravitational potential jointly generated by the model's components is used
to examine the galaxy's chemical composition within action space. The
observational data probably cover all parts of action space that are
populated by stars.  The overwhelming majority of stars have angular momentum
$J_\phi>0$ implying that they were born in the Galactic disc. High-$\alpha$
stars dominate in a region that is sharply bounded by $J_\phi\la
J_\phi(\hbox{solar})$. Chemically the model is defined by giving each stellar
component a Gaussian distribution in ([Fe/H],[Mg/Fe]) space about a mean that
is a linear function of the actions. The model's 47 dynamical
parameters are chosen to maximise the likelihood of the data given the model
in 72 three-dimensional velocity spaces while its  70 chemical parameters are
similarly chosen in five-dimensional
chemo-dynamical space. The circular speed falls steadily from $237\kms$ at $R=4\kpc$ to
$218\kms$ at $R=20\kpc$. Dark matter contributes half the radial force on the
Sun and has local density $0.011\msun\pc^{-3}$, there being
$24.5\msun\pc^{-2}$ in dark matter and $26.5\msun\pc^{-2}$ in stars within
$1.1\kpc$ of the plane.
\end{abstract}

\def\Jd{J_{\rm d}} \def\Jv{J_{\rm v}}
\def\Jt{\widetilde J}
\def\Jdo{J_{\rm d0}} \def\Jvo{J_{\rm v0}}
\def\Jro{J_{\rm r0}} \def\Jzo{J_{\rm z0}} \def\Jpo{J_{\phi0}}
\section{Introduction} \label{sec:intro}

ESA's Gaia satellite provides locations and space velocities for tens of millions of
stars \citep{GaiaEDR3general,GaiaDR3general}.
In anticipation of the arrival of Gaia astrometry, several teams around the
world have been accumulating the spectra of millions of stars at higher
resolution than Gaia can achieve and using these spectra to derive the stars'
chemical compositions, which are expected to yield insight into our Galaxy's
history.  The APOGEE survey \citep{Majewski2017} is particularly powerful in
this respect because, being based on the mid-infrared H band, it can probe
the disc nearer the plane and over a wider radial range than other surveys, which
are more strongly restricted by dust.

Since the middle of the 20th century we have known that the ages and chemical
compositions of stars vary systematically with their locations and velocities
\citep{Roman1950,Roman1999,ELS1962,GilmoreWyse1998,Fuhrmann2011}. With the emergence of the
theory of nucleosynthesis \citep{BBFH1957} and models of the
chemical evolution of the ISM \citep{Tinsley1980,Pagel1997} conviction grew that by
studying the chemodynamical structure of our Galaxy we should be able to
trace its history \citep{FrBH02}. Over the last two decades efforts to realise this goal
have taken two lines of attack. One line centres on simulations of galaxy
formation that include gas, stars and dark matter, usually in some sort of
cosmological context \citep[e.g.][]{Brook2004,Grand2017}. Another
line of attack models the Galaxy as a series of annuli within which stars
form from gas that they simultaneously enrich
\citep{Matteucci1989,Chiappini1997,SchoenrichB2009a,SchoenPJM,Sharma2021,ChenHayden2022}.
Strengths of the latter line of attack include the ability to fit models to
the very detailed data now available for our Galaxy and to develop
understanding of how specific physical processes, such as radial migration
and the late arrival of type Ia supernovae manifest themselves in
observational data. 

The central premise of \citet[hereafter SB09]{SchoenrichB2009a} and much
prior work is that all
disc stars were born on nearly circular orbits in the plane from gas that is
azimuthally well mixed, so its metallicity [Fe/H] and $\alpha$-abundance
[Mg/Fe] are functions of Galactocentric radius $R$ and look-back time $\tau$.
Since the gross chemistry of stellar atmospheres evolves little, to a good
approximation it then follows that the location ([Fe/H],[Mg/Fe]) of a star in
the chemical plane is an (a priori unknown) function of its birth radius
$R_{\rm b}$ and
age.  SB09 modelled the functions [Fe/H]$(R_{\rm b},\tau)$ and
[Mg/Fe]$(R_{\rm b},\tau)$ by adopting a radial profile of star formation and
following the production of heavy elements and the dispersal of these
elements by radial flows and winds. This effort lead to predictions for the
number and chemistry of stars born at each radius and time.

Fluctuations in the Galaxy's gravitational potential cause the orbits of
stars to drift \citep[e.g.][]{BinneyLacey1988}.  \cite{SellwoodB2002}
divided this drift
into (i) `blurring', which is the drift away from circular orbits to more
eccentric and inclined orbits, and (ii) `churning', by which  stars change
their angular momenta without increasing their random velocities.  By
adjusting the intensity of blurring and churning, SB09
were able to match the distribution of solar-neighbourhood stars in chemical
space. Remarkably, the observed bimodality of the chemical distribution emerged
naturally in a model in which the star-formation rate declined monotonically
with time and radius. The bimodality was a consequence of the rapid decline
in [Mg/Fe] about a gigayear after the start of star formation as type Ia supernova
set in.

The methodology of SB09 has been extended in various directions.
\cite{ChenHayden2022} updated their work by (a) using recent nucleosynthetic
yields, and (b) comparing the model predictions at 24 locations $(R,|z|)$ in
the Galaxy rather than just in the solar neighbourhood. This important latter
step was made possible by DR14 of the APOGEE survey. \cite{Sharma2021} had
previously fitted models to these data but instead of deriving chemical
compositions from a model of star formation and nucleosynthetic yields, they
specified a functional form for [Fe/H]$(R_{\rm b},\tau)$ that contains
parameters to be fitted to the data.  They further assumed that [Mg/Fe] is a
function of [Fe/H] and age that has a specified functional form, and fitted
the form's parameters to the data. 

Both \cite{Sharma2021} and \cite{ChenHayden2022} followed SB09 in adopting a
Schwarzschild DF $f(E_R,J_\phi,E_z)$ \citep[e.g.][\S4.4.3]{GDII}. Several
powerful arguments favour use of DFs that are functions $f(\vJ)$ of the
action integrals rather than the approximate energies $E_R$ and $E_z$
\citep[e.g.][]{JJBPJM16}, so \cite{SaJJB15:EDF} reformulated SB09 in terms of
action-based DFs. Specifically, they assumed that in the absence of churning,
the DF of each coeval cohort of disc stars would have the form of the
quasi-isothermal DF that was introduced by \cite{JJBPJM11:dyn}.  Blurring was
represented by the radial and vertical velocity dispersion parameters of the
DF being functions of age, and the cohort's current DF was obtained by
convolution of this quasi-isothermal DF with a kernel that represented
diffusion in $J_\phi$. Unfortunately at that time model-data comparisons were
only possible in the solar neighbourhood, but the work of \cite{Sharma2021},
in which a model was successfully fitted to the wide-ranging APOGEE data, is
methodologically similar to \cite{SaJJB15:EDF}.

These studies approach the data with a clear preconception of our Galaxy's
history. Our aim here is to be more data-driven: regardless of history, how
are the Galaxy's stars distributed in `chemodynamical space' -- the
five-dimensional space spanned by the actions, [Fe/H] and [Mg/Fe]?  Logically
this question should precedes the question of {\it why} stars are distributed as
they are. A map of this distribution would be an invaluable descriptor of our
Galaxy that would transcend theories of galaxy formation.

\citet[hereafter BV23]{BinneyVasiliev2023} used the {\it AGAMA} software
package \citep{AGAMA} to fit to Gaia DR2 data a model in which both stars and
dark matter were represented by DFs of the form $f(\vJ)$ and moved in the
potential $\Phi(\vx)$ that they and interstellar gas jointly generate. Here
we revise and extend this model: we revise it by modelling the Galaxy's bulge
as a fat disc rather than a spheroid; we update it by fitting to data from
APOGEE and the third rather than the second Gaia data release, and we extend it by
assigning a chemical model to each of its stellar components.

In Section \ref{sec:data} we describe the stellar sample that we have
analysed. Section~\ref{sec:AAchem} first examines how the means of [Mg/Fe]
and [Fe/H] vary with location in action space and then studies the stellar
density and mean values of the actions in four regions of action space within
which particular stellar components are expected to dominate.
Section~\ref{sec:modelling} introduces a scheme for modelling the Galaxy's
chemodynamical structure: Section~\ref{sec:EDFstructure} explains how we
extend a model based on standard DFs $f(\vJ)$ to a chemodynamical model;
Section~\ref{sec:DFcomponents} defines the functional forms we have assumed
for $f(\vJ)$; Sections~\ref{sec:initial_params} to \ref{sec:min_lnL} specify
the values of the model parameters from which data-fitting commenced, explain
our strategy for dealing with the survey's selection function, and define the
likelihood that the search maximises and the maximising procedure.
Section~\ref{sec:results} describes the model fitted to the data and
discusses the quality of the fit it provides.  Section~\ref{sec:Besancon}
compares the scope of the present model to that of the widely used Besan\c con
model, and Section~\ref{sec:conclude}
sums up and identifies next steps.  Appendix~\ref{app:validate} validates our
procedure for maximising a likelihood.

\section{The data}\label{sec:data}

From the 17th data release of the Sloan Digital Sky Survey \citep{APOGEE17}
we selected data for stars that have Gaia DR3 astrometry
\citep{GaiaEDR3general,GaiaDR3general} and parameters from the StarHorse
catalogue \citep{StarHorse,StarHorse2,StarHorse3}. We removed stars with a probability $>0.5$ of
belonging to a globular cluster according to \cite{VasilievBaumgardt2021},
stars with the {\tt STAR\_WARN} bit (7) of the {\tt ASPCAP\_FLAG} set, and
stars with a StarHorse distance with an uncertainty larger than $0.75\kpc$.
We further required the astrometric `fidelity' parameter of
\cite{Rybizki2022} to exceed $0.5$.  Finally, the sample was restricted to
giants by requiring $\log g<3.5$ and $T_{\rm eff}<5500$. We convert
heliocentric data to galactocentric coordinates assuming the Sun's
phase-space coordinates are $(R,z)=(8.27,0.025)\kpc$ \citep{Gravity2022} and
$(V_R,V_z,V_\phi)=(14,7,251)\kms$ \citep{Schoenrich2012,ReidBrunthaler2020}.

\begin{figure}
\centerline{\includegraphics[width=.8\hsize]{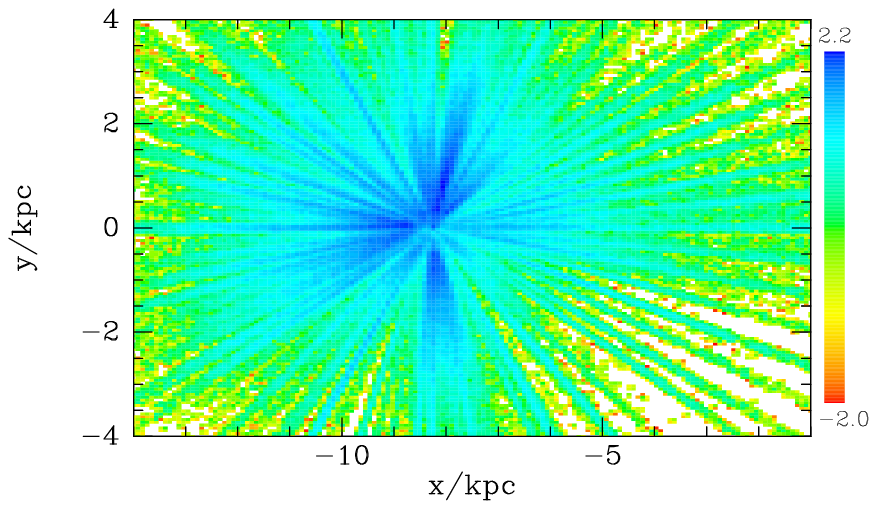}}
\centerline{\includegraphics[width=.8\hsize]{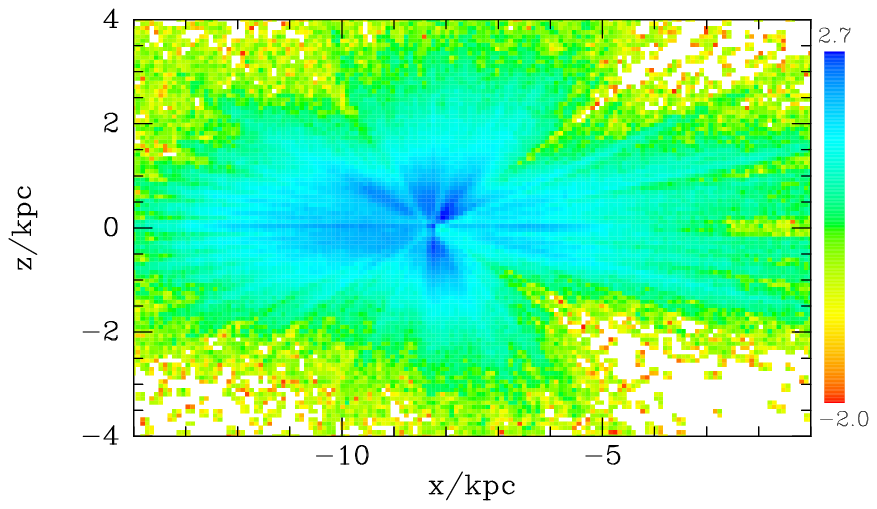}}
\caption{The spatial distribution of the stellar sample: projections along
the $z$ axis (upper panel) and the $y$ axis (lower panel). The colour scale
is (base 10) logarithmic.}\label{fig:space}
\end{figure}

Fig.~\ref{fig:space} shows the spatial distribution of the sample's stars
projected onto the $xy$ plane and $xz$ planes in the upper and lower panels,
respectively.  The upper panel shows very clearly the rapid variation of the
selection function inevitable in a pencil-beam spectroscopic survey. Also
evident is the strong bias towards the Sun inevitable in a magnitude-limited
survey.  Notwithstanding these regrettable characteristics, the figure shows
that the survey provides good coverage of the Sun's side of the Galaxy below
$|z|\sim4\kpc$ in the radial range $1\kpc\le R\le 14\kpc$.

\begin{figure}
\includegraphics[width=\hsize]{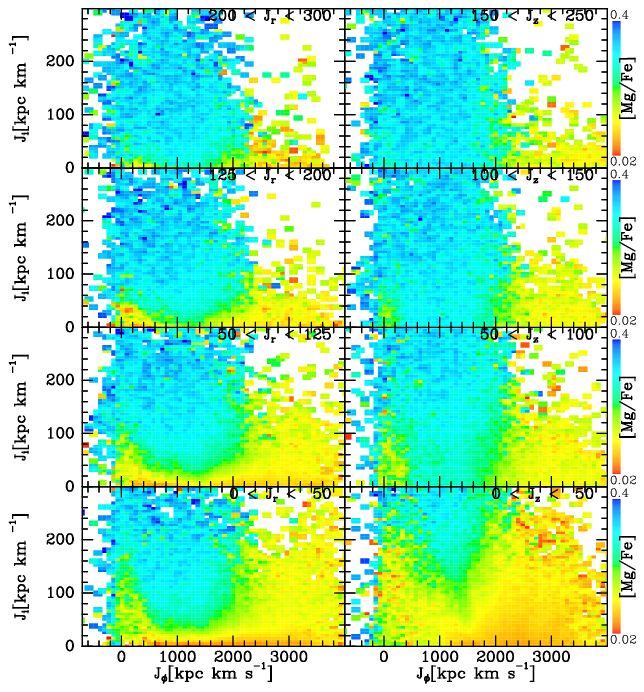}
 \caption{Mean values of [Mg/Fe] of stars as a function of position in action
space. Values are shown for four bands in $J_r$ (left column) and $J_z$ as
marked in the upper right of each panel.} \label{fig:LzFeBasic}
\end{figure}

\begin{figure}
\includegraphics[width=\hsize]{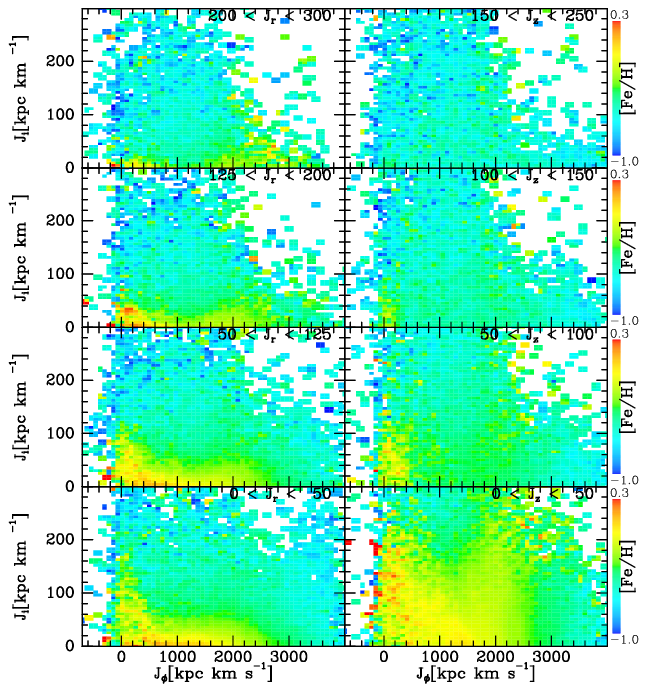}
 \caption{Mean values of [Fe/H] of stars as a function of position in action
space.  Values are shown for four bands in $J_r$) (left column) and $J_z$ as
marked in the upper right of each panel.} \label{fig:LzFeBasicFe}
\end{figure}

\begin{figure}
\centerline{\includegraphics[width=.95\hsize]{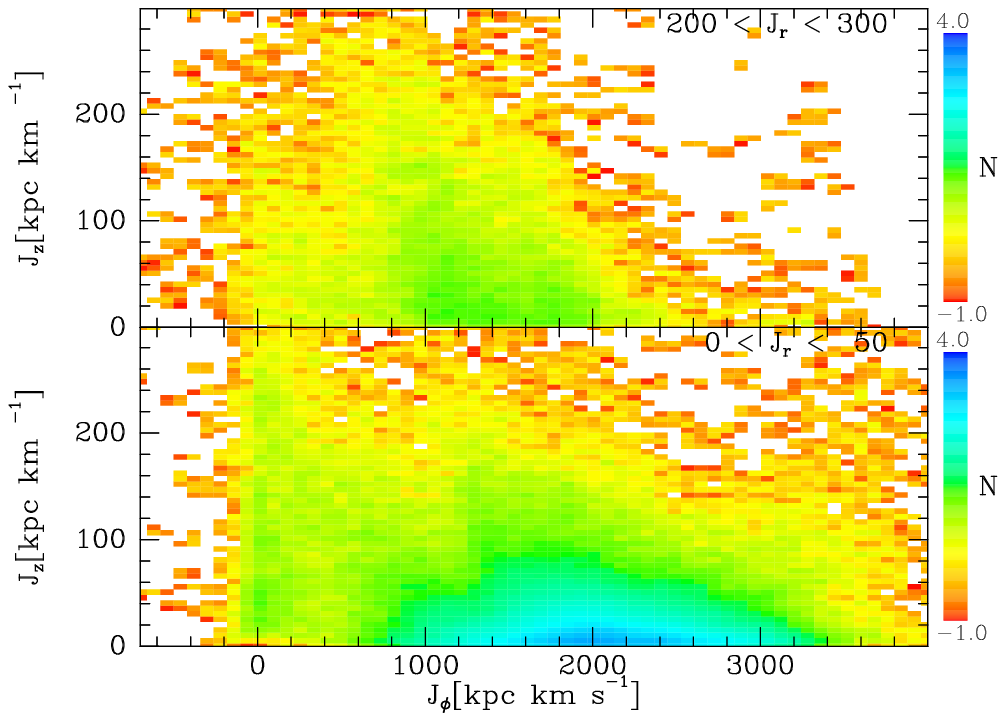}}
\caption{The number of stars in each cell shown in the top and bottom panels 
of the left columns of Figs.~\ref{fig:LzFeBasic} and \ref{fig:LzFeBasicFe}.} \label{fig:LzFeN}
\end{figure}

\section{Action-space chemistry}\label{sec:AAchem}

We computed the giants' phase-space coordinates and from them computed the
actions $J_r$, $J_z$ and $J_\phi$ in the gravitational potential of the
self-consistent Galaxy model that is presented in Section \ref{sec:results}
below.  In this potential $217\,863$ stars are bound and 97 unbound; actions
cannot be computed for unbound stars, so these stars were eliminated from the
sample.

\subsection{Mean values of [Fe/H] and [Mg/H]}
 
The panels of Fig.~\ref{fig:LzFeBasic} show the mean value of [Mg/Fe] in
cells in action space, while Fig.~\ref{fig:LzFeBasicFe} shows mean values of
[Fe/H]. The left columns show projections onto the $(J_\phi,J_z)$ plane
grouped by their values of $J_r$, while the right columns show projections
onto the $(J_\phi,J_r)$ plane grouped by values of $J_z$. In the left columns
the values of $J_r$ increase from the bottom panel upwards, while in the
right columns the values of $J_z$ increase upwards  -- the relevant range in $J_r$ or $J_z$ is shown at
top-right of each panel.

Stars on orbits that are either nearly circular or in the plane contribute,
respectively, to the bottom left or right pair of panels in each figure,
while stars on orbits that are either highly eccentric or highly inclined
contribute to the top pair of panels.

Fig.~\ref{fig:LzFeN} shows the number of stars contributing to the bottom and
top left panels of Figs.~\ref{fig:LzFeBasic} and \ref{fig:LzFeBasicFe}. These
numbers are strongly influenced by APOGEE's selection function.  In
particular, the lower panel for low $J_r$ shows a strong concentration around
the Sun's location.  The plots of mean [Mg/Fe] and [Fe/H] in
Figs.~\ref{fig:LzFeBasic} and \ref{fig:LzFeBasicFe} show no sign of this
bias.

Note the extraordinarily wide coverage in $J_\phi$ -- at low $J_z$ there are
stars with $J_\phi$ down to zero from a maximum value that exceeds twice the
Sun's value of $J_\phi$ ($\sim2000\H$). This wide coverage of $J_\phi$ at
small $J_r,J_z$ is possible because APOGEE extends to low Galactic latitude
$b$ and covers an unprecedentedly wide range in Galactic radius $R$, so it
includes stars on near circular orbits at a wide range of radii.

In every panel of Figs.~\ref{fig:LzFeBasic} to \ref{fig:LzFeN}
the populated region has a sharp left edge that lies just to the left of the
line $J_\phi=0$. If stars were in fact strictly confined to $J_\phi>0$,
observational errors would still cause some stars to scatter to $J_\phi<0$.
Indeed, a small over-estimation of the distance $s$ to a star at
$\ell\simeq0$ and $R\ll R_0$ can move a star from the near to the far side of
the Galactic centre without much effect on its apparent velocity, and thus
reverse the sign of its measured $J_\phi$. The sharpness of the left
boundaries of the populated regions in Figs.~\ref{fig:LzFeBasic} to
\ref{fig:LzFeN} attest to the accuracy of the StarHorse distances used
to compute $\vJ$. 

The near total confinement of stars to $J_\phi\ge0$ is a clear indication
that the overwhelming majority of stars were born in a disc within our Galaxy
-- stars accreted from other galaxies would end up on orbits with both signs
of $J_\phi$ with roughly equal probability. A significant part of the stellar
halo is thought to comprise such accreted stars, and as a consequence the
halo shows negligible net rotation. The stellar halo should be dominant at
small $J_\phi$ and significant $J_r$ and/or $J_z$. Hence the sharpness of the
boundary at $J_\phi=0$ in all the panels of Figs.~\ref{fig:LzFeBasic} to
\ref{fig:LzFeN} implies that the stellar halo contributes rather few stars to
the sample. It may well be that halo stars, being metal-poor and thus
weak-lined, have trouble passing our spectroscopic quality cuts.
{\rd Moreover, our requirement that a star's distance uncertainty should not
exceed $0.75\kpc$ will have removed many distant halo stars.  A better
requirement might be upper limits on the uncertainties in $\ln R$ and $\ln
|z|$ because what really matters is the fractional uncertainties in $R$ and
$|z|$ not the absolute uncertainty in distance.}

There are several remarkable features of  Fig.~\ref{fig:LzFeBasic}:
\begin{itemize}

\item Along the $J_\phi$ axis of the bottom left panel of
Fig.~\ref{fig:LzFeBasic} there is a narrow orange region indicative of solar
[Mg/Fe]. This is the low-$\alpha$ disc. Above it a blue-shaded region of high
[Mg/Fe] fills the interior of a U centred on $J_\phi\sim1200\H$. To
left and right as well as below, this region transitions sharply to yellow
shades indicative of lower [Mg/Fe]. The blue region is part of the
high-$\alpha$ disc.

\item In this bottom-left panel, to the right of $J_\phi\simeq2000\H$
yellow shades extend to the highest populated values of $J_z\sim150\H$.
This phenomenon clearly shows that low-$\alpha$ stars, even ones on
significantly inclined orbits,  dominate at $J_\phi\ga2000\H$. It
implies that the high-$\alpha$ disc has a remarkably
sharp outer edge at roughly $R_0$.

\item As one proceeds up the left column of Fig.~\ref{fig:LzFeBasic} through
samples of stars with increasing $J_r$, the orange/yellow region of the thin,
low-$\alpha$ disc yields ground to the blue high-$\alpha$ region. This
phenomenon indicates that the high-$\alpha$ disc extends down to $J_z=0$; at
low $J_r$ it is overwhelmed by the low-$\alpha$ disc, but comes to the fore
as $J_r$ increases because its DF decreases less rapidly with increasing
$J_r$.

\item Similar trends are evident as one proceeds up the right column of
Fig.~\ref{fig:LzFeBasic} through  samples of stars with increasing $J_z$,
except that in the range $500\H<J_\phi<2000\kms\kpc$ the yellow shades
of low-$\alpha$ stars recede more rapidly:  in fact they have already
disappeared from the sample with $50\H<J_z<100\H$. This
phenomenon indicates that the main body of the  low-$\alpha$ disc covers a
wider range in $J_r$ than in $J_z$, as is natural in a `thin' disc.

\item The V of green colours that pushes down at $J_\phi\simeq1500\H$
in the bottom right panel of Fig.~\ref{fig:LzFeBasic} suggests that within
the low-$\alpha$ disc $\ex{J_r}$ has a minimum just interior to the Sun.

\item Whereas in the bottom left panel of Fig.~\ref{fig:LzFeBasic} brown
shades are confined to a narrow band above the $J_\phi$ axis, in the bottom-right
panel at $J_\phi>2000\H$ they extend to $J_r\ga50\H$. This
indicates that the outer low-$\alpha$ disc is radially hot.

\item In the  top two panels of the right column of
Fig.~\ref{fig:LzFeBasic}, blue colours indicative of high-$\alpha$ extend
right down to the $J_\phi$ axis except at very low $|J_\phi|$ and
$J_\phi>2200\H$. This indicates that at $J_z\ga100\H$ the
high-$\alpha$ disc dominates outside the bulge and the outer disc.
\end{itemize}

Fig.~\ref{fig:LzFeBasicFe} shows mean values of [Fe/H] in the format used to
display [Mg/Fe] in Fig.~\ref{fig:LzFeBasic}. Blue shades now imply low
metallicity, so the high-$\alpha$ disc with $\ex{\hbox{[Fe/H]}}\sim-0.7$ is
coloured cyan, wile the metal-rich, inner low-$\alpha$ disc is coloured
yellow. With this colour scheme very similar patterns are observed in
Fig.~\ref{fig:LzFeBasicFe} to those discussed above for
Fig.~\ref{fig:LzFeBasic}: high-$\alpha$ often implies low-metallicity. The
most striking difference occurs in the region $J_\phi>2000\H$: in
Fig.~\ref{fig:LzFeBasicFe} this region is largely green because the outer
low-$\alpha$ disc is metal poor. However in the bottom right panel of
Fig.~\ref{fig:LzFeBasicFe} a tongue of yellow is evident at $J_\phi\sim2000\H$
that signals that the low-$\alpha$ disc becomes radially excited before it
turns metal-poor.

The bottom-left panel of Fig.~\ref{fig:LzFeBasicFe} (for $J_r<50\H$) shows
[Fe/H$]>0$ is confined to small $J_z$ except at small $|J_\phi|$. The
bottom-right panel shows that at small $J_z$ and $|J_\phi|$, [Fe/H$]>0$
occurs predominantly at large $J_r$. These facts are consistent with the idea
that metal-rich stars populate the kind of tightly bound, eccentric,
low-inclination orbits that form the backbone of the bar. In the bottom right
panel of Fig.~\ref{fig:LzFeBasicFe} (for $0<J_z<50\H$) in the range
$1400\H<J_\phi<2200\H$ a yellow-brown ridge reaches upwards: this is the
signature of a vertically thin annulus of fairly metal-rich stars on
eccentric orbits.

{\rd Figs.~\ref{fig:LzFeBasic} and \ref{fig:LzFeBasicFe} bear comparison with
Fig.~31 of \cite{Recio-BlancoGaia2023}, which shows the distribution of more
than 5 million stars in the $(J_\phi,J_r)$ plane. The chemistry of these
stars was determined from Gaia radial velocity spectra, and  in four of its panels,
cells are coloured by median [M/H] or median  [$\alpha$/Fe]. The left
panels plot over a wide range in $(J_\phi,J_r)$ while the right panels zoom
into the most densely populated part of the plane :
$1000\kpc\kms<J_\phi<3000\kpc\kms$ and
$J_r<100\kpc\kms$. One of the zoomed panels shows steep drops in [M/Fe]
along two lines that slope up and to the left, intersecting the $J_\phi$ axis
at $J_\phi\simeq2000$ and $2130\kpc\kms$. One of these features is probably
related to the drop apparent in the lower-right panels of
Fig~\ref{fig:LzFeBasicFe} but the correspondence is not clear. The
bottom-right panel of the figure in the Gaia paper shows a
decline in [$\alpha$/Fe] around $J_\phi=2000\kpc\kms$, but again the line of
steepest descent slopes up and to the left rather than running straight up, and it
isn't as sharp as that seen in Fig.~\ref{fig:LzFeBasic} (and in
Fig.~\ref{fig:global} below). In this connection, note that the Gaia paper
tracks [$\alpha$/Fe] through lines of calcium rather than magnesium. }

\begin{figure}
\centerline{\includegraphics[width=.9\hsize]{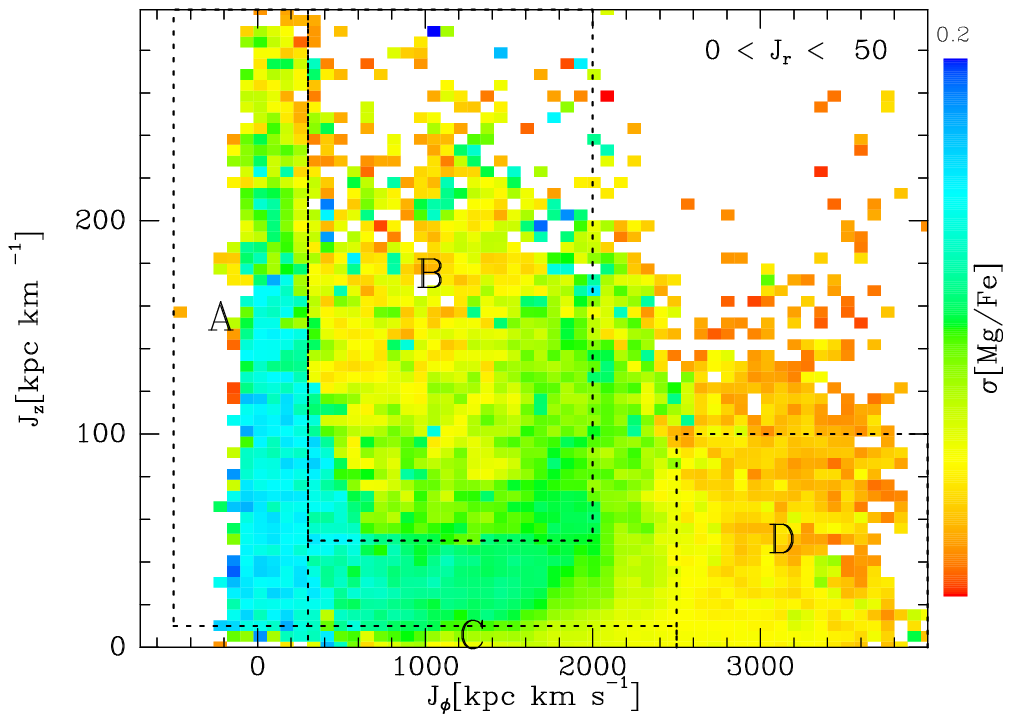}}
\caption{The standard deviation in [Mg/Fe] in a slice of action space. The
dotted lines mark the regions for which chemical pdfs are presented in
Fig.~\ref{fig:LzFePatch}.}\label{fig:rects}
\end{figure}

\begin{figure*}
\centerline{\includegraphics[width=.33\hsize]{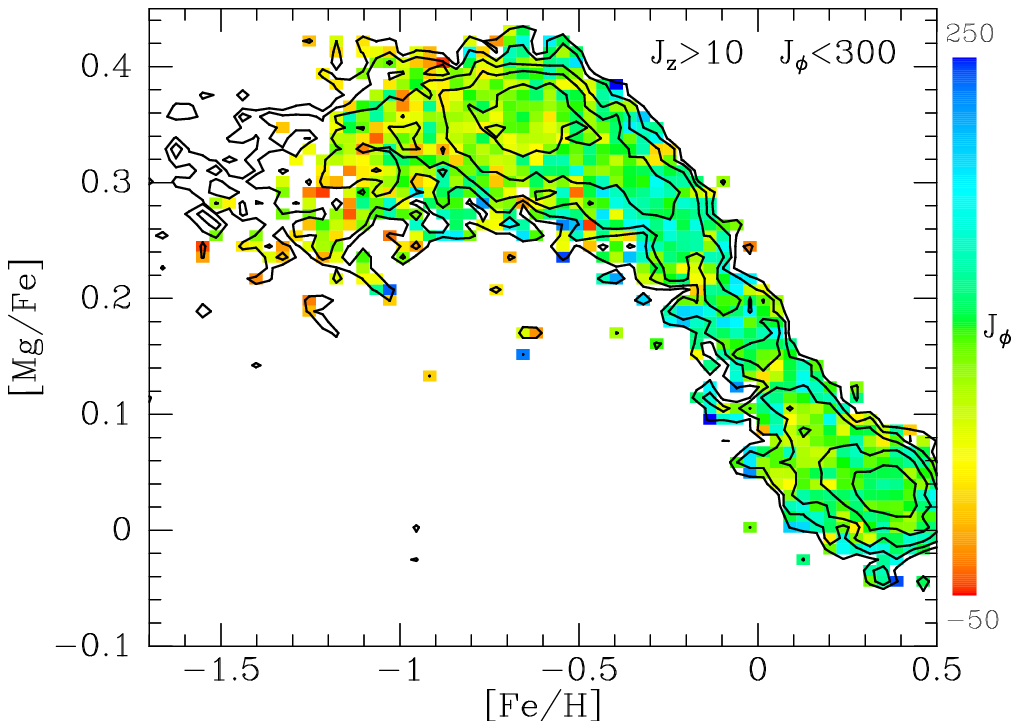}
\includegraphics[width=.33\hsize]{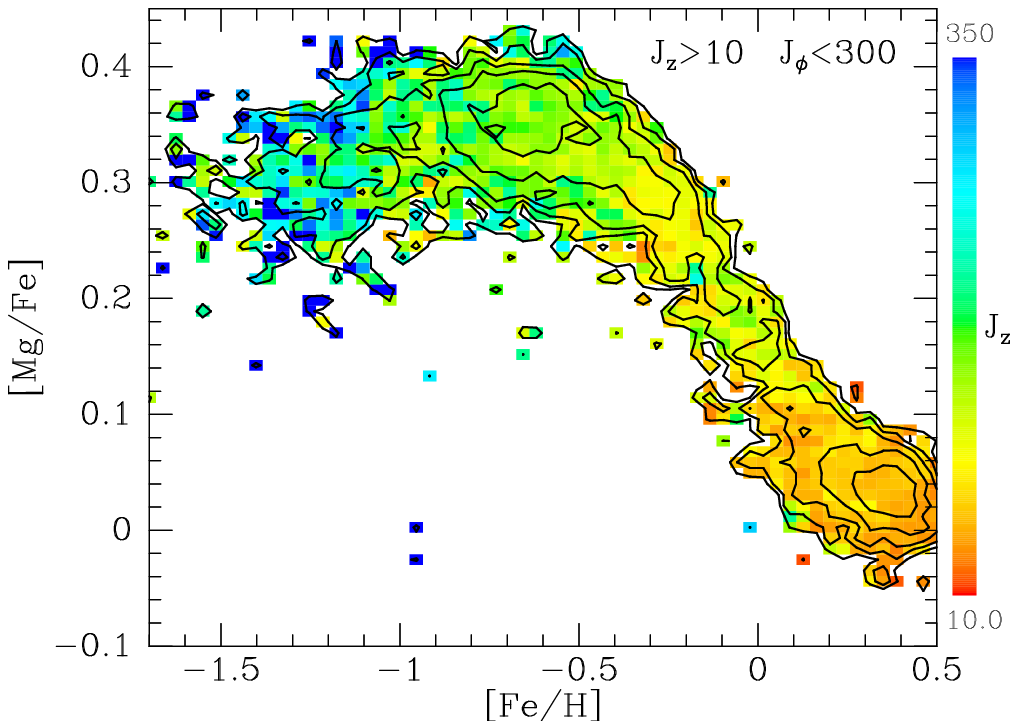}\includegraphics[width=.33\hsize]{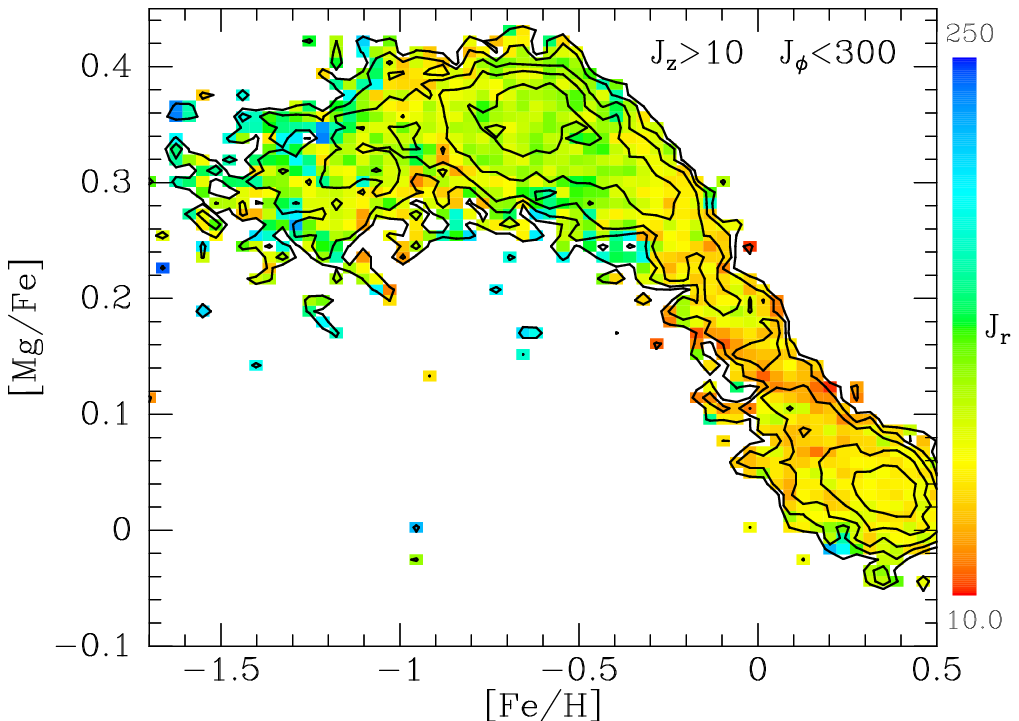}}
\centerline{\includegraphics[width=.33\hsize]{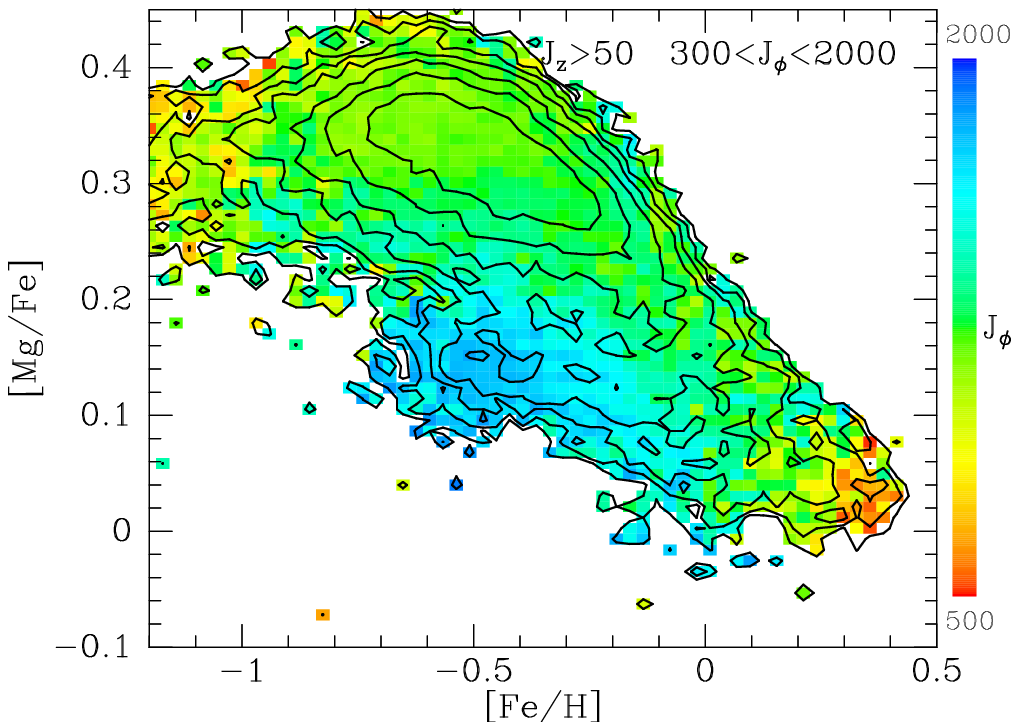}
\includegraphics[width=.33\hsize]{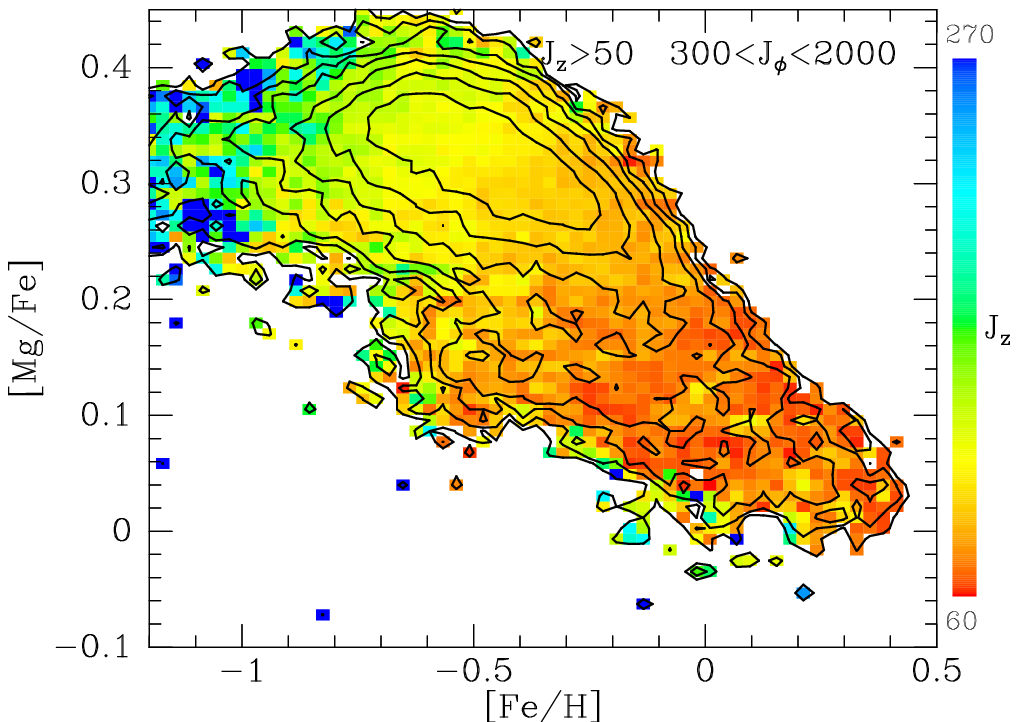}\includegraphics[width=.33\hsize]{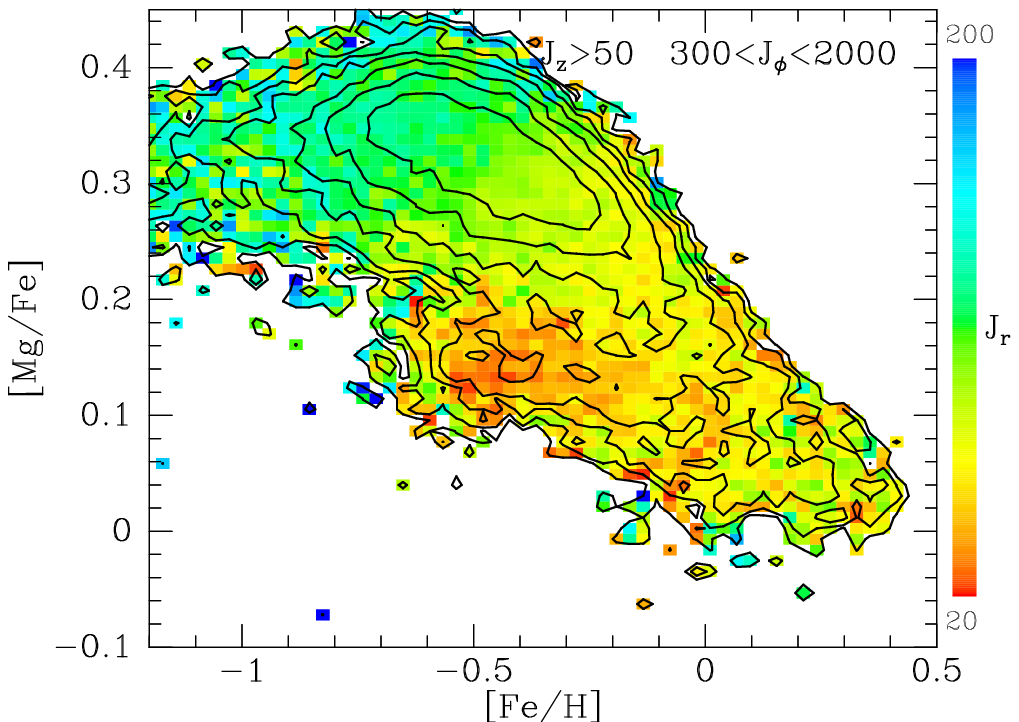}}
\centerline{\includegraphics[width=.33\hsize]{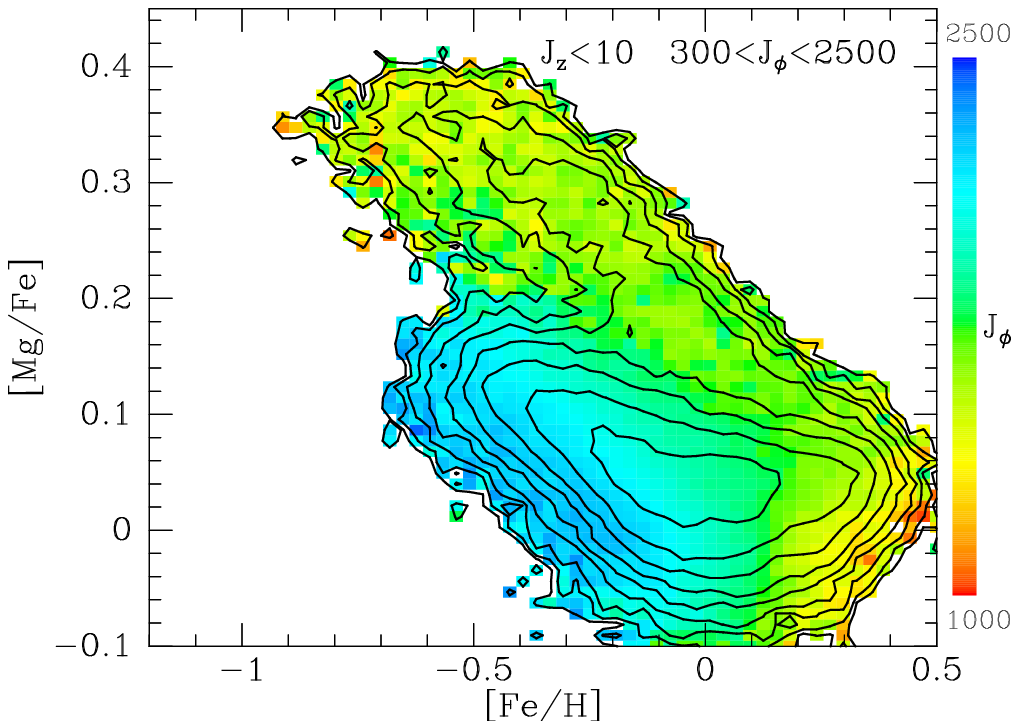}
\includegraphics[width=.33\hsize]{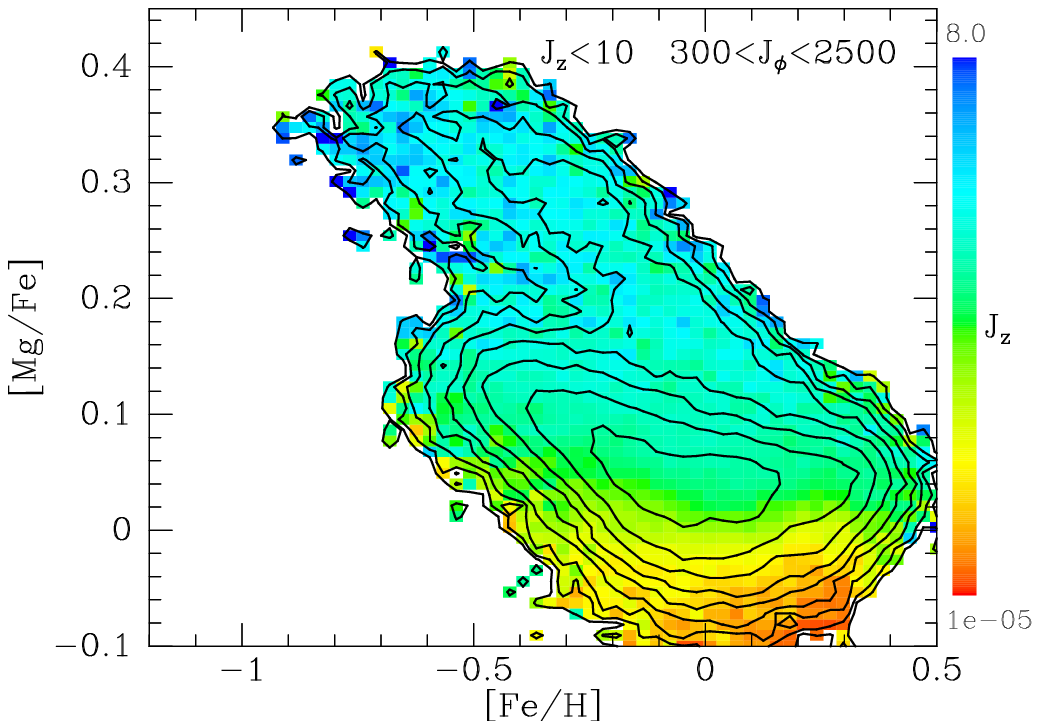}\includegraphics[width=.33\hsize]{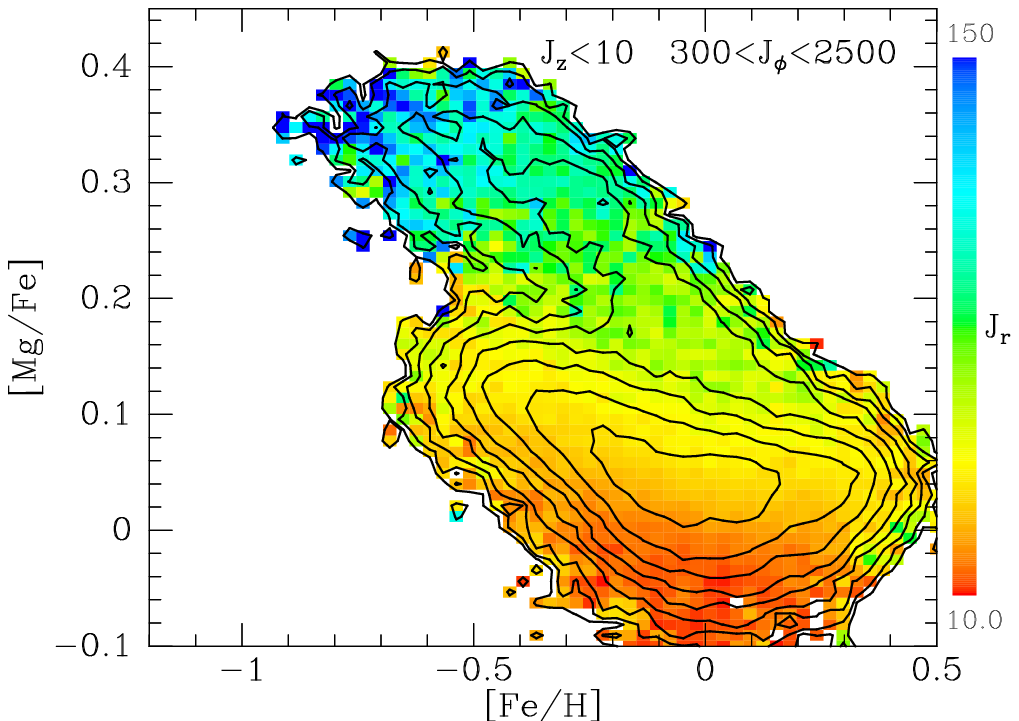}}
\centerline{\includegraphics[width=.33\hsize]{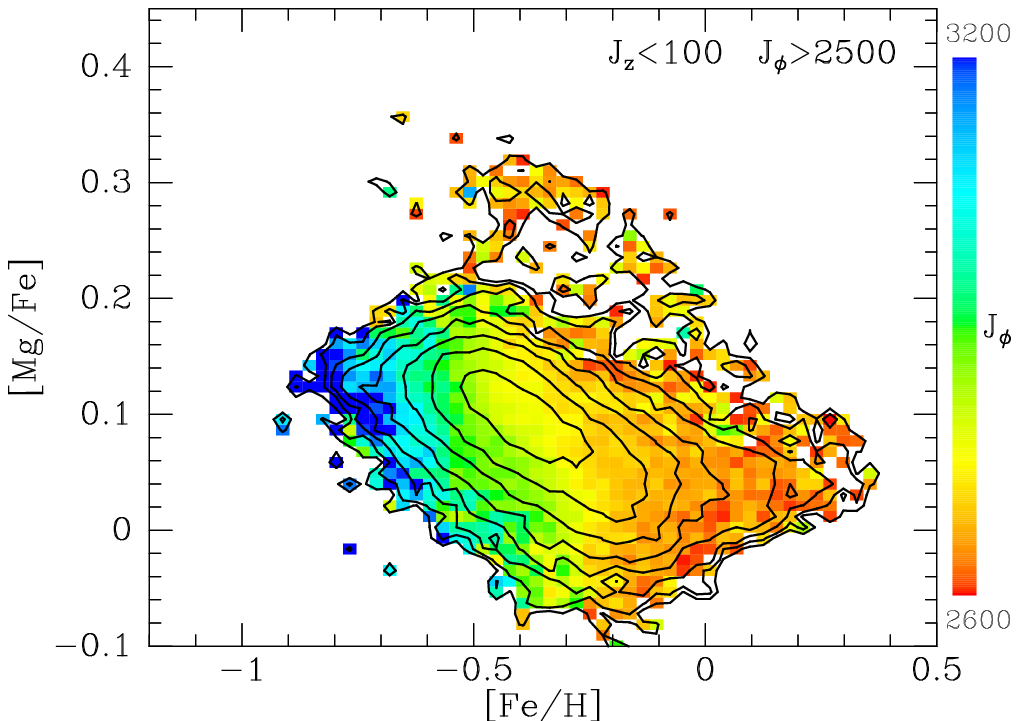}
\includegraphics[width=.33\hsize]{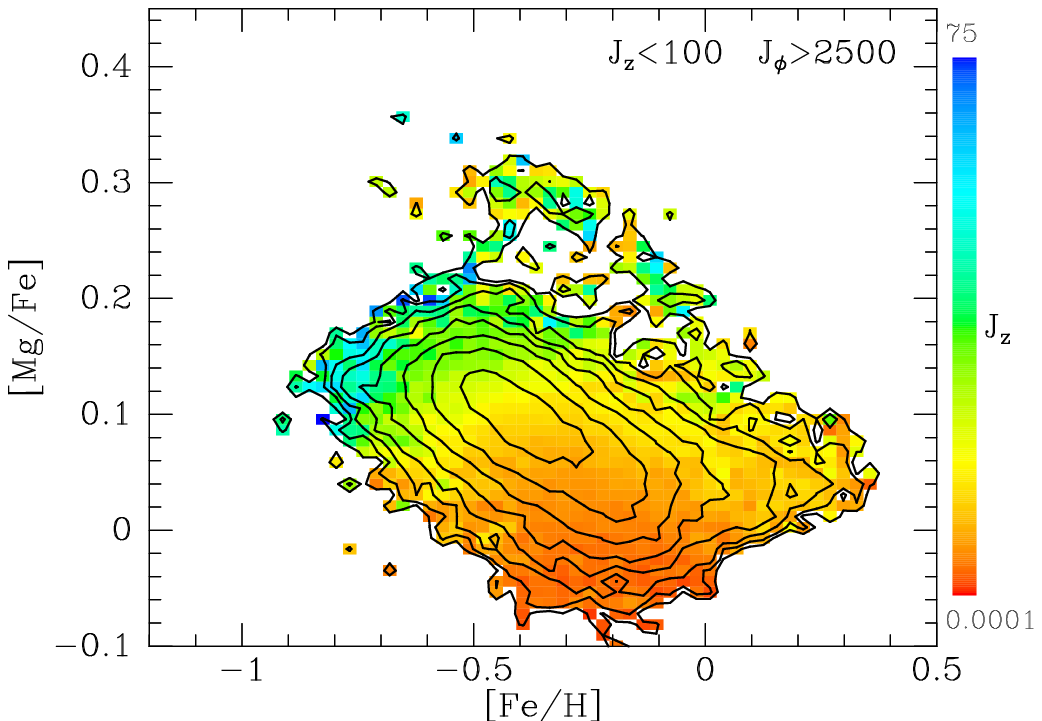}\includegraphics[width=.33\hsize]{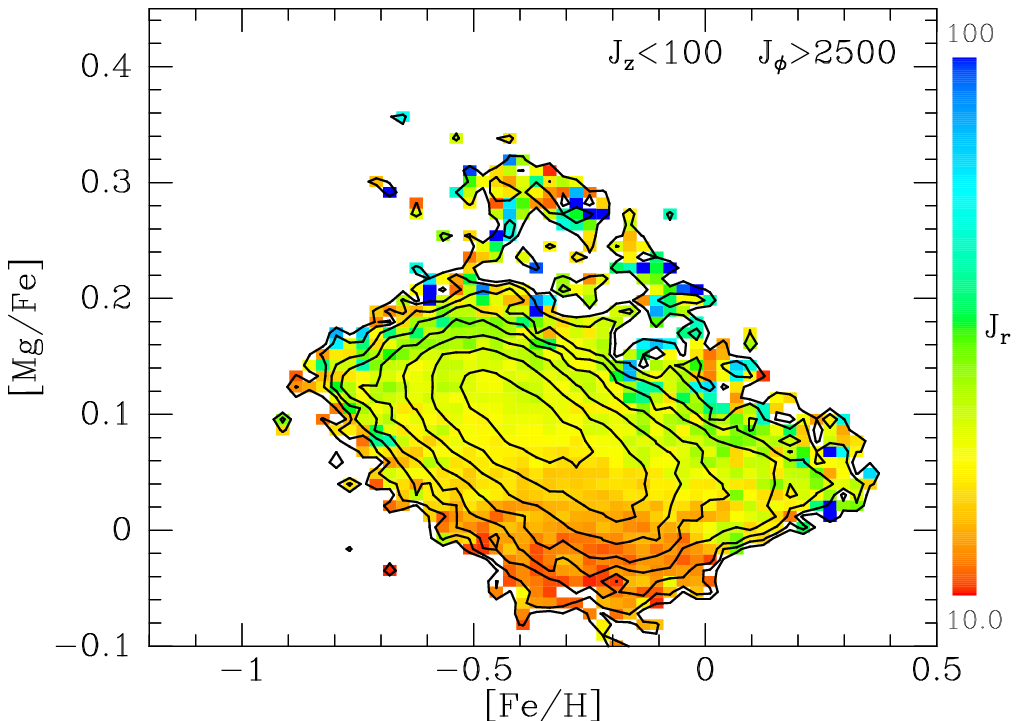}}
\caption{Contours show the chemistry of the four regions A--D marked by dotted
lines in Fig.~\ref{fig:rects}. The top row is for
the bulge/halo region A around the $J_z$ axis. The second row is for the
high-$\alpha$ region B. The third row is for the
thin-disc  region C just above the $J_\phi$ axis, and the bottom row is for the outer, flaring
low-$\alpha$ disc, D. The panels are coloured by mean $J_\phi$, $J_z$ and
$J_r$ in the left, centre and right columns, respectively. Note that while
the [Fe/H] axis covers the same range $(-1.2,0.5)$ in the lower three rows,
in the top row it extends to metal-poorer stars: [Fe/H$]=-1.7$. 
}\label{fig:LzFePatch}
\end{figure*}

\subsection{The ([Fe/H], [Mg/Fe]) plane}\label{sec:FeMgplane}

Figs.~\ref{fig:LzFeBasic} and \ref{fig:LzFeBasicFe} show just mean values of
[Mg/Fe] and [Fe/H].  To gain further insight we now explore distributions in
the ([Fe/H],[Mg/Fe]) plane for the four regions of the $(J_\phi,J_z)$ plane
that are marked in
Fig.~\ref{fig:rects} over a plot of the standard deviation in [Mg/Fe] in the
slice of action space with smallest $J_r$. The standard deviation tends to be
large where components with differing chemistry overlap. 

The first of these regions, labelled A, lies along the $J_z$ axis at
$J_z>10\H$ and $J_\phi<300\H$. Stars in this region are on highly
inclined orbits, so they dominate the stellar density near the $z$ axis. We
call this the bulge/halo region. 
The second region, B, occupies the heart of the high-$\alpha$ region of
action-space -- $J_z>50\H$ and $300<J_\phi<2000\H$. 
The third region, labelled C, lies just
above the $J_\phi$ axis at $J_\phi>300\H$. Stars in this region are on
nearly planar orbits, so will be predominantly thin-disc
stars. The fourth
region, D, at $J_\phi>2500\H$, is dominated by the outer, flaring disc.

The contours in Fig.~\ref{fig:LzFePatch} show the chemical structure of these
four regions: the bulge/halo, high-$\alpha$ disc, thin-disc and outer disc
regions are depicted by rows from top to bottom. In the left, centre and
right columns the colours show mean values of $J_\phi$, $J_z$ and $J_r$,
respectively. It should be borne in mind that the stellar densities
contoured, unlike the mean values plotted in Fig.~\ref{fig:LzFeBasic},
reflect the APOGEE selection function. Most significantly, stars near the Sun
are over-represented.

\begin{itemize}
\item The top row of Fig.~\ref{fig:LzFePatch} shows that the bulge/halo
region is indeed a
superposition of two components, one centred on
([Fe/H],[Mg/Fe]$)\simeq(-0.65,0.35)$ and the other on
([Fe/H],[Mg/Fe]$)\simeq(0.36,0.03)$. The first population has a tail that
extends to [Fe/H$]\sim-1$ at slightly lower [Mg/Fe].  It is probably a
mixture of the high-$\alpha$ disc and the stellar halo with the former
dominating. The metal-rich, low-$\alpha$ component is presumably the bulge.
The mean values of $J_\phi$ and $J_r$ shown by the colours vary little with
chemistry, although there is a marginal tendency for $\ex{J_\phi}$ to
decrease with [Fe/H] and to be negative at [Fe/H]$<-1$.  The central panel
shows that $\ex{J_z}$ increases systematically with decreasing [Fe/H]. That
is, $\ex{J_z}$ is lowest in the bulge, highest in the halo and intermediate
in the high-$\alpha$ disc. 

\item The second row in Fig.~\ref{fig:LzFePatch} shows the chemistry of the
high-$\alpha$ region. The dominant component is centred on
([Fe/H],[Mg/Fe]$)\simeq(-0.49,0.33)$, with a tail sloping down to solar
chemistry (0,0) and beyond. Within the dominant, high-$\alpha$ component, the
left panel shows that any
gradient in $\ex{J_\phi}$ is weak, although at [Fe/H]$<-1$ there is a slight
tendency for $\ex{J_\phi}$ to decrease with [Fe/H].
The central panel shows
a systematic increase in $\ex{J_z}$ with decreasing [Fe/H] within the
high-$\alpha$ component but no evidence of an analogous gradient in the
low-$\alpha$ component. 

\item The thin-disc region shown in the third row of
Fig.~\ref{fig:LzFePatch} comprises a dominant component centred on
([Fe/H],[Mg/Fe]$)\simeq(-0.06,0.036)$ that must be the main body of the
low-$\alpha$ disc. On top we see a significant tail
from the high-$\alpha$ disc. Thus the high-$\alpha$ disc makes a significant
contribution to the thin-disc region.  The right panel of the third row
shows that in the dominant component, $\ex{J_r}$ increases with [Mg/Fe] --
this is likely the signature of stochastic heating: stars with larger [Mg/Fe]
are older and kinematically hotter. Another effect of age increasing with
[Mg/Fe] is the widening spread in [Fe/H] with increasing [Mg/Fe]: older stars
have migrated further with the consequence that in the over-represented
solar-neighbourhood group an older population displays a wider spread in
[Fe/H].  The colours in the left panel of the third row indicate that the
dominant component has a clear gradient in $\ex{J_\phi}$, which must be a
manifestation of the familiar metallicity gradient in the disc
\citep[e.g.][and references therin]{MendezDelgado2022}. There is no
sign of an analogous gradient in the high-$\alpha$ component.  Surprisingly,
the third row shows a tendency for $\ex{J_z}$ to increase with [Mg/Fe] --
given that stars contributing to this row have by construction
$J_z<10\H$, it is puzzling to observe significant variation of
$\ex{J_z}$. The signal is unmistakable nonetheless.  Stars with exceptionally
low $J_z$ must lie very close to the plane, so they are either very close to
the Sun or are significantly extincted. Could high extinction artificially
lower their measured values of [Mg/Fe]?

Taken together the second and third rows confirm the
presence of two populations that coexist at many points in action space.  The
low-$\alpha$ population dominates at low $J_z$ and the high-$\alpha$
population dominates at high $J_z$. 

The first and second rows show that stars metal-poorer than
$\hbox{[Fe/H]}\simeq-0.8$ are only encountered at [Mg/Fe$]\ga0.2$. Moreover,
these stars are confined to $J_z\ga200\H$ and $J_r\simeq100\H$.

\item The bottom panels of  Fig.~\ref{fig:LzFePatch} show the chemistry of
the  outer, flaring disc. It is almost entirely accounted for by a single
population centred on ([Fe/H],[Mg/Fe]$)\simeq(-0.4,0.1)$. This low-$\alpha$ population
shows the gradient in $\ex{J_\phi}$ that we encountered in the inner
low-$\alpha$ disc, so we have every reason to suppose that this population is
just an extension of the dominant component in the third row (the
low-$\alpha$ disc). Moreover, the central and right panels of the fourth row
show the same trends in $\ex{J_z}$ and $\ex{J_r}$ we encountered in the
third row, and attributed to increasing age and stochastic heating with
[Mg/Fe].

\end{itemize}

Perhaps the most striking aspects of Fig.~\ref{fig:LzFePatch} are two `dogs
that didn't bark': (i) the absence of stars at low-metallicity and
low-$\alpha$ in the top row (bulge/halo region) and (ii) the complete
disappearance of the high-$\alpha$ population between the second and fourth
rows (high-$\alpha$ and outer disc regions). Another notable result is the
contrast between the strong gradients in $\ex{J_\phi}$, $\ex{J_z}$ and
$\ex{J_r}$ in the low-$\alpha$ disc and the extremely weak gradients in mean
actions in the high-$\alpha$ component.

Fig.~\ref{fig:low-high} reveals key differences in how low- and high-$\alpha$
stars are distributed in action space by plotting projections of the sample
onto the $(J_\phi,J_z)$ plane in the upper, and the $(J_\phi,J_r)$ plane in
the lower, pairs of panels. The absence of high-$\alpha$ stars at
$J_\phi>3000\H$ is striking, as is the extent to which the low-$\alpha$ stars
extend to high $J_z$ at both small and large $J_\phi$ while being restricted
to lower $J_z$ at intermediate $J_\phi$.  In $J_r$, the pattern of the
low-$\alpha$ stars is very different: the populated region reaches highest in
$J_r$ at intermediate $J_\phi$ even though stars with large
$|J_\phi-J_{\phi\odot}|$ can reach the Sun, and thus boost their chances of
entering the sample, only if they have large $J_r$.  The wide spread in $J_r$
at $J_\phi\simeq J_{\phi\odot}$ may be the result of resonant scattering by
spirals, while the wide spread in $J_z$ at large $J_\phi$ might be a legacy
of tidal interactions with objects in the dark halo.

{\rd It is worth noting how much the bottom panel of Fig.~\ref{fig:low-high}
differs from similar plots of stellar density in the $(J_\phi,J_r)$ plane for stars
near the Sun \citep[e.g.][]{HuntBubBovy2019,TrickRix2019}: here we see
no lines associated with resonances. To see these features, the sample must be
restricted to a narrow band in radius or distance from the Sun because (as
Fig.~\ref{fig:low-high}  demonstrates)
resonant orbits are not more- or less-populated than neighbouring
non-resonant orbits. They nonetheless appear to be so if we impose a
sufficient selection in angle variables $\vtheta$ because resonant terms in
the Hamiltonian shuffle stars between the actions that are the axes of the
plots. This motion is correlated with $\vtheta$, so at some angles the
density of stars has been enhanced at particular actions, while at other
angles the density at these actions has been diminished. A sample that is
strongly selected in radius or distance from the Sun is biased in both
$\theta_r$ and $\theta_\phi$ and thus reveals the effects of perturbing terms
in a way that the broadly selected sample of Fig.~\ref{fig:low-high} does not.

The sample of stars with high-quality RVS spectra analysed by
\cite{Recio-BlancoGaia2023} covers quite a wide range in $R$ and distance
from the Sun but it is much more concentrated around the Sun that the APOGEE
sample. This may be why it shows ridges in the $(J_\phi,J_r)$. As the Gaia
paper  remarks, it is not evident that its ridges
are the same as those presented by \cite{HuntBubBovy2019} and
\cite{TrickRix2019}. The ridges of the Gaia paper are most evident in its
plot of median values of $z_{\rm max}$, the greatest distance a star moves
from the plane.  Fig.~\ref{fig:Jzbar} is a plot of the mean value of $J_z$
that ought to be closely related to a plot of mean values of $z_{\rm max}$.
Any ridges analogous to those in the Gaia paper are extremely indistinct.}

\section{Modelling the data}\label{sec:modelling}

We now turn to the construction of a chemodynamical model of our Galaxy that
reproduces as closely as possible the trends discovered in the last section.

\subsection{EDF structure}\label{sec:EDFstructure}

\cite{SaJJB15:EDF} introduced the concept of an {\it extended distribution
function} (EDF), that is a density of stars in the five-dimensional space spanned
by $\vJ$ and chemistry,
\[
\vc\equiv\big(\hbox{[Fe/H],\,[Mg/Fe]}\big).
\]
An excellent way of summarising the chemodynamical structure of our Galaxy
would be to decompose it into components that individually have simple EDFs.
Specifically, we seek components that have analytic dynamical DFs $f(\vJ)$
that are extended to EDFs by multiplication by an analytic probability
density $P(\vc|\vJ)$ that gives the probability that a star with actions
$\vJ$ has the chemistry $\vc$.  {\it Any} EDF $F(\vc,\vJ)$ can be written as
a product $f(\vJ)P(\vc|\vJ)$ -- simply define
$f(\vJ)\equiv\int\d^2\vc\,F(\vc,\vJ)$ and $P(\vc|\vJ)\equiv
F(\vc,\vJ)/f(\vJ)$ and is trivial to show that $\int\d^2\vc\,P=1$ --
but we want to define our components so 
$P(\vc|\vJ)$ is analytic. 

{\rd Non-negativity is a fundamental property of a probability density function
(pdf), and multi-Gaussian expansions are widely used to approximate
non-negative functions analytically. In this spirit we assume that
$P(\vc|\vJ)$ can  be approximated by a Gaussian distribution in $\vc$ with
mean and dispersion depending on $\vJ$.}

The general Gaussian two-dimensional probability density can be written
\[\label{eq:PGauss}
P(\vc|\vJ)={\surd\det(\vK)\over{2\pi}}\exp\left(\fracj12(\vc-\vc_\vJ)^T\cdot\vK\cdot(\vc-\vc_\vJ)\right),
\]
where $\vK$ is a $2\times2$ symmetric
matrix and the subscripts imply dependence on $\vJ$ -- in order to limit the
number of parameters to determine, we assume that $\vK$ is independent of
$\vJ$. A simple assumption is that $\vc_\vJ$ depends linearly on $\vJ$:
\[\label{eq:defC}
\vc_\vJ=\vc_0+\vC\cdot(\vJ-\vJ_0),
\]
 where $\vc_0$ is a two-component object and $\vC$ is a $2\times3$ matrix.
Without loss of generality, we can choose the reference actions
$\vJ_0=(0,0,V_{\rm c}R_0)$ to be similar to those of the Sun, with the
implication that $\vc_0$ becomes the mean chemistry of stars in the given
component that are on solar-type orbits.

\begin{figure}
\centerline{\includegraphics[width=.9\hsize]{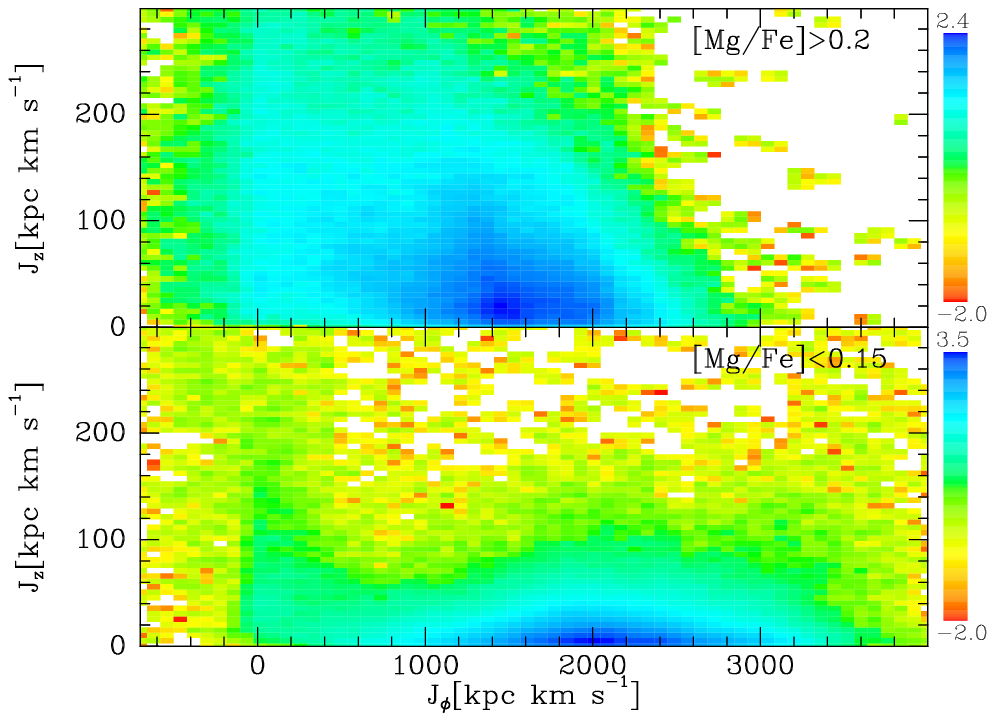}}
\centerline{\includegraphics[width=.9\hsize]{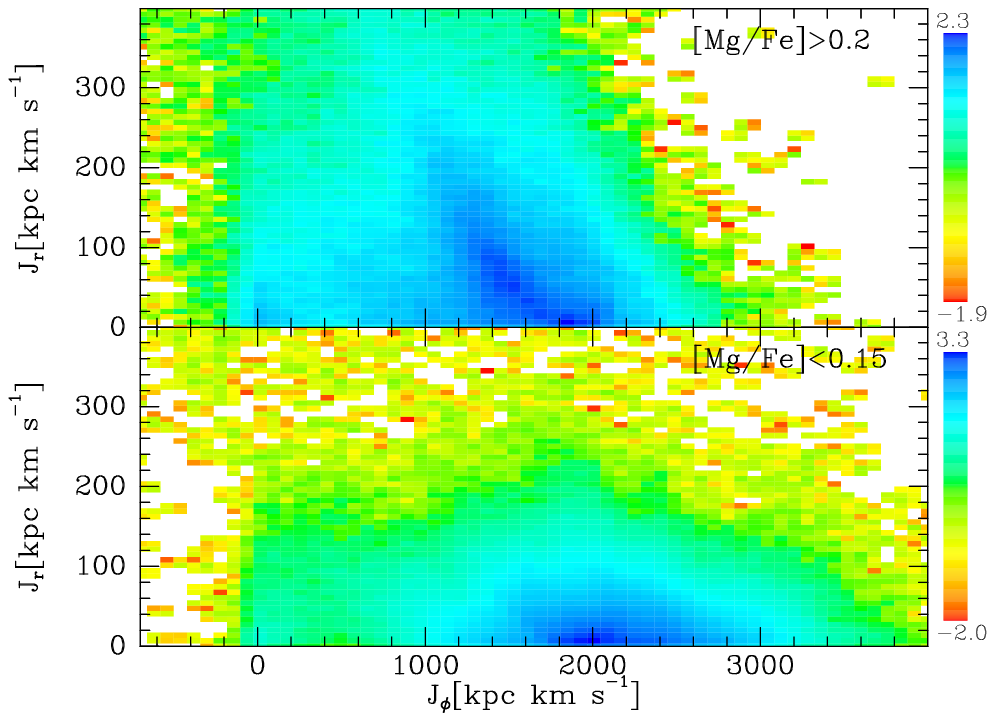}}
\caption{The action-space distributions of low- and high-$\alpha$ stars. The
colour scale gives the base-10 logarithm of the star density when projected
onto the $(J_\phi,J_z)$ plane in the upper panels and the $(J_\phi,J_r)$
plane in the lower panels.  In each pair the upper panel is for [Mg/Fe$]>0.2$
and the lower panel is for [Mg/Fe$]<0.15$. }\label{fig:low-high}
\end{figure}

\begin{figure}
\centerline{\includegraphics[width=.9\hsize]{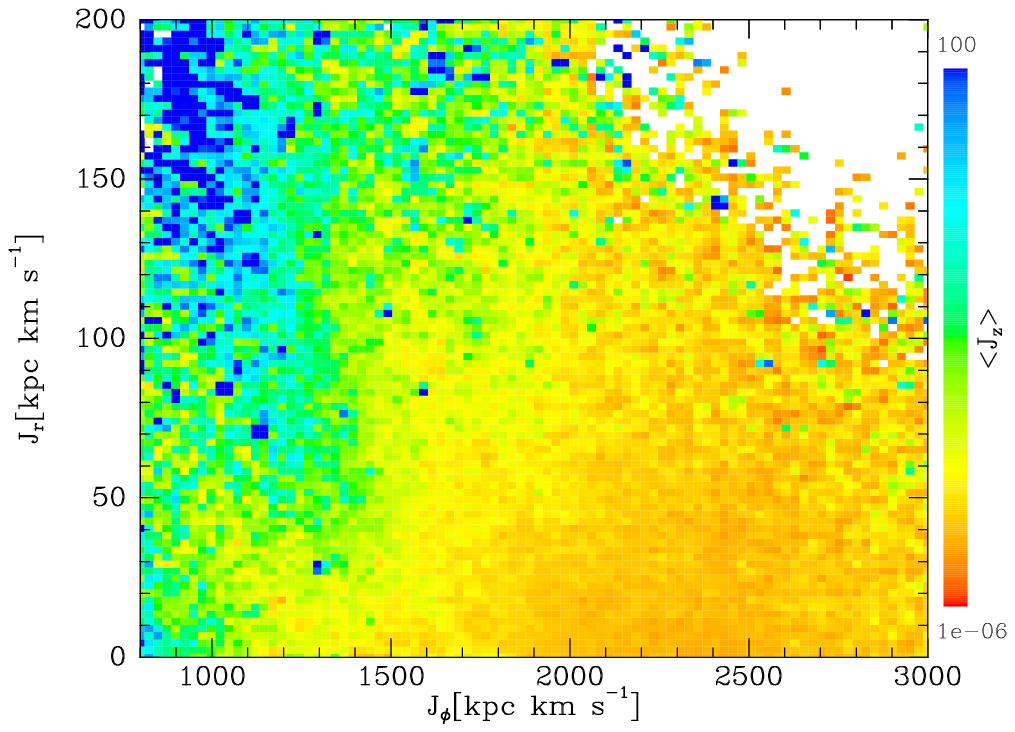}}
\caption{The mean value of $J_z$ within the $(J_\phi,J_r)$
plane.}\label{fig:Jzbar}
\end{figure}

\subsection{DFs of the components}\label{sec:DFcomponents}

We follow BV23 in modelling the stellar component of our Galaxy with six DFs
$f(\vJ)$. Five of these are instances of a generalisation of the exponential
DF introduced by \cite{AGAMA}. The overall DF is  a product of factors
\[
f(\vJ)=f_\phi(J_\phi)f_r(J_r,J_\phi)f_z(J_z,J_\phi)f_{\rm int}(J_\phi)f_{\rm
ext}(J_\phi)
\]
The function $f_r$ controls the breadth of the distribution in $J_r$ and thus
the velocity dispersions $\sigma_R$ and $\sigma_\phi$ near the equatorial
plane. The function $f_z$ similarly controls the breadth of the distribution
in $J_z$ and hence controls both the thickness of the disc and the velocity
dispersion $\sigma_z$. The other three factors control the disc's radial
profile. On its own, the factor $f_\phi$ generates a roughly exponentially
declining surface density $\Sigma(R)\simeq\exp(-R/R_\d)$. The factor $f_{\rm
int}$ tapers this profile towards the centre. At a theoretical level, this
factor is motivated by the notion that the central portion of the disc has
morphed into the bar/bulge leaving a depression at the centre of the
surviving disc. At an empirical level, modelling the young disc through the
observed distribution of OB stars, \cite{LiBinneyYoungD} concluded that an
exponential that extends right to the centre predicts too many stars at small
radii.  The factor $f_{\rm ext}$ truncates the disc at some outer radius.
This feature is motivated by the sharp transition from blue to yellow at
$J_\phi\simeq2000\H$ in Fig.~\ref{fig:LzFeBasic}, which suggests that the
high-$\alpha$ disc has a sharp outer edge.  These factors take the form
\begin{align}
f_i(J_\phi)&={E\over E+1/E}
\end{align}
where
\[
E=\begin{cases}\exp\big[(J_\phi-J_{\rm int})/D_{\rm int}\big]&\hbox{for }
i={\rm int},\cr
\exp\big[-(J_\phi-J_{\rm ext})/D_{\rm ext}\big]&\hbox{for }
i={\rm ext}.
\end{cases}
\]
Here the action $J_{\rm int}$ determines the characteristic radius of the
central depression and $D_{\rm int}$ is a parameter that determines the
sharpness of its boundary. Similarly, the action $J_{\rm ext}$ determines the
disc's outer truncation radius and $D_{\rm ext}$ sets the sharpness of the
cutoff there.  If $D_i$ is set to a negative value, $f_i=1$ is returned,
ensuring that there is no central depression or radial truncation.

The functional forms adopted for $f_r$ and $f_z$ are essentially
the same as those adopted by BV23:
\[
f_i=x_i\e^{-x_iJ_i}
\]
where
\[
x_i\equiv{(J_{\rm v}/J_{\phi0})^{p_i}\big/ J_{i0}}\quad\hbox{for }i=r,z.
\]
Here $J_{\phi0}$ is the action that sets the disc's characteristic scale
length and the action $J_{i0}$ sets the velocity dispersions  $\sigma_R$ and $\sigma_z$.
The exponent $p_i$ controls the radial variation of these dispersions: the
larger $p_i$ is, the faster the dispersions fall with radius. The action
$J_{\rm v}$ ,
a surrogate for energy, is taken to be
\[
J_{\rm v}\equiv J_r+J_z+J_\phi+J_{\rm v0},
\]
where $J_{\rm v0}$ is a constant that controls the way the dispersions vary
at small radii as discussed by BV23. The factor $f_\phi$ has the form
\[
f_\phi(\vJ)=\begin{cases}
0&\hbox{when $J_\phi<0$}\cr
{\displaystyle{M\over(2\pi)^3}{J_{\rm d}\over J_{\phi0}^2}\e^{-J_{\rm
d}/J_{\phi0}}}&$otherwise$.
\end{cases}
\]
where
\[
J_{\rm d}\equiv J_r+J_z+J_\phi+J_{\rm d0},
\]
with $J_{\rm d0}$ a constant that controls the way the surface density
varies at small radii, as discussed by BV23. These formulae for $f_r$, $f_z$
and $f_\phi$ are the same as those in BV23 except that here  $J_{\rm v}$ and
$J_{\rm d}$ depend on $J_r$ and $J_z$ in addition to $J_\phi$. Since disc
stars typically have  $J_\phi\gg J_r,J_z$ the additional dependence of
$J_{\rm v}$ and $J_{\rm d}$ is generally insignificant. It does, however,
yield more plausible velocity distributions in the neighbourhood of
$V_\phi=0$.

We follow BV23 in modelling the low-$\alpha$ disc as a superposition of three
discs, presumed to be of increasing age and velocity dispersion: the young,
middle-aged and old discs. Away from their centres, these discs
are simple exponentials, but they may have central depressions in their
surface density consistent with the bar/bulge having formed out of them. 
BV23
modelled the bulge by a spheroidal DF. Figs.~\ref{fig:LzFeBasic},
\ref{fig:LzFeBasicFe} and \ref{fig:low-high} indicate that this was a poor
choice by suggesting that the bulge is almost entirely confined to
$J_\phi>0$, as is natural in a component that formed from a thin disc.
Therefore we model the bulge as a truncated exponential, in which $J_{\phi0}$
is small and $J_{r0}$ and $J_{z0}$ are large.

The high-$\alpha$ disc is assumed to be a radially truncated exponential.  The
stellar halo is modelled by the same non-rotating spheroidal DF used by BV23.
This probably provides a poor representation of the truth, but since the halo
contributes very little to the APOGEE data, we defer improvement of its DF to
future work.

\subsection{Initial parameters}\label{sec:initial_params}

The chemodynamical model we are fitting has a large number of parameters. An
automated search for a good model through a high-dimensional space is
unlikely to succeed if started from a random location. Hence we started by
hand-fitting the DFs to the dynamical data in the manner described by BV23.
The resulting model differed from that of BV23 principally because the data
employed extended from $R=1\kpc$ to $14\kpc$ rather than the narrower range
$R_0\pm3\kpc$. 

The starting parameters of the DFs were largely taken to be those determined
by BV23. Experiment showed that central tapers in the surface densities of
disc components tend to yield circular-speed curves that rise more slowly than the
data imply, so in the final model only the young disc has a central taper. 
For the high-$\alpha$ disc, the sharp
colour transition in Fig.~\ref{fig:LzFeBasic} around $J_\phi=2000\H$
motivated the choice $J_{\rm ext}=2000\kms\kpc$, $D_{\rm ext}=200\H$.

\begin{table*}
\caption{Initial values of the chemical pdfs (for fitted values see
Table~\ref{tab:chem} below).  The units of $\theta$ are degrees while $x_0$, $y_0$,
$\sigma_x,\sigma_y$ are given in dex. The values quoted for the gradient
matrices $\vC$ (eqn \ref{eq:defC}) are in dex per $\hbox{Mpc}\kms$.
$C_{1,J_r}\equiv C_{\hbox{[Fe/H]},J_r}$ while 
$C_{2,J_r}\equiv C_{\hbox{[Mg/Fe]},J_r}$, etc.}\label{tab:start_chem} 
\begin{tabular}{lccccccccccc}
Component 	 & $\theta$	 & $x_0$	 & $y_0$	 & $\sigma_x$	 & $\sigma_y$	& $C_{1,J_r}$	& $C_{1,J_z}$	& $C_{1,J_\phi}$	& $C_{2,J_r}$ 	& $C_{2,J_z}$ 	& $C_{2,J_\phi}$\cr 
 \hline 
young disk	&$-7$	&$-0.06$	&$0.036$	&$0.1$	&$0.035$	&$0$	&$0$	&$-0.29$	&$0$	&$0$	&$0$\cr
middle disk	&$-7$	&$-0.1$	&$0.04$	&$0.12$	&$0.03$	&$0$	&$0$	&$-0.29$	&$0$	&$0$	&$0$\cr
old disk	&$-8$	&$-0.4$	&$0.1$	&$0.16$	&$0.04$	&$0$	&$0$	&$-0.29$	&$0.1$	&$0.1$	&$0$\cr
high-$\alpha$ disk	&$-8$	&$-0.5$	&$0.33$	&$0.2$	&$0.05$	&$0$	&$-10$	&$0$	&$0$	&$0$	&$0$\cr
stellar halo	&$-3$	&$-1.1$	&$0.3$	&$0.15$	&$0.05$	&$0$	&$0$	&$-0.29$	&$0$	&$0$	&$0$\cr
bulge	&$-6$	&$0.4$	&$0.03$	&$0.3$	&$0.04$	&$0$	&$0$	&$0$	&$0$	&$0$	&$0$\cr

\end{tabular}
\end{table*}

The other fresh choices required for the DFs were all the parameters of the
bulge's DF.  The bar/bulge extends out to $R\simeq3\kpc$, which corresponds
to $J_{\rm ext}\simeq800\H$, so $J_{\rm ext}$ should be a value of this
order; we started from $J_{\rm ext}=1000\H$ and $D_{\rm
ext}=200\H$. The bulge is a hot component, so we started from
$J_{r0}=100\H$ and $J_{z0}=50\H$.

Table~\ref{tab:start_chem} lists the initial values of the chemical
parameters. On the left we have the tilt $\theta$ of the Gaussian ellipsoids
with respect to the chemical axes\footnote{The angle $\theta$ is defined such
the $P(\vc)\propto\e^{-(x^2/\sigma_x^2+y^2/\sigma_y^2)/2}$, where
$x=\delta\hbox{[Fe/H]}\cos\theta+\delta\hbox{[Mg/Fe]}\sin\theta$ and
$y=-\delta\hbox{[Fe/H]}\sin\theta+\delta\hbox{[Mg/Fe]}\cos\theta$.} and the means $x_0,y_0$ and dispersions
$\sigma_x,\sigma_y$ of those Gaussians at the Sun's action-space location.
As we proceed from young disc to old and high-$\alpha$, the mean value of
[Fe/H] is presumed to decline while that of [Mg/Fe] is presumed to increase.
The dispersions $\sigma_x$ and $\sigma_y$ increase along this sequence
consistent with the evidence that stars diffuse from circular orbits as they
age.  The bulge is assumed to be metal-rich and $\alpha$-normal, with a large
dispersion $\sigma_x$.

The gradients of chemistry in action space are listed on the right of
Table~\ref{tab:start_chem}. Most values are set to zero. Exceptions are the
coefficients $C_{1,J_\phi}$ that set the radial metallicity gradients in the
thin disc and the stellar halo -- these are set to a value similar to that,
$-0.31\pm0.02\dex$ per Mpc$\kms$, that was inferred for d[Fe/H]/d$J_\phi$ by
\cite{Spina2022} for open clusters. Other non-zero values are of $C_{2,J_r}$
and $C_{2,J_z}$ for the old disc; these imply that [Mg/Fe] increases with
eccentricity and inclination, and a value for the high-$\alpha$ disc's
$C_{1,J_z}$ that implies strongly decreasing [Fe/H] with inclination.

\subsection{Managing the APOGEE selection function}\label{sec:managingSF}

As remarked above, the contours in Fig.~\ref{fig:LzFePatch} depend on
APOGEE's very complex, dust-dependent, selection function (SF) in addition
to the EDF. Our approach to circumventing this problem is as follows. Our
basic assumption is that the SF is independent of both velocity and chemistry
but depends strongly on position. The independence of velocity is clear;
less so is the independence of $\vc$ since colour criteria were involved in
the selection of stars.  Nonetheless any dependence of the SF on $\vc$ must
be tiny by comparison with the dependence on $\vx$.

Since our model is axisymmetric and symmetric in $z$, velocity distributions
are functions of $(R,|z|)$ only. So we bin the real stars in the 72 bins in
$(R,|z|)$ space that are specified by Table~\ref{tab:bdys}\footnote{It is
strictly neither necessary nor desirable to bin the real stars in space -- we
could sample velocity space at the location of each real star, but that would
be computationally expensive. Grouping stars into bins merely reduces the
computational cost of generating mock stars, and our goal in choosing bins is
to cover real space as evenly as possible with the smallest number of bins.}
and determine the barycentre $\vX_\alpha$ of the stars in the $\alpha$th bin
-- let $N_\alpha$ be the number of stars in this bin. Then, given a trial set
of components and their EDFs, we determine the density
$\rho_{i\alpha}=\int\d^3\vv,f_i$ contributed by the $i$th component at
$\vX_\alpha$ and let 
\[
k_{i\alpha}\equiv n{\rho_{i\alpha}\over\sum_j\rho_{j\alpha}}
\]
be $n\sim5$ times the fraction of real stars predicted to be contributed by the $i$th
component. Now we create $k_{i\alpha}N_{\alpha}$ mock stars by sampling
velocity space at $\vX_\alpha$ under the velocity distribution specified by
$f_i(\vJ)$. Each chosen velocity $\vV_{i\alpha}$ corresponds to an action
$\vJ_{i\alpha}$, and we use this to choose a chemistry $\vc_{i\alpha}$ by sampling the
Gaussian $P(\vc|\vJ_{i\alpha})$. When this is done, each spatial bin contains
$nN_\alpha$ mock stars in addition to the $N_\alpha$ real stars. The mock
stars, unlike the real stars, have known component memberships.

{\rd BV23 should perhaps have stressed the novelty of their approach to
mapping a galaxy's gravitational field. Competing methods, using the Jeans
equations \citep[e.g.][]{Read2017} or Schwarzschild modelling
\citep[e.g][]{vdBea08,Zhu2018}, hinge on knowledge of the density $\rho(\vx)$
of a stellar component. BV23 {\it predict} the densities of components, but
they they use as an input only the density profile $\rho(z)$ in the solar
cylinder. The basic physical principle of the BV23 approach is most easily
understood in the context of a one-dimensional problem. Let $(x,v)$ be the
system's phase-space coordinates and let $\Phi(0)=0$ with $\d\Phi/\d x\ge0$,
so $x=0$ is the bottom of the potential well.  Then at $x=0$ stars of every
energy $E$ are present, and at $x$ we find only stars with $E\ge\Phi(x)$.
Moreover, the velocity distribution in the vicinity of $v=0$ at location $X$
follows from the shape of the velocity distribution in the vicinity of
$v=\sqrt{2\Phi(X)}$ at $x=0$. In words: the potential relates velocity
distributions at different locations, so it's logically possible to recover
it from those velocity distributions without knowing the stellar density
$\rho(\vx)$. When the DF is isothermsl, $f=F\e^{-\beta E}$, the velocity
distribution, $n(\vv)=F\e^{-\beta\Phi(\vx)}\e^{-\beta v^2/2}$, is the same
Gaussian at all locations and $\Phi$ cannot be recovered without knowing
$\rho(\vx)$. Fortunately, no stellar system can have an isothermal DF because
the latter includes many unbound stars.
  }

\subsection{Defining the likelihood}\label{sec:def_lnL}

Given a model potential $\Phi(\vx)$, a set of DFs and a model chemistry, we
compute the log likelihood $\ln L$ of the data as follows. We
distribute each bin's mock stars over a grid in velocity space by the
cloud-in-cell algorithm \citep[e.g.][Box 2.4]{GDII}.  The velocity grid has
$n_g^3$ cells covering the cuboid with boundaries at $\pm V_{R\rm max}$, $\pm
V_{z\rm max}$ and $V_{\phi\rm min}$ to $V_{\phi\rm max}$, where $V_{R\rm
max}=2.5\sigma_R$, $V_{z\rm max}=2.5\sigma_z$, $V_{\phi\rm min}=-50\kms$ and
$V_{\phi\rm max}=\ex{V_\phi}+2.5\sigma_{\rm phi}$. Here the grid size $n_g$
is proportional to the cube root of the number of real stars and $\sigma_R$
etc are the standard deviations of the components of the velocities of the
real stars:  $n_g$ ranged up to 23 with average value $9.2$.

The mass assigned to each grid cell is then divided by the number of mock
stars in the relevant spatial bin. Then we use the cloud-in-cell algorithm to
determine the mock-star density at the location of each real star and compute
the contribution $\ln L_{\alpha\,\rm dyn}$ of the $\alpha$th spatial bin to
the overall dynamical log likelihood as the sum over stars of the logarithms
of these densities. In Appendix~\ref{app:validate} we show that the maximum
value of $\ln L_{\alpha\,\rm dyn}$ under variation of the masses assigned to
cells, subject to a fixed total mass on the grid, is achieved when the mass of mock
stars in each cell coincides with the mass that would be obtained by
distributing the real stars over the grid in the same way that the mock stars
are distributed.

The dynamical log likelihood is the sum
\[
\ln L_{\rm dyn}={1\over N_{\rm real}}\sum_{{\rm bins}\,\alpha}\ln L_{\alpha\,\rm dyn}
\]
over spatial bins normalised by the number of real stars. This normalisation
prevents the likelihood being dominated by grid cells that lie close to the Sun and
are in consequence heavily populated: sparsely populated cells far from the
Sun are powerful probes of the Galaxy's structure.

{\tabcolsep 3.5pt
\begin{table}
\caption{Boundaries (in kiloparsecs)  between spatial bins}\label{tab:bdys}
\begin{tabular}{lccccccccccccc}
$|z|$&0&0.3&0.7&1&1.5&2&3\cr
$R$&0.5&1.5&3&4&5&6&7&8&9&10&11.5&13&14.5\cr
\end{tabular}
\end{table}
}

The computation of the chemical contribution to the log likelihood is
similar. We use the cloud-in-cell algorithm to distribute the mock stars over
a five-dimensional grid $([\hbox{Fe/H}],[\hbox{Mg/Fe}],J_\phi,J_r,J_z)$. Then the
mass assigned to each cell is divided by the number of mock stars, and $\ln
L_{\rm chem}$ is computed as the mean of the logarithm of the grid density at
the location of each real star. The final log likelihood is
\[
\ln L=\ln L_{\rm dyn}+\ln L_{\rm chem}.
\]

An indication of how closely a model approximates the data can be obtained by
distributing real rather than mock stars over the grids and then computing
$\ln L$ as before by determining the grid density at the locations of the
real stars. We call this the perfect-fit value of $\ln L$. The 
perfect-fit and actual-fit values  of $\ln L_{\rm dyn}$ for the chosen model
are $-6.460$ and
$-6.573$, while the corresponding values of $\ln L_{\rm chem}$ are $-9.7737$
and $-9.7875$.

We will see below that the algorithm chooses surprisingly large values
for some dispersions $\sigma_y$ and gradient coefficients $C_{ij}$. We tested
the robustness of thee choices by including in the quantity to be maximised a
Gaussian prior term proportional to $-\sigma_y^2$ and/or $-C_{ij}^2$. These
priors had little effect.

\subsubsection{A shortcut}

The procedure just described assumes that the mock stars are drawn from the
current DFs. It is expedient to be able to compute $\ln L$ from mock stars
obtained by sampling a DF $f_0$ that differs slightly from the current DF, $f$.
To do this we weight each mock star by $f(J)/f_0(J)$, the ratio of the star's
current probability density to the probability density when it was randomly
drawn. The sum of these weights is the number of `effective' mock stars and
this is the number used to normalise the masses in cells after distributing
mock stars on a grid.  The need
to re-sample was determined by monitoring the largest values of
$f(\vJ)/f_0(\vJ)$.  There is no need to re-sample after varying the chemical
model.

\subsection{Likelihood maximisation}\label{sec:min_lnL}

We have used the Nelder-Mead algorithm encapsulated in the \agama\ function
{\tt findMinNdim} to minimise $-\ln L$. We alternated a few hundred steps in
which the chemical model was fixed while the DFs were varied with a few
hundred steps in which the DFs were fixed and the chemical model was varied.
The code sought to maximise the sum of $\ln L_{\rm dyn}$ and $\ln L_{\rm
chem}$ regardless of which parameters were being varied because $\ln L_{\rm
chem}$ is useful when changing the DFs: it is sensitive to the relative
masses of components and indicates whether a deficiency of stars at some
phase-space location should be remedied by increasing the mass of the
high-$\alpha$ disc or the old disc, for example.

When the DFs are varied without updating the potential $\Phi(\vx)$, the fit
quality is invariant under multiplication of the masses $m_\alpha$ of every
component's DF by a common factor: with $\Phi$ fixed, the fit depends only on
the relative masses of components. So the Nelder-Mead algorithm varied the
DFs with the total stellar mass held constant.

After a few hundred Nelder-Mead adjustments of the DFs, the DF parameters
were reviewed and potentially changed by hand before updating $\Phi$
to self-consistency. Then a new sample of mock stars was drawn.

\begin{figure}
\centerline{\includegraphics[width=.8\hsize]{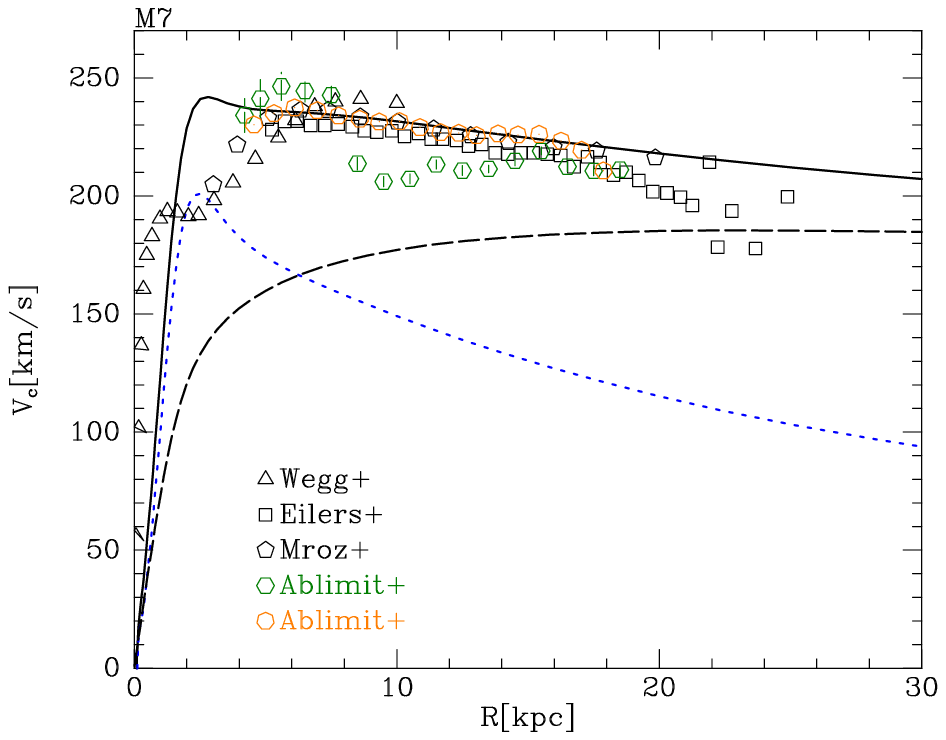}}
\caption{Circular speed  of the potential with prior estimates.}\label{fig:Vc}
\end{figure}

\begin{figure}
\centerline{\includegraphics[width=.8\hsize]{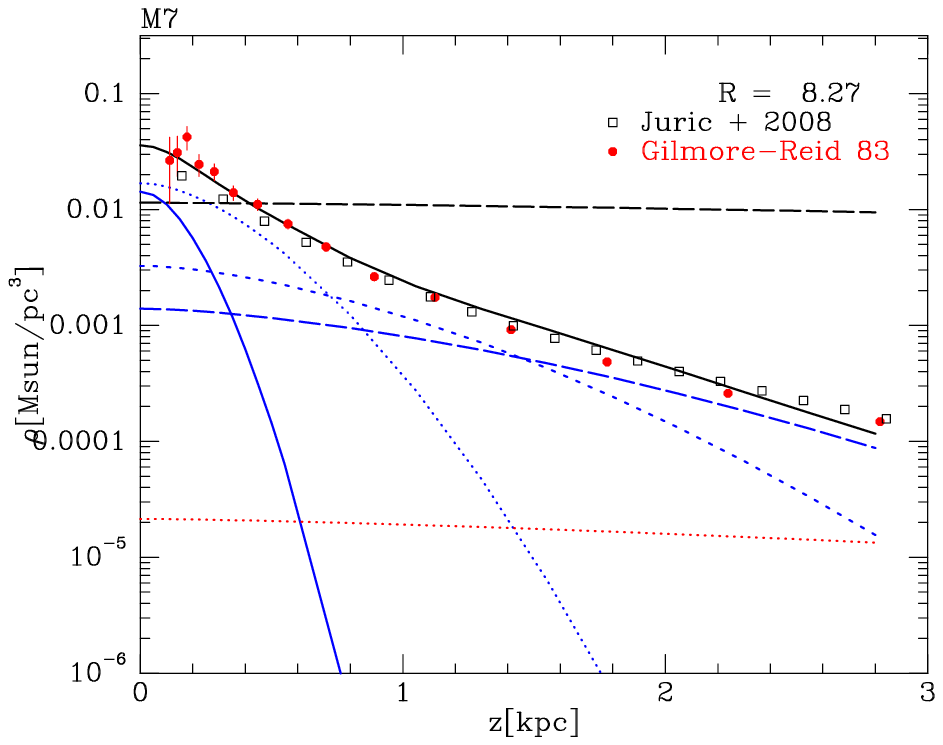}}
\caption{Vertical density profiles at $R_0$.}\label{fig:vertProfile}
\end{figure}

\begin{figure}
\centerline{\includegraphics[width=.8\hsize]{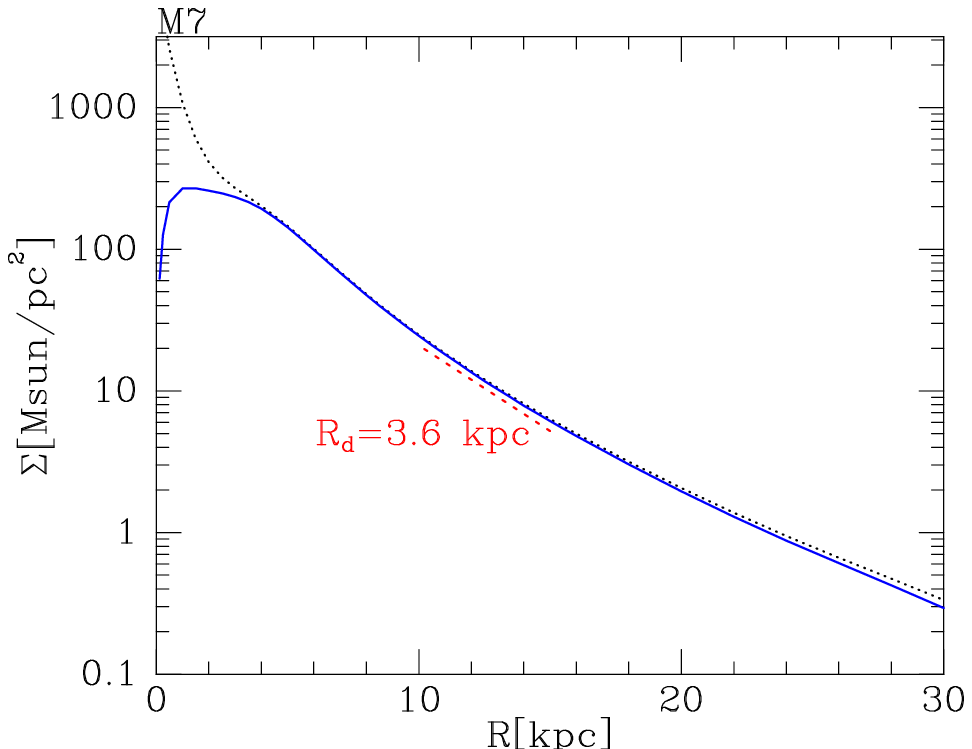}}
\caption{Full curve: surface density of disc stars. Dotted curve: surface
density of all stars. Red dashed curve: exponentially falling surface density
with scale length $R_\d=3.6\kpc$.}\label{fig:sdProfile}
\end{figure}

\begin{figure}
\centerline{\includegraphics[width=.8\hsize]{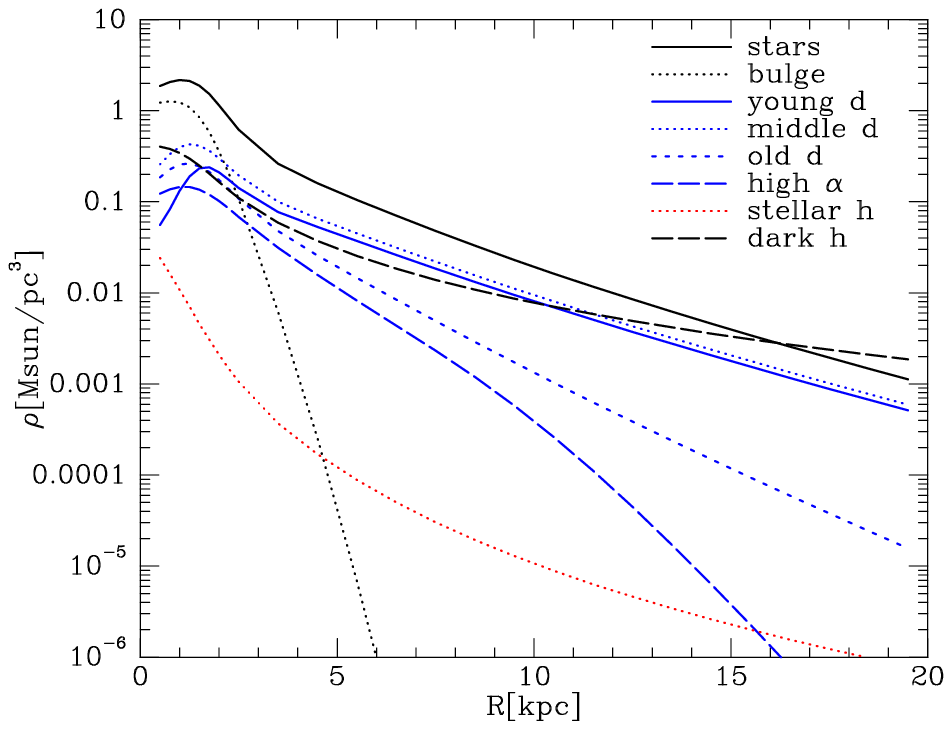}}
\centerline{\includegraphics[width=.8\hsize]{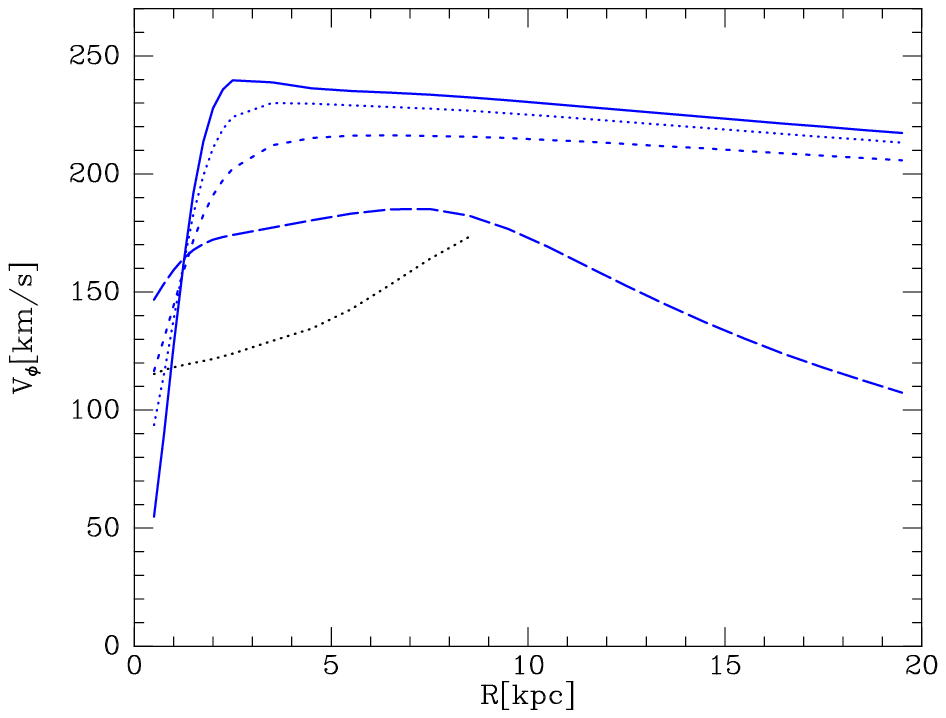}}
\caption{Upper panel: contributions to the density at $(R,0)$. Lower panel:
rotation rates of the components at $(R,0)$.}\label{fig:radProfile}
\end{figure}

\begin{figure*}
\centerline{
\includegraphics[height=.99\hsize]{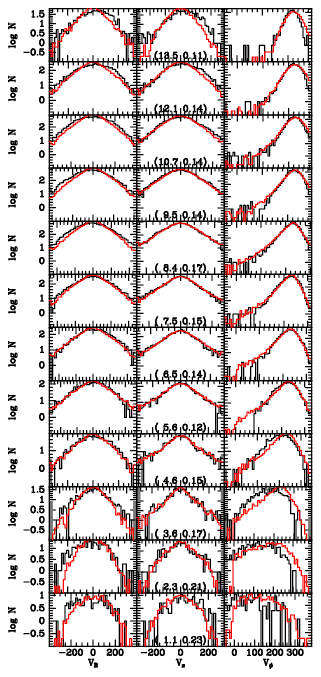}
\includegraphics[height=.99\hsize]{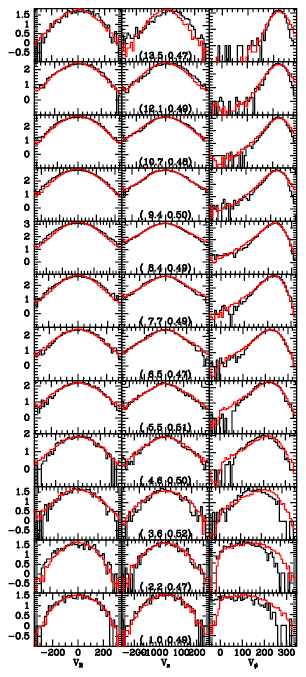}
}
 \caption{Velocity distributions within 24 spatial cells. The left column
shows the distributions in $V_R$, $V_z$ and $V_\phi$ in the cells that cover
the equatorial plane and have Galactocentric radii $R$ that increase from
$1.1\kpc$ at the bottom to $13.5\kpc$ at the top. The right column shows the
corresponding velocity distributions for the adjacent cells -- these have
barycentres at $|z|\sim0.5\kpc$. The numbers in the panels for $V_z$ give the
exact coordinates of the relevant barycentre. In each panel the velocity
scale covers $\pm3\sigma_i$, where $\sigma_i$ is the standard deviation of
the data histogram -- the numbers along the bottom refer only to the bottom
row of panels.}\label{fig:oned0}
\end{figure*}

\begin{figure*}
\centerline{
\includegraphics[height=.99\hsize]{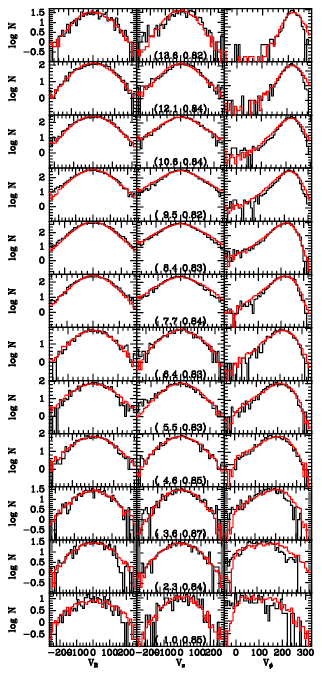}
\includegraphics[height=.99\hsize]{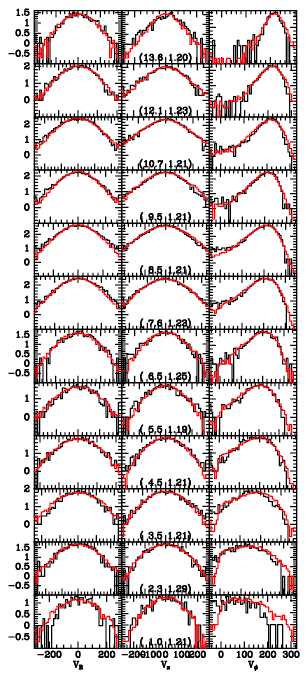}
}
\caption{As Fig.~\ref{fig:oned0} but for cells with barycentres at $|z|\sim
0.8\kpc$
(left column) and $|z|\sim1.2\kpc$.}\label{fig:oned1}
\end{figure*}

\begin{figure*}
\centerline{
\includegraphics[height=.99\hsize]{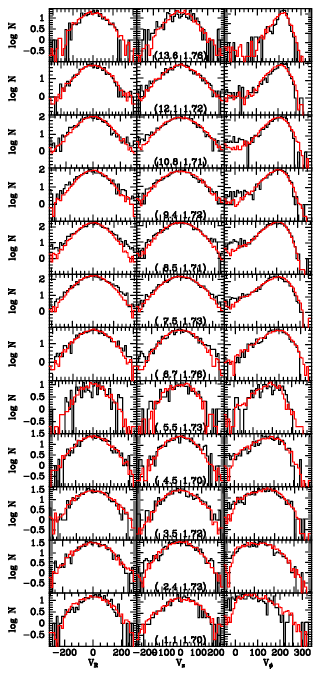}
\includegraphics[height=.99\hsize]{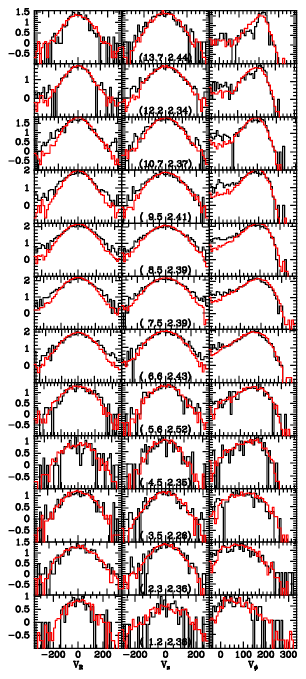}
}
\caption{As Fig.~\ref{fig:oned0} but for cells with barycentres at $|z|\sim
1.7\kpc$
(left column) and $|z|\sim2.4\kpc$.}\label{fig:oned2}
\end{figure*}

\section{results}\label{sec:results}

Regardless of whether it is the DFs or the chemical model that is being
adjusted, the Nelder-Mead algorithm yields a sequence of $\ln L$ values that
increases rapidly at first and then more and more slowly. So one does not
have the impression that it has located a global maximum. Hence the model
presented here has the status of fitting the data about as well as any model
we have tried rather than being a definitive model.

The full curve in Figs~\ref{fig:Vc} shows the circular-speed curve $V_{\rm
c}(R)$ of the final potential; also shown are the contributions to $V_{\rm
c}$ from the baryons (blue dotted curve) and dark matter (black dashed curve)
along with several previous estimates of $V_{\rm c}(R)$ derived from tracers
presumed to be on near-circular \citep{Mroz2019,Ablimit2020}, and a study of
$23\,000$ red giants by \cite{Eilers2019}. At $R\la4\kpc$, where circular
orbits are prohibited by the bar, the model curve deviates significantly from
the triangular points from \cite{Wegg2013}, which describe the axisymmetric
component of a non-axisymmetric potential. Clearly we cannot expect an
axisymmetric model to reproduce the data for this region, but the extent of
the conflict with earlier work suggests that our bulge is too massive and
insufficiently compact.  At $R\ga4\kpc$ the model curve in Fig.~\ref{fig:Vc}
runs just above most of the points from previous studies. This is a
consequence of taking $V_{\phi\odot}=251\kms$ for the Sun: when
$V_{\phi\odot}$ is increased, the black (data) curves in the right columns of
Figs.~\ref{fig:oned0} to \ref{fig:oned2} move to the right and the
potential has to be adjusted to make the red (model) curves follow them. Our
value for $V_{\phi\odot}$ follows directly from the observations of Sgr A*
and the assumption that this black hole can define the Galactic rest frame
\citep{ReidBrunthaler2020,Gravity2022}. Since the model predicts $V_{\rm
c}(R_0)=234\kms$ the Sun's peculiar velocity is $V_{\phi\odot}=17\kms$. These
values are consistent with the values $V_{\rm c}(R_0)=238\pm9\kms$ and
$V_{\phi\odot}=12\pm9\kms$ reported by \cite{Schoenrich2012}.

The full curve in Fig.~\ref{fig:vertProfile} shows the model density of stars
as a function of $|z|$ at the solar radius, while the points describe the
observational estimates of this by \cite{GiRe83} and \cite{Juea08}, which can
be moved vertically because they relate to star density rather than mass
density. The agreement is satisfactory. The black dashed line shows the
almost constant density of dark matter, while the blue curves show the
density profiles of the young disc (full curve), the middle disc (dotted
curve), the old disc (short-dashed curve) and the high-$\alpha$ disc
(long-dashed curve). The red dotted curve shows the tiny contribution from
the stellar halo. The dark halo makes the largest contribution to the density
at $|z|>400\pc$.

Comparison with Figs 14 and 15 in BV23 shows that the differences between
the circular-speed curves of the two models are confined to the bar-bulge
region $R\la5\kpc$ -- the curve of the BV23 model rises much more steeply at
$R<1\kpc$.  The BV23 discs are less massive, yield a shorter overall radial
scale length $R_\d$, and have smaller scale heights -- this is especially
true of the old disc, for which the $J_{z0}$ parameter has jumped from 5 to
$24\H$. At $R>10\kpc$ the old and high-$\alpha$ discs of BV23
have much lower $\sigma_R$ and significantly lower $\sigma_z$, changes that
are clearly mandated by the new data.  The local densities of dark matter are
almost identical in the two models.

The full blue curve in Fig.~\ref{fig:sdProfile} shows the surface density of
disc stars as a function of radius while the black dotted curve shows the
surface density of all stars. At $R\ga5\kpc$ the surface density falls nearly
exponentially with scale length $R_\d\simeq3.6\kpc$ (marked by the red dashed
curve). This is a significantly longer scale-length than was inferred by
\cite{Roea03} from the 2MASS survey. Recently, \cite{Robin2022} inferred a
scale-length parameter $H_\rho=2900\kpc$ for the DFs of the thin disc
components, which is best compared to the values of
$J_{\phi0}/V_c(R_0)\simeq4.27\kpc$ for the young and middle discs and
$2.1\kpc$ for the old disc.  \cite{BovyRix2012} derived scale lengths for
mono-abundance populations in APOGEE and found that these increased steadily
from $R_\d<2\kpc$ to $R_\d>4\kpc$ as [Mg/Fe] decreases and [Fe/H] increases.
The results in Table~\ref{tab:discDFs} are qualitatively consistent with that
early work.

The full curve in the upper panel of Fig.~\ref{fig:radProfile} shows as a
function of radius the density of stars at $z=0$, while the blue curves show
the corresponding densities of the disc components. The contributions of the
bulge and the stellar halo are shown by the black and red dotted curves. The
density of the dark halo is shown by the black long-dashed curve.

The blue curves in the lower panel of Fig.~\ref{fig:radProfile} show the mean
streaming velocities at $z=0$ of the three thin-disc components and of the
high-$\alpha$ disc. Asymmetric drift causes the streaming velocity to fall
further and further below the circular speed as the velocity dispersions
increase. The black dotted curve shows the rising mean streaming velocity in
the bulge. 

\begin{figure*}
\centerline{
\includegraphics[height=.6\hsize]{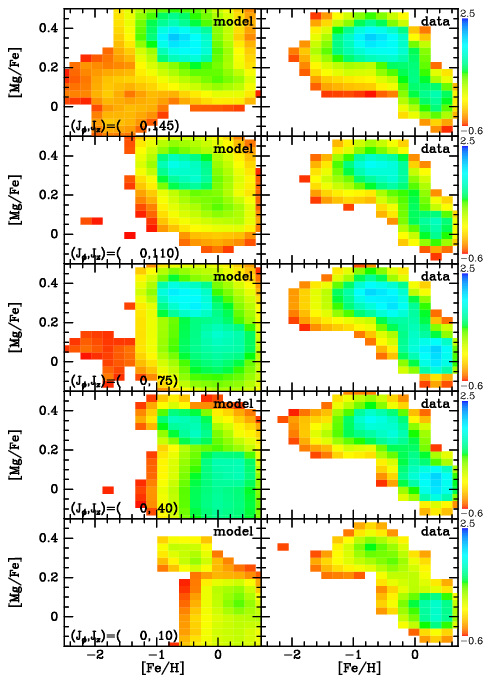}
\includegraphics[height=.6\hsize]{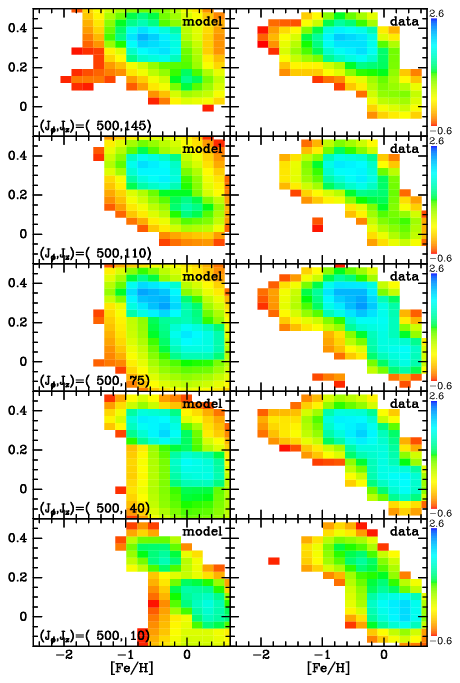}
}
\caption{The chemical composition of stars that lie in 10 cells in action
space. The left column is for cells with $J_\phi\in(-200,200)\H$ and $J_z$
in intervals that increase from bottom to top; specifically $(0,20)$,
$(25,55)$, $(60,90)$, $(95,125)$ and $(130,160)\H$. The right column is for
$J_\phi\in(300,700)\H$ and the same intervals in $J_z$. Within each column the
right panels show the distribution of real stars, while the left panels show
mock stars.}\label{fig:plaque0}
\end{figure*}

\begin{figure*}
\centerline{
\includegraphics[height=.6\hsize]{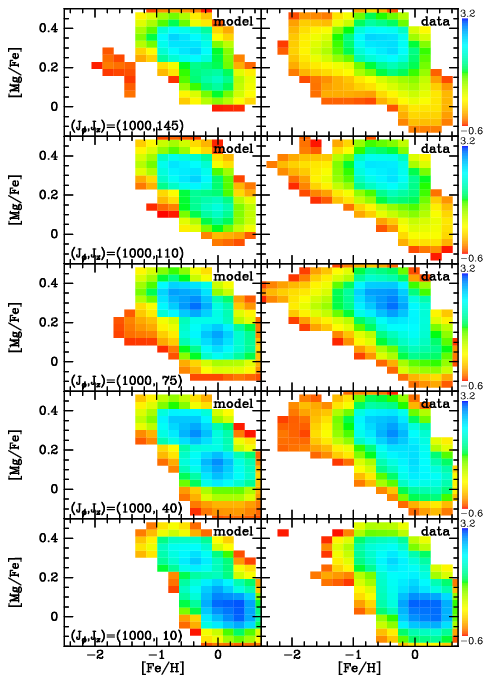}
\includegraphics[height=.6\hsize]{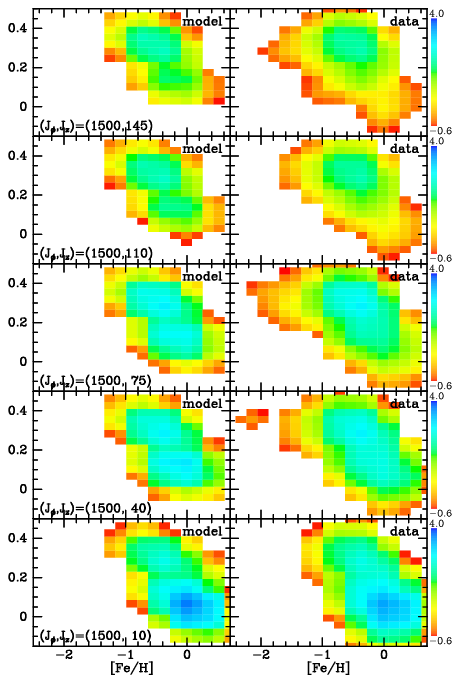}
}
\caption{The same as Fig.~\ref{fig:plaque0} but for
$J_\phi\in(800,1200)\H$ (left column) 
$J_\phi\in(1300,1700)\H$ (right column).}\label{fig:plaque1}
\end{figure*}

\begin{figure*}
\centerline{
\includegraphics[height=.6\hsize]{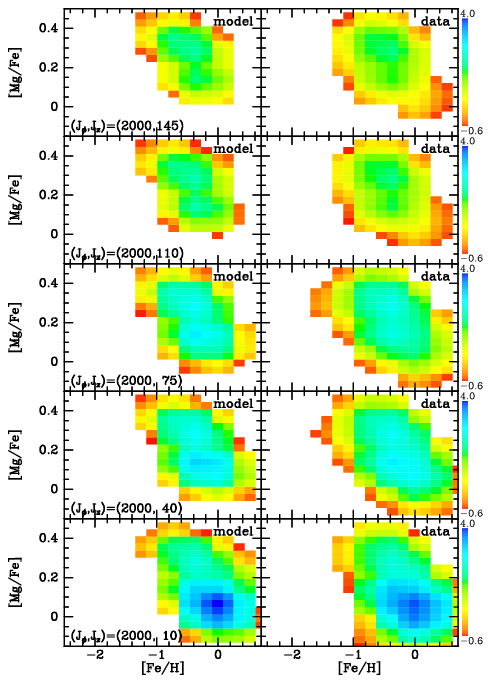}
\includegraphics[height=.6\hsize]{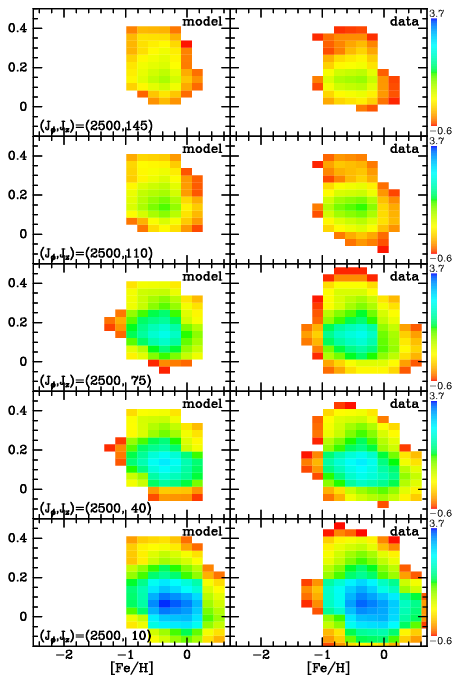}
}
\caption{The same as Fig.~\ref{fig:plaque0} but for
$J_\phi\in(1800,2200)\H$ (left column) 
$J_\phi\in(2300,2700)\H$ (right column).}\label{fig:plaque2}
\end{figure*}

Tables \ref{tab:discDFs} to \ref{tab:chem} list the model's parameters, while
Figs \ref{fig:oned0} to \ref{fig:plaque2} show the resulting fits to data.
These figures show that this model fits many aspects of the data very well
while showing shortcomings in other aspects.

\subsection{Fits to kinematics}\label{sec:fit_kin}

Figs.~\ref{fig:oned0} to \ref{fig:oned2} show, from left to right in each
column, distributions of $V_R$, $V_z$ and $V_\phi$ marginalised over the
other two velocity components for the 72 spatial bins that were used to
compute $\ln L$. In the panels for $V_R$ and $V_z$ the
histograms cover $\pm3\sigma$ where $\sigma$ is the standard deviation, while
the histograms of $V_\phi$ cover $-50\kms$ to $\ex{V_\phi}+3\sigma$. From
bottom to top of each column the radius of the spatial bin increases, while
different columns correspond to different values of  $|z|$: the numbers at
the bottom of each $V_z$ histogram give the mean values of $R$ and $z$ of
stars in the
relevant bin. The black histograms show the APOGEE DR17 $+$ Gaia DR3 data
while the red histograms show the distributions of mock stars.

\begin{table*}
\caption{Parameters of the disc DFs. Masses are in units of $10^{10}\msun$
and actions in kpc$\kms$.}\label{tab:discDFs}
\begin{tabular}{lcccccccccccc}
Component 	&$M$	&$J_{\phi0}$	&$J_{r0}$	&$J_{z0}$	&$J_{\rm int}$	&$D_{\rm int}$	&$J_{\rm ext}$	&$D_{\rm ext}$	&$p_r$	&$p_z$	&$J_{\rm v0}$	&$J_{\rm d0}$\cr
 \hline 
young disk	&$0.45$	&$977.9$	&$2.806$	&$1.296$	&$186.9$	&$278$	&--	&--	&$-0.76$	&$-0.23$	&$152.1$	&$102.1$\cr
middle disk	&$1.2$	&$1030$	&$22.82$	&$3.24$	&--	&--	&--	&--	&$-0.23$	&$-0.7$	&$146.4$	&$731.9$\cr
old disk	&$0.87$	&$508$	&$47.14$	&$24.04$	&--	&--	&--	&--	&$0.034$	&$-0.043$	&$132.7$	&$733.9$\cr
high-$\alpha$ disk	&$0.766$	&$399$	&$116.4$	&$64.6$	&--	&--	&$2212$	&$207.8$	&$0.1$	&$0.17$	&$150$	&$40$\cr
bulge	&$1.27$	&$127.5$	&$122.2$	&$34.21$	&--	&--	&$611.2$	&$217.5$	&$0.82$	&$-0.13$	&$150$	&$20$\cr

\end{tabular}
\end{table*}

\begin{table*}
\caption{Parameters of the spheroidal DFs. Masses are in units of $10^{10}\msun$
and actions in kpc$\kms$}\label{tab:haloDF}
\begin{tabular}{lcccccccc}
Component 	&$M$	& $J_0$	& $J_{\rm core}$	& $J_{\rm cutoff}$	& $\alpha_{\rm in}$	& $\alpha_{\rm out}$	& $F_{\rm in}$	& $F_{\rm out}$\cr
\hline
dark halo	&$94$	&$1e+04$	&$100$	&$2e+04$	&$1.6$	&$2.7$	&$1.4$	&$1.2$\cr
stellar halo	&$0.03969$	&$583.2$	&$6.148$	&$1e+05$	&$1.6$	&$4.2$	&$1.8$	&$1.2$\cr

\end{tabular}
\end{table*}

The largest discrepancies between data and model occur in $V_\phi$ at small
$R$ and $|z|$ (bottom right of each column in Fig.~\ref{fig:oned0}). These
bins are dominated by the Galactic bar, so it is natural that our
axisymmetric model fails to model the data well. More surprising is how well
the model fits the $V_R$ and $V_z$ distributions for these bins, and
even provides passable fits to the $V_\phi$ distributions above $|z|=0.7\kpc$
(lower right of the columns of Figs~\ref{fig:oned1}) and \ref{fig:oned2}).

The model generally provides good fits to the $V_\phi$ histograms at
$R\ga4\kpc$. This result implies that the potential correctly gives the
circular speed $V_{\rm c}(R)$ because, as BV23 explained, even a small
mismatch between a model's  circular speed and the true circular speed
leads to an unmistakable displacement of the red and black curves in the
$V_\phi$ plots.  If some $V_\phi$ histograms for $R>4\kpc$ show
discrepancies, it is because the model under-predicts stars with
$V_\phi\sim0$. 

In a few of the histograms for $V_R$ and $V_z$ (see the top of
Fig.~\ref{fig:oned0}) the red model curve fits the black data curve on one
side much better than on the other. Since the red curves are by construction
left-right symmetric (up to shot noise), such one-sided fits point to
deviations from equilibrium in the data caused by spiral structure or tidal
interactions, for example \citep[e.g.][]{McMillan2022,Khanna2023}.

\begin{table*}
\caption{Parameters of the fitted chemical pdfs.  
The units of $\theta$ are degrees while $x_0$, $y_0$,
$\sigma_x,\sigma_y$ are given in dex. The values quoted for the gradient
matrices $\vC$ (eqn \ref{eq:defC}) are in dex per $\hbox{Mpc}\kms$.
$C_{1,J_r}\equiv C_{\hbox{[Fe/H]},J_r}$ while 
$C_{2,J_r}\equiv C_{\hbox{[Mg/Fe]},J_r}$, etc.
The conventional radial
metallicity gradient, $\d\hbox{[Fe/H]}/\d R\simeq C_{1,J_\phi}\times V_{\rm
c}\simeq-0.00036\times234\simeq-0.084\,\hbox{dex}\kpc^{-1}$ for the middle
disc.}\label{tab:chem} 
\begin{tabular}{lccccccccccc}
Component 	 & $\theta$	 & $x_0$	 & $y_0$	 & $\sigma_x$	 & $\sigma_y$	& $C_{1,J_r}$	& $C_{1,J_z}$	& $C_{1,J_\phi}$	& $C_{2,J_r}$ 	& $C_{2,J_z}$ 	& $C_{2,J_\phi}$\cr 
 \hline 
young disk	&$-6.79$	&$0.00119$	&$-0.0206$	&$0.0996$	&$-0.00664$	&$-0.345$	&$9.24$	&$-0.327$	&$1.97$	&$23.9$	&$0.0142$\cr
middle disk	&$-6.79$	&$-0.0509$	&$0.0407$	&$0.0496$	&$0.00192$	&$0.0566$	&$-0.648$	&$-0.363$	&$0.142$	&$1.79$	&$0.0243$\cr
old disk	&$-7.33$	&$-0.276$	&$0.133$	&$0.112$	&$0.0223$	&$1.08$	&$-0.892$	&$-0.265$	&$-0.262$	&$0.236$	&$0.0099$\cr
high-$\alpha$ disk	&$-8$	&$-0.453$	&$0.317$	&$0.144$	&$0.00613$	&$0.897$	&$-0.792$	&$0.0665$	&$-0.25$	&$0.0669$	&$-0.0173$\cr
stellar halo	&$-3$	&$-1.09$	&$0.281$	&$0.509$	&$0.128$	&$0.897$	&$2.49$	&$0.565$	&$0.538$	&$0.24$	&$0.154$\cr
bulge	&$-6$	&$0.416$	&$0.0884$	&$0.407$	&$0.0942$	&$0.811$	&$-6.52$	&$0.057$	&$2.48$	&$4.65$	&$0.248$\cr

\end{tabular}
\end{table*}

\subsubsection{Parameters of the DFs}

Tables~\ref{tab:discDFs} and \ref{tab:haloDF} give the parameters of the
fitted DFs. The middle disc is the most massive disc component: with
$1.2\times10^{10}\msun$ it is 50 percent more massive than the old and
high-$\alpha$ discs and nearly three times as massive as the young disc. The
bulge with $1.27\times10^{10}\msun$ is slightly more massive than the middle
disc. The mass, $\sim4\times10^8\msun$, of the stellar halo is
gravitationally negligible. As discussed in Section~\ref{sec:AAchem}, our sample
may be biased against very metal-poor stars, so the mass we report for the
stellar halo is likely an underestimate -- for the halo within $100\kpc$
\cite{DeasonVasilyJason2019} estimate luminosity
$L=9.4\pm2.4\times10^8L_\odot$ leading to mass
$M=1.4\pm0.4\times10^9\msun$ at least twice our value.

The young and middle discs have comparable scale actions
$J_{\phi0}\sim1000\H$, while the old and high-$\alpha$ discs have much
smaller scale actions, $\sim500$ and $400\H$, respectively. In the
context of inside-out growth of galaxies, this result is to be expected.  The
parameter values $J_{\rm int}\sim190\H$ and $D_{\rm
int}\sim280\H$ for the young disc imply that it does not differ
strongly from a pure exponential. The high-$\alpha$ disc is quite sharply
radially truncated near $R_0$: $J_{\rm ext}\simeq2200$, $D_{\rm
ext}\simeq210\H$.

The scale actions $J_{r0}$ and $J_{z0}$ that control the in-plane and
vertical velocity dispersions  increase, as expected, along the sequence young
disc, middle disc, old disc, high-$\alpha$ disc. The biggest jump in $J_{r0}$
occurs between the young and the middle disc, while the biggest jump in
$J_{z0}$ occurs between the middle and old disc, so the disc with the largest
velocity anisotropy is the middle disc. Such anisotropy is the signature of
heating by spiral arms \citep[e.g.][]{BinneyLacey1988}.

The value of $J_{r0}$ for the bulge is very similar to that of the
high-$\alpha$ disc, while the bulge's value for $J_{z0}$ is only half that of
the high-$\alpha$ disc.

The values of the parameter $p_r$ that controls the radial gradient of the
in-plane velocity dispersions increases systematically along the sequence
young -- high-$\alpha$ disc. This result implies that the radial gradient in
$\sigma_R$ steepens along this sequence. 

The parameter $p_z$ that controls the radial gradient in $\sigma_z$ does not
vary systematically along the disc sequence. The middle disc has the most
negative value of $p_z$ and therefore the weakest radial gradient, while the
high-$\alpha$ disc has the steepest radial gradient.

The bulge DF's values $J_{\phi0}=127\H$ and $J_{\rm ext}=611\H$ ensure that
the bulge is compact. Its in-plane dispersions are similar to those of the
high-$\alpha$ disc ($J_{r0}=122\H$) but it has much smaller vertical
dispersions and extent because $J_{z0}=34$ versus $64\H$ for the
high-$\alpha$ disc. For some reason $p_r=0.82$ has been set large and
positive (causing $\sigma_R$ to decline steeply with $z$) while $p_z=-0.13$
causes $\sigma_z$ to decline much more slowly with $R$.  Consequently the bulge is
most anisotropic at small radii. The bulge mass, $1.27\times10^{10}\msun$ may
be compared with the value $1.43\pm0.18\times10^{10}\msun$ that
\cite{Portail2015} estimated from made-to-measure modelling of data for
red-clump giants.

The total stellar mass is $4.60\times10^{10}\msun$, on the lower side of
recent estimates. At $R_0$ the stellar
density is $0.036\msun\pc^{-3}$ in the plane falling to $0.0024\msun\pc^{-3}$ at
$z=1.1\kpc$.

At $R_0$ the density of dark matter is $0.011\msun\pc^{-3}$ in the plane
falling to $0.0095\msun\pc^{-3}$ at $z=1.1\kpc$ in agreement with most recent
estimates \citep[see Fig.~1 of][for a review]{deSalas2021}. The vertical
component of the gravitational acceleration at $R_0$ satisfies
\begin{align}
K_z(R_0,1.1\kpc)&=1.84\kms\Myr^{-1}\cr
{K_z(R_0,1.1\kpc)\over2\pi G}&=66.4\msun\pc^{-2}
\end{align}
The actual surface density of stars and dark matter within $1.1\kpc$ of the
plane is slightly lower than the naive estimate based on $K_z$ because the
gravitational field has a large, position dependent, radial component. It is
\[
\Sigma(R_0,1.1\kpc)=63.9\msun\pc^{-2}.
\]
and comprises $26.5\msun\pc^{-2}$ in stars, $24.7\msun\pc^{-2}$ in dark
matter and $12.6\msun\pc^{-2}$ in gas.

In Table~\ref{tab:haloDF} the mass of the dark halo is given as
$0.94\times10^{12}\msun$ but the great majority of this mass lies outside the
volume for which we have data. The scale radius $r_{\rm s}$ of the dark halo
is set by the parameter $J_{\phi0}$, which was arbitrarily set to
$10\,000\H$, so the volume modelled lies inside $r_{\rm s}$, where, in the
absence of the stars, the dark-matter density would satisfy $\rho\sim
r^{-2}$. As explained by BV23, the actual density profile of the dark halo is
more complex on account of the pull of the stars.

\subsection{Fits to chemistry}\label{sec:fit_chem}

Figs.~\ref{fig:plaque0} to \ref{fig:plaque2} show predicted (left panels)
and observed distributions in the ([Fe/H],[Mg/Fe]) plane within 30 bins in
the $(J_\phi,J_z)$ plane -- the actions at the centre of the bin are given at the bottom of
the bin's model panel. As one proceeds up each column, the mean value
of $J_z$ increases, so the bulge or thin disc dominates the bottom panels of each
column. As one proceeds from column to column, the mean value of $J_\phi$
increases, so the typical Galactocentric radius of stars increases from the
left column of Fig.~\ref{fig:plaque0} to the right column of
Fig.~\ref{fig:plaque2}. Fig.~10 of \cite{Eilers2022} show similar data in a
slightly different representation by plotting $\sqrt{J_r^2+J_z^2}$ vertically
rather than $J_z$.

In Fig.~\ref{fig:plaque0} the largest discrepancies between data and model occur around $J_\phi=0$.
This region of action space will be dominated by the stellar halo and the
bulge and we have no confidence in the functional forms we have have adopted
for their DFs, so it is not surprising that the model performs least well
there. In fact it is encouraging that the model does capture the main features
of the chemistry even there: in particular a strongly populated ridge that slopes from
([Fe/H],[Mg/Fe])$\simeq(0.4,0.05)$ up towards $(-0.08,0.38)$, which has
conentrations towards each end that vary in importance with $J_z$.

At $J_\phi\sim500\H$ and $J_z<50\H$ (bottom of the right column of
Fig.~\ref{fig:plaque0}) the model under-populates the principal ridge at high
[Fe/H] and low [Mg/Fe].

Fig.~\ref{fig:plaque1}, which covers $750\H<J_\phi<1750\H$, shows
good agreement between model and data at all values of $J_z$ except the
largest: at the top of he left column the model struggles to reproduce a
population of metal-rich, low-$\alpha$ stars.  This issue with a deficit of
metal-rich, low-$\alpha$ stars at high $J_z$ is the only significant
shortcoming of the fits shown in Fig.~\ref{fig:plaque2} for the largest angular
momenta.  Data for the stars that the model fails to provide is sparse and most
liable to observational error, so it is not clear how significant
the deficit in the model is.

\begin{figure}
\includegraphics[width=\hsize]{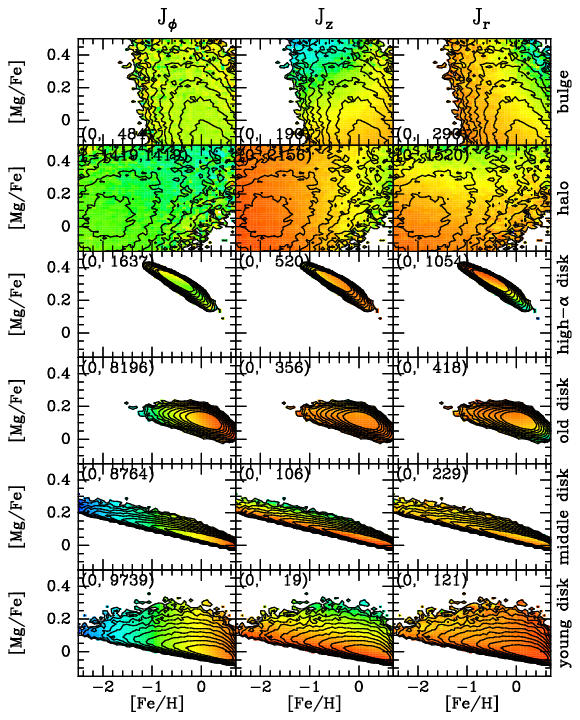}
\caption{The components in chemical space.  Each row shows
the structure of the component listed on the right-hand edge. Contours show
the density of stars of the given chemistry while colours show, from left to
right, the mean 
values of $J_\phi$, $J_z$ and $J_r$. The numbers in brackets give the values
in $\!\H$ corresponding to red and blue hues. The density increases by a
factor 2 between adjacent contours.}\label{fig:extraplot}
\end{figure}

\subsubsection{Chemical parameters}

Table~\ref{tab:chem} gives the values of the parameters of the chemical model
that yields the fits shown in Figs.~\ref{fig:plaque0} to \ref{fig:plaque2},
while each row of Fig.~\ref{fig:extraplot} displays the chemodynamical
structure of the component listed along the figure's right-hand edge.
Contours show the density of stars with a given chemistry that one obtains by
sampling the DF without any selection function and then using the chemical
model of Table~\ref{tab:chem} to assign a chemistry to each selected star.
This plot differs from all our other plots in being representative of what
the model claims is out there, rather than what APOGEE can see. In the left
column each pixel is coloured by the mean of the $J_\phi$ values of the stars
in that pixel, while the centre and right columns are coloured by the mean
values of $J_z$ and $J_r$, respectively. The numbers in brackets in each
panel show the values (in $\!\H$) associated with red and blue. 

{\rd 
The contours in the top row of Fig.~\ref{fig:extraplot} show that the main
body of the bulge is modelled as a very metal-rich, roughly $\alpha$-normal
structure. From this there extends a tail of stars with progressively higher
[Mg/Fe] and slightly lower [Fe/H]. The ridge line is similar to the standard
evolutionary trajectory of a population that has a high star-formation rate
\citep[e.g.][]{SchoenrichB2009b}. The colours in the top row show that in the
bulge chemistry barely depends on $J_\phi$ but [Mg/H] increases
with $J_z$ and [Fe/H] increases with $J_r$.

The second row in Fig.~\ref{fig:extraplot} shows that the stellar halo is
modelled as a metal-poor structure ($x_0=-1.1,\,y_0=0.28$) widely scattered
over the chemical plane ($\sigma_x=0.51,\,\sigma_y=0.13$). Its chemistry
depends little on $J_\phi$ but significantly on $J_z$ and $J_r$ in the sense
of [Fe/H] increasing with $J_z$ and [Mg/H] increasing with $J_r$ -- the
reverse of the dependencies in found the bulge.  The stellar halo is now
believed to owe much to the `Enceladus' merger event, which formed a mildly
counter-rotating, relatively metal-rich, radially anisotropic and flattened
component of the halo \citep{Belokurov2018,Helmi2018,MyeongEvans2018}.
According to this picture the mean value of $J_r$ should increase with [Fe/H]
while the mean value of $J_z$ should decrease with [Fe/H]. These are the
trends seen in the bulge and not those Fig.~\ref{fig:extraplot} implies for
the halo.

The third row of Fig.~\ref{fig:extraplot} shows that the high-$\alpha$ disc
covers a narrow range in chemistry, all at large [Mg/H] as we expect of a
component that formed before type Ia supernovae became important. The
narrowness of its footprint reflects a notably small dispersion
$\sigma_y=0.0061$, 24 times smaller than its dispersion in $x$.
Table~\ref{tab:chem} shows that [Fe/H] depends only weakly on $J_\phi$ and in the
sense that metallicity {\it increases} withe $J_\phi$ and therefore radius.
\cite{SchoenPJM} discuss the origin of this `inverse metallicity gradient'.
[Fe/H] depends quite strongly on $J_r$ and $J_z$ but in opposing directions:
increasing with $J_r$ ad decreasing with $J_z$.

The bottom three rows of Fig.~\ref{fig:extraplot}, for the thin-disc
components, show distributions that slope in the same sense as the bulge and
thick disc but more gradually so they do not reach such large values of
[Mg/Fe]. In these components the dominant dependence of chemistry on actions
is upon $J_\phi$ and the value
$C_{1,J_\phi}=-0.363\times10^{-3}\hbox{\,dex}/\H$ given for the middle disc
implies $\d\hbox{[Fe/H]}/\d R\simeq-0.085\,\hbox{dex}\kpc^{-1}$, consistent
with the values $-0.077\pm0.013$ to $-0.059\pm0.012\,\hbox{dex}\kpc^{-1}$
deduced by \cite{MendezDelgado2022} for interstellar N and O, respectively.

The young and middle discs extend to much lower [Fe/H] than the old disc,
which is a priori surprising.  The reason is that the these discs have lager
scale actions $J_{\phi0}$ so they are predicted to extend to larger radii
and higher values of $J_\phi$.  Given the negative value of
$\d\hbox{[Fe/H]}/\d J_\phi$, the outer (largely unobserved) parts of these
discs are predicted to be very metal-poor. It is likely that
$|\d\hbox{[Fe/H]}/\d J_\phi|$ diminishes for $J_\phi$ much larger than the
solar value, with the consequence that $\ex{\hbox{[Fe/H]}}$ does not fall to
the low values predicted under our assumption of constant $\d\hbox{[Fe/H]}/\d
J_\phi$.

The middle, old and high-$\alpha$  discs have $C_{1,J_r}>0$ implying that
metallicity increases with eccentricity. Radial migration, combined with
radial gradients in metallicity and $J_r$ can generate this correlation: at a
given value of $J_\phi$ there are stars that migrated outwards and inwards.
On average, the former will have higher metallicities and radial actions than
the latter, thus inducing a correlation between metallicity and $J_r$ at
given $J_\phi$. In the  young disc, by contrast, [Fe/H] decreases with $J_r$.

The chemistry of the young disc depends strongly on $J_r$ and very strongly
on $J_z$. In part this may reflect the smaller range in actions within this
cool component. Fig.~\ref{fig:extraplot} shows that [Mg/H] increases strongly
with $J_z$ and decreases with $J_r$. The young disc's value of $\sigma_y=0.0066$ is
small so significant values of [Mg/H] are reached when $J_z$ is abnormally
large. This fact accounts for the sharp lower boundary of the thin disc's
footprint in the bottom panels of Fig.~\ref{fig:extraplot}.

All thin-disc components  have $C_{2,J_\phi}>0$, implying an outward
increase in [Mg/Fe], while for the high-$\alpha$ discs $C_{2,J_\phi}<0$. Thus
the signs of both $C_{1,J_\phi}$ and $C_{2,J_\phi}$ reverse as one passes to
the high-$\alpha$ disc from the  thin disc. Given that high [Mg/Fe] is the
signature of a high star-formation rate, it is natural for $C_{2,J_\phi}$ to
be negative in the absence of radial migration.

The discs' values of $C_{2,J_r}$ decrease along the young -- old
sequence so [Mg/Fe] increases with eccentricity in the young disc and
decreases with eccentricity in the old and high-$\alpha$ discs.  $C_{2,J_z}$
is positive but decreasing along the young -- high-$\alpha$ sequence. 
A picture in which [Mg/Fe]
increases with age and stars are secularly scattered away from planar,
circular orbits implies that [Mg/Fe] should increase with eccentricity and
inclination. Table~\ref{tab:chem} shows that it does increase with
inclination but in the older components it is flat or decreasing with eccentricity.
}

\begin{figure*}
\centerline{\includegraphics[width=.4\hsize]{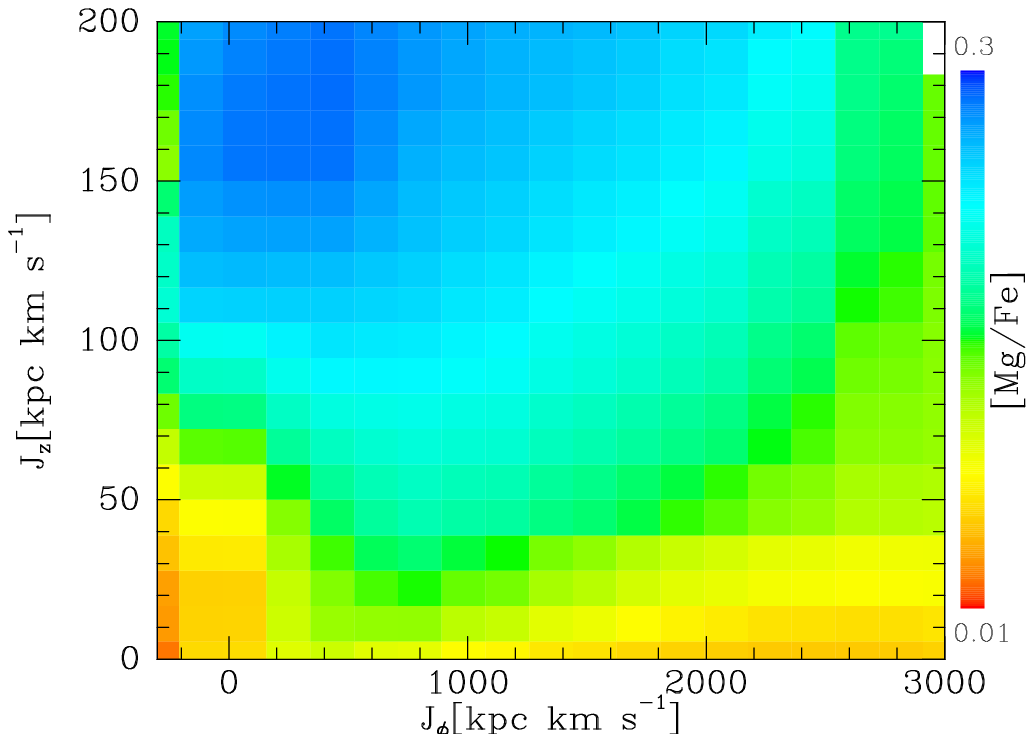}\ 
\includegraphics[width=.4\hsize]{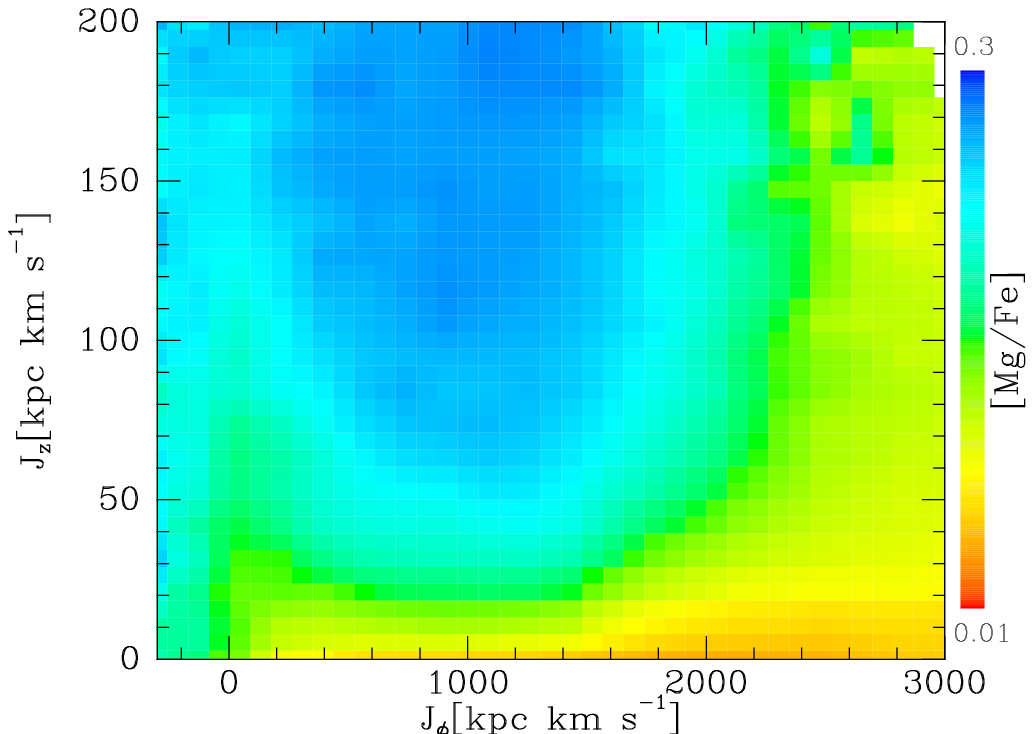}
}
\centerline{\includegraphics[width=.4\hsize]{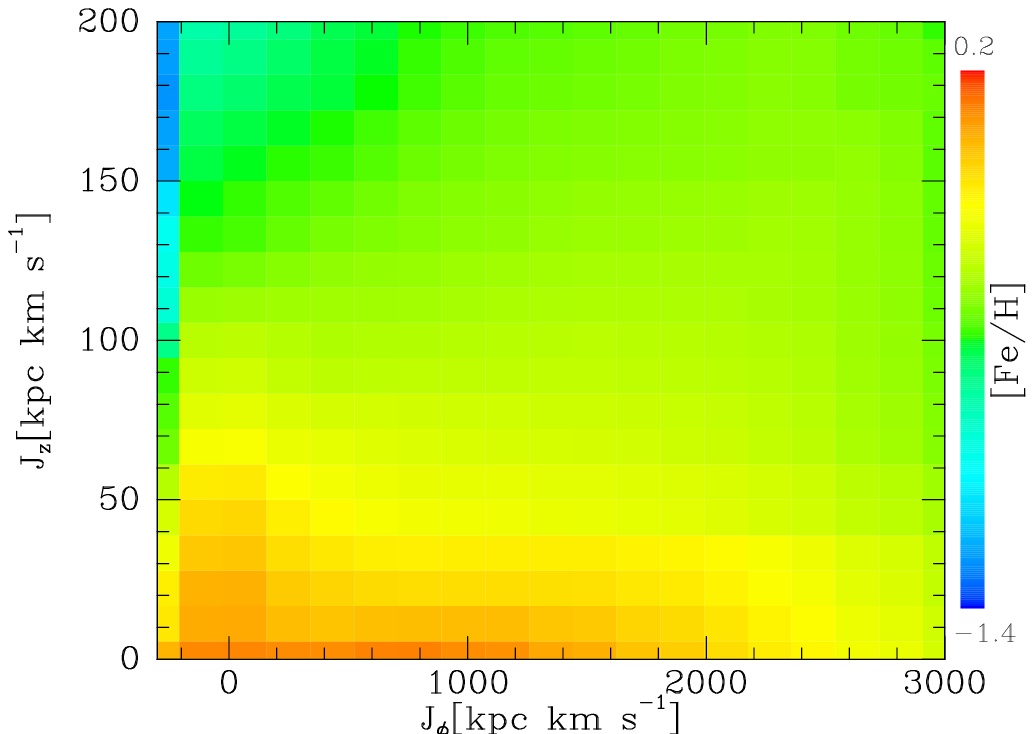}\ 
\includegraphics[width=.4\hsize]{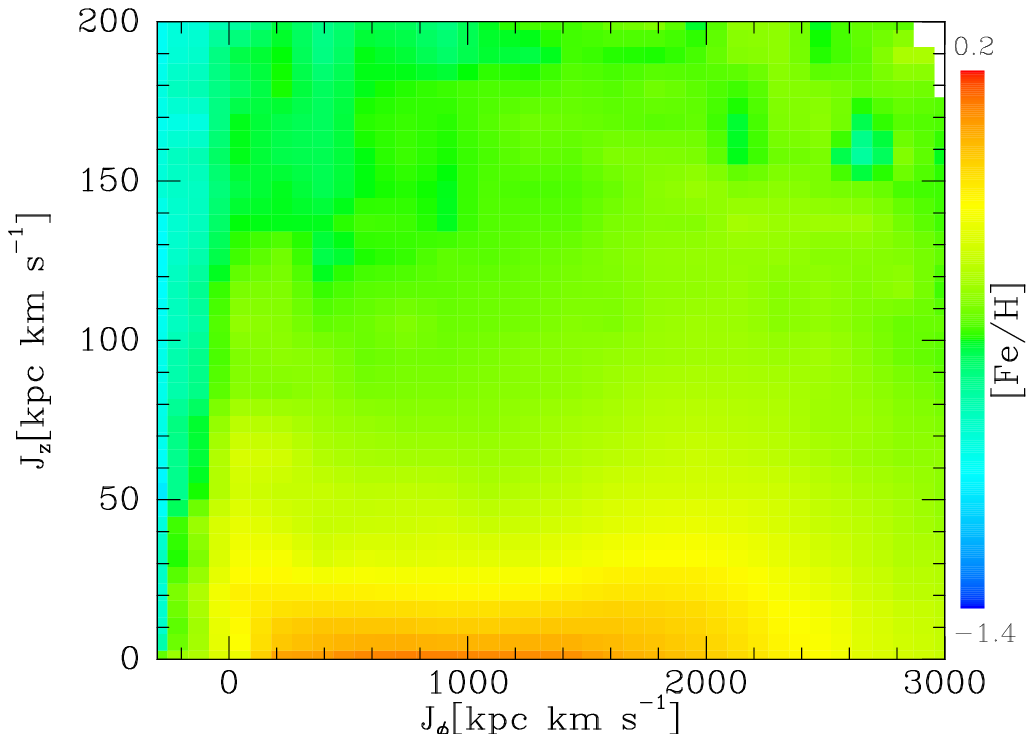}
}
\caption{Mean values of [Mg/Fe] (upper panels) and [Fe/H] (lower panels) in
the $(J_\phi,J_z)$ plane from the observations (right panels) and as
predicted by the model (left panels).}\label{fig:global}
\end{figure*}

The left columns of Figs~\ref{fig:LzFeBasic} and \ref{fig:LzFeBasicFe} show
the mean values of observed [Mg/Fe] and [Fe/H] split into tranches of $J_r$.
The right panels of Fig.~\ref{fig:global} show the same data but regardless
of $J_r$. The U-shaped boundary to the (blue) region dominated by the
high-$\alpha$ disc is now spectacularly evident, and in the lower plot the
metal-rich thin disc is seen to be bounded below by $J_\phi\simeq200\H$ and
above by $J_\phi\simeq2000\H$, interestingly coinciding with the right-hand
boundary of high-$\alpha$ disc.

The left panels of Fig.~\ref{fig:global} show the model's version of these
plots. The principal features are captured but the match is far from perfect.
Most notably, in the upper left plot the boundary of the U-shaped region is
insufficiently sharp {\rd and [Mg/H] is too low at $J_\phi\la0$. In the lower left
plot [Fe/H] is too high at low $J_z$ and $J_\phi\simeq0$.}

\section{Comparison with the Besan\c con model}\label{sec:Besancon}

A natural question is how the present model compares with the `Besan\c con
Galaxy model' (BGM), which has recently been updated by \cite{Robin2022} to
self-consistency in the context of Gaia DR3 astrometry. Major differences
include: (i) In the BGM the bulge, the stellar halo and the dark halo were
represented by pre-defined density distributions rather than DFs. The BGM
bulge is spherical while the two haloes are ellipsoidal; (ii) In the BGM the
integrals used are $(E,J_\phi,I_3)$ rather than $\vJ$; (iii) The BGM was
fitted to sky-plane velocities $(V_\ell,V_b)$ computed using inverse
parallaxes rather than StarHorse distances, and line-of-sight velocities were
not used; (iv) The BGM was fitted to (a) the $(V_\ell,V_b)$ distributions of
stars within $100\pc$ of the Sun viewed in 36 directions, and (b) the
$(V_\ell,V_b)$ and parallax distributions of more distant stars viewed along
26 lines of sight. Fits were made to the $0.1,\,0.25,\,0.5,\,0.75$ and $0.9$
quantiles in $V_\ell$ and $V_b$ rather than to densities in three-dimensional
velocity space at 72 locations in $(R,|z|)$. {\rd(v) Rather than ignoring age
distributions as we have done, the BGM embodies them by being fitted to
colour-magnitude diagrams at various locations;} (vi) Modelling the parallax
distributions required engagement with the Gaia scanning law, and adoption of
both a dust model and a luminosity functions for each DF. (vi) The Galaxy's
circular-speed curve was an input to the BGM model whereas here it follows
from the fits to the kinematics.

While there are significant differences in the data employed and how the
model parameters were fitted, the two models have similar scope: both are
axisymmetric, self-consistent dynamical models from which mock samples can be
drawn by specifying selection criteria. The present model does not come with
pre-defined luminosity functions but the new version of \agama\ provides for
each DF to include a luminosity function appropriate to the component's age
and chemistry. The new version defines classes for lines of sight and dust
models, so magnitude-limited samples can be drawn for any line of
sight.

\section{Conclusions}\label{sec:conclude}

The APOGEE spectroscopic survey combined with Gaia astrometry casts a
brilliant light on the chemodynamical structure of our Galaxy. From DR17 of
the Sloan Digital Sky Survey we selected just under 218\,000 giant stars that
are unlikely to belong to globular clusters and have reliable distances in
the Bayesian StarHorse catalogue \citep{StarHorse2}. The selected
stars cover the radial range $0.5<R<14.5\kpc$ and lie within $4\kpc$ of the
plane. 

We have avoided engaging with the complex selection function of the APOGEE
survey by examining the kinematic and chemical distributions within 72
spatial bins that extend from $R=1\kpc$ to $R=14\kpc$ at $|z|<4\kpc$.

We have used these distributions to update and extend the self-consistent
dynamical model of BV23. This model is defined by distribution functions
$f(\vJ)$ for six stellar components and the dark halo, together with a
prescribed surface density of gas. The gravitational potential that these
eight ingredients jointly generate was computed iteratively and the resulting
predictions for the kinematics of stars at 72 spatial locations were compared
with the data.

The functional forms adopted for the DFs differ from those used by BV23. A
major difference is that here the bulge is assumed to be a fat disc rather
than a spheroidal component. Another difference is that here the
high-$\alpha$ disc is taken to be radially truncated. A minor difference is
in the way the variables $J_\d$ and $J_{\rm v}$ introduced by \cite{AGAMA}
are defined.

Another difference with BV23 is in how the kinematic data are fitted: whereas
BV23 fitted by hand histograms of $V_R$, $V_z$  and $V_\phi$ marginalised over the
other two components of velocity, we have used the Nelder-Mead algorithm to
fit the predicted densities of stars in 72 three-dimensional velocity spaces.

In a major extension of BV23 we have added a chemical dimension to the model
by assigning to each stellar DF a probability density in the chemical space
$\vc=(\hbox{[Fe/H]},\hbox{[Mg/Fe]})$. These probability densities are
Gaussians with mean values that are linear functions of $\vJ$. 

We examined the variation in action space of the mean values of [Mg/Fe] and
[Fe/H]. Plots of this distribution (Figs.~\ref{fig:LzFeBasic} and
\ref{fig:global}) show a sharp
boundary at $J_\phi=0$ that is a testament to the accuracy of the StarHorse
distances we have used. It also implies that within the (wide) spatial region
covered by the data, few bulge stars counter-rotate,
so the bulge should not be modelled as a spheroidal component but as a hot
disc. In addition to the sharp boundary at $J_\phi=0$, plots of
$\ex{\hbox{[Mg/Fe]}}$ show a sharp transition at $J_\phi\simeq2000\H$ that
we interpret as radial truncation of the high-$\alpha$ disc.

{\rd Devising a mathematical formula for the density of objects in a
five-dimensional space is challenging. We have broken the task into dynamical
and chemical parts. We consider that the dynamical problem has been
satisfactorily solved as regards disc stars, but less satisfactorily as
regards the bulge and the spheroidal components. Given a DF, one has to
quantify the density of its stars in chemical space, which the data show to
be a function of actions: $C_{1,J_\phi}$ is the only gradient coefficient
conventionally considered, but Table~\ref{tab:chem} shows that it is by no
means the only significant coefficient, even bearing in mind that $J_\phi$
spans a wider range than the other two actions.  The simplest ansatz for a
probability density is a Gaussian, but the Gaussian's properties must be
continuous functions of the actions. We have implemented the simplest,
non-trivial scheme, namely similar Gaussians centred on means that depend
linearly on the actions. Even though this scheme is very basic, we ended up
needing 70 parameters for the chemical pdf.

Alternatively, we could have
followed \cite{SaJJB15:EDF} and used a model inspired by the notion that disc
stars diffuse from circular orbits. We were discouraged from pursuing this
idea by unsuccessful attempts to model the  APOGEE  data published in
\cite{Hayden2015} by the approach of \cite{SaJJB15:EDF}, but the proposal
might be worth revisiting.
A radically different scheme would be to represent the logarithm of the star
density
as a sum of basis functions, such as wavelets. A successful scheme would
combine flexibility, a low parameter count and intrinsic non-negativity.
Given the very basic nature of our ansatz, the fits to data shown in
Figs.~\ref{fig:plaque0} to \ref{fig:plaque2} may be counted a success.
  
 We followed BV23 in representing the age distribution in the thin
disc with three distinct discs, rather than following \cite{JJB12:dfs} and
\cite{SaJJB15:EDF} in using a
single thin-disc DF that includes an age parameter. Given the limited success
of the discrete approach, a return to continuous variation with age may be
wise. The challenge is to model in a flexible yet parameter-poor way the
dependence upon age of both the dynamical and the chemical parameters.
}

It is likely that the replacement of the halo and bulge DFs by better
functional forms would make it possible to obtain better fits even without
restructuring the chemical model. DFs that are not confined to one sign of
$J_\phi$ are subject to subtle constraints, as will be explained elsewhere
(Binney in preparation). The parameters of the functional form used for the
stellar and dark-matter DFs cannot be freely varied if unphysical kinematics
are to be avoided, with the consequence that in the model neither component
is as radially biased as it probably should be. The data imply that the bulge
like the discs lies overwhelmingly at $J_\phi>0$ but its DF should surely not
vanish completely at $J_\phi<0$, as it does in the model. This artificial
vanishing of the bulge DF may be responsible for the poor match of the
model's circular-speed curve to the points from \cite{Wegg2013} in
Fig.~\ref{fig:Vc}. Including in the modelling some observational bias against
low-metallicity stars may also improve fits.

{\rd Our `to do' list is as follows: (i) replace the quality cut on distances
by cuts based on uncertainties in $\ln R$ and $\ln|z|$, and possibly relax
quality cuts on low-metallicity stars in order to reduce biases against halo
stars; (ii) replace the DFs of dark and stellar haloes and explore more
anisotropic DFs for these components; (iii) try to replace the three
thin-disc DFs by a single DF that includes age-dependent dynamics and
chemistry; (iv) add age data.  
}

A model such as ours can be used to make innumerable predictions regarding
orbits around the Galaxy and the density of stars of given chemistry at any
location in phase-space; here we have shown only a tiny selection of such
predictions. For example, it would be interesting to produce plots like Fig.~10 of
\cite{Eilers2022} or to fit the functional forms of \cite{Lian2022} to the
spatial structures of our low- and high-$\alpha$ discs and compare with the
values Lian et al obtained.  Using the \agama\ package it is easy to make
predictions in seconds by drawing from the model samples of stars with known
component memberships.  In
addition to yielding mock observations, such samples can be used as initial
conditions for an N-body simulation. \cite{AGAMA} showed that simulations
started in this way form precisely equilibrium systems, and subtle departures
from equilibrium can be explored by slightly shifting the phase-space
coordinated returned by \agama\ before advancing the N-body model in time.

\section*{DATA AVAILABILITY}
The code that generates Galaxy  models and fits the present chemical models can be downloaded from the \agama\
website https://github.com/GalacticDynamics-Oxford/Agama.

\def\physrep{Phys.~Reps}
\def\apss{Ast.~Phys.~Sp.~Sci.}
\def\fcp{Fund.~Cosmic~Phys.}
\bibliographystyle{mn2e} \bibliography{/u/tex/papers/mcmillan/torus/new_refs}

\begin{thebibliography}{68}
\expandafter\ifx\csname natexlab\endcsname\relax\def\natexlab#1{#1}\fi

\bibitem[{{Abdurro'uf} {et~al}\mbox{.}(2022){Abdurro'uf}, {Accetta}, {Aerts},
  {Silva Aguirre}, {Ahumada}, {Ajgaonkar}, {Filiz Ak}, {Alam}, {Allende
  Prieto}, {Almeida}, {Anders}, {Anderson}, {Andrews}, {Anguiano},
  {Aquino-Ort{\'\i}z}, {Arag{\'o}n-Salamanca}, {Argudo-Fern{\'a}ndez}, {Ata},
  {Aubert}, {Avila-Reese}, {Badenes}, {Barb{\'a}}, {Barger},
  {Barrera-Ballesteros}, {Beaton}, {Beers}, {Belfiore}, {Bender}, {Bernardi},
  {Bershady}, {Beutler}, {Bidin}, {Bird}, {Bizyaev}, {Blanc}, {Blanton},
  {Boardman}, {Bolton}, {Boquien}, {Borissova}, {Bovy}, {Brandt}, {Brown},
  {Brownstein}, {Brusa}, {Buchner}, {Bundy}, {Burchett}, {Bureau}, {Burgasser},
  {Cabang}, {Campbell}, {Cappellari}, {Carlberg}, {Wanderley}, {Carrera},
  {Cash}, {Chen}, {Chen}, {Cherinka}, {Chiappini}, {Choi}, {Chojnowski},
  {Chung}, {Clerc}, {Cohen}, {Comerford}, {Comparat}, {da Costa}, {Covey},
  {Crane}, {Cruz-Gonzalez}, {Culhane}, {Cunha}, {Dai}, {Damke}, {Darling},
  {Davidson}, {Davies}, {Dawson}, {De Lee}, {Diamond-Stanic}, {Cano-D{\'\i}az},
  {S{\'a}nchez}, {Donor}, {Duckworth}, {Dwelly}, {Eisenstein}, {Elsworth},
  {Emsellem}, {Eracleous}, {Escoffier}, {Fan}, {Farr}, {Feng},
  {Fern{\'a}ndez-Trincado}, {Feuillet}, {Filipp}, {Fillingham}, {Frinchaboy},
  {Fromenteau}, {Galbany}, {Garc{\'\i}a}, {Garc{\'\i}a-Hern{\'a}ndez}, {Ge},
  {Geisler}, {Gelfand}, {G{\'e}ron}, {Gibson}, {Goddy}, {Godoy-Rivera},
  {Grabowski}, {Green}, {Greener}, {Grier}, {Griffith}, {Guo}, {Guy},
  {Hadjara}, {Harding}, {Hasselquist}, {Hayes}, {Hearty}, {Hern{\'a}ndez},
  {Hill}, {Hogg}, {Holtzman}, {Horta}, {Hsieh}, {Hsu}, {Hsu}, {Huber},
  {Huertas-Company}, {Hutchinson}, {Hwang}, {Ibarra-Medel}, {Chitham}, {Ilha},
  {Imig}, {Jaekle}, {Jayasinghe}, {Ji}, {Johnson}, {Jones}, {J{\"o}nsson},
  {Katkov}, {Khalatyan}, {Kinemuchi}, {Kisku}, {Knapen}, {Kneib}, {Kollmeier},
  {Kong}, {Kounkel}, {Kreckel}, {Krishnarao}, {Lacerna}, {Lane}, {Langgin},
  {Lavender}, {Law}, {Lazarz}, {Leung}, {Leung}, {Lewis}, {Li}, {Li}, {Lian},
  {Liang}, {Lin}, {Lin}, {Lin}, {Lintott}, {Long}, {Longa-Pe{\~n}a},
  {L{\'o}pez-Cob{\'a}}, {Lu}, {Lundgren}, {Luo}, {Mackereth}, {de la Macorra},
  {Mahadevan}, {Majewski}, {Manchado}, {Mandeville}, {Maraston},
  {Margalef-Bentabol}, {Masseron}, {Masters}, {Mathur}, {McDermid}, {Mckay},
  {Merloni}, {Merrifield}, {Meszaros}, {Miglio}, {Di Mille}, {Minniti},
  {Minsley}, {Monachesi}, {Moon}, {Mosser}, {Mulchaey}, {Muna}, {Mu{\~n}oz},
  {Myers}, {Myers}, {Nadathur}, {Nair}, {Nandra}, {Neumann}, {Newman},
  {Nidever}, {Nikakhtar}, {Nitschelm}, {O'Connell}, {Garma-Oehmichen}, {Luan
  Souza de Oliveira}, {Olney}, {Oravetz}, {Ortigoza-Urdaneta}, {Osorio},
  {Otter}, {Pace}, {Padilla}, {Pan}, {Pan}, {Parikh}, {Parker}, {Peirani},
  {Pe{\~n}a Ram{\'\i}rez}, {Penny}, {Percival}, {Perez-Fournon},
  {Pinsonneault}, {Poidevin}, {Poovelil}, {Price-Whelan}, {B{\'a}rbara de
  Andrade Queiroz}, {Raddick}, {Ray}, {Rembold}, {Riddle}, {Riffel}, {Riffel},
  {Rix}, {Robin}, {Rodr{\'\i}guez-Puebla}, {Roman-Lopes},
  {Rom{\'a}n-Z{\'u}{\~n}iga}, {Rose}, {Ross}, {Rossi}, {Rubin}, {Salvato},
  {S{\'a}nchez}, {S{\'a}nchez-Gallego}, {Sanderson}, {Santana Rojas},
  {Sarceno}, {Sarmiento}, {Sayres}, {Sazonova}, {Schaefer}, {Schiavon},
  {Schlegel}, {Schneider}, {Schultheis}, {Schwope}, {Serenelli}, {Serna},
  {Shao}, {Shapiro}, {Sharma}, {Shen}, {Shetrone}, {Shu}, {Simon}, {Skrutskie},
  {Smethurst}, {Smith}, {Sobeck}, {Spoo}, {Sprague}, {Stark}, {Stassun},
  {Steinmetz}, {Stello}, {Stone-Martinez}, {Storchi-Bergmann}, {Stringfellow},
  {Stutz}, {Su}, {Taghizadeh-Popp}, {Talbot}, {Tayar}, {Telles}, {Teske},
  {Thakar}, {Theissen}, {Tkachenko}, {Thomas}, {Tojeiro}, {Hernandez Toledo},
  {Troup}, {Trump}, {Trussler}, {Turner}, {Tuttle}, {Unda-Sanzana},
  {V{\'a}zquez-Mata}, {Valentini}, {Valenzuela}, {Vargas-Gonz{\'a}lez},
  {Vargas-Maga{\~n}a}, {Alfaro}, {Villanova}, {Vincenzo}, {Wake}, {Warfield},
  {Washington}, {Weaver}, {Weijmans}, {Weinberg}, {Weiss}, {Westfall}, {Wild},
  {Wilde}, {Wilson}, {Wilson}, {Wilson}, {Wolf}, {Wood-Vasey}, {Yan}, {Zamora},
  {Zasowski}, {Zhang}, {Zhao}, {Zheng}, {Zheng}, \& {Zhu}}]{APOGEE17}
{Abdurro'uf} {et~al.}, 2022, \apjs, 259, 35

\bibitem[{{Ablimit} {et~al}\mbox{.}(2020){Ablimit}, {Zhao}, {Flynn}, \&
  {Bird}}]{Ablimit2020}
{Ablimit} I., {Zhao} G., {Flynn} C., {Bird} S.~A., 2020, \apjl, 895, L12

\bibitem[{{Anders} {et~al}\mbox{.}(2022){Anders}, {Khalatyan}, {Queiroz},
  {Chiappini}, {Ard{\`e}vol}, {Casamiquela}, {Figueras}, {Jim{\'e}nez-Arranz},
  {Jordi}, {Mongui{\'o}}, {Romero-G{\'o}mez}, {Altamirano}, {Antoja}, {Assaad},
  {Cantat-Gaudin}, {Castro-Ginard}, {Enke}, {Girardi}, {Guiglion}, {Khan},
  {Luri}, {Miglio}, {Minchev}, {Ramos}, {Santiago}, \&
  {Steinmetz}}]{StarHorse2}
{Anders} F. {et~al.}, 2022, \aap, 658, A91

\bibitem[{{Belokurov} {et~al}\mbox{.}(2018){Belokurov}, {Erkal}, {Evans},
  {Koposov}, \& {Deason}}]{Belokurov2018}
{Belokurov} V., {Erkal} D., {Evans} N.~W., {Koposov} S.~E., {Deason} A.~J.,
  2018, \mnras, 478, 611

\bibitem[{{Binney}(2012)}]{JJB12:dfs}
{Binney} J., 2012, \mnras, 426, 1328

\bibitem[{{Binney} \& {Lacey}(1988)}]{BinneyLacey1988}
{Binney} J., {Lacey} C., 1988, \mnras, 230, 597

\bibitem[{{Binney} \& {McMillan}(2011)}]{JJBPJM11:dyn}
{Binney} J., {McMillan} P.~J., 2011, \mnras, 413, 1889

\bibitem[{{Binney} \& {McMillan}(2016)}]{JJBPJM16}
{Binney} J., {McMillan} P.~J., 2016, \mnras, 456, 1982

\bibitem[{{Binney} \& {Tremaine}(2008)}]{GDII}
{Binney} J., {Tremaine} S., 2008, {Galactic Dynamics: Second Edition}.
  Princeton University Press

\bibitem[{{Binney} \& {Vasiliev}(2023)}]{BinneyVasiliev2023}
{Binney} J., {Vasiliev} E., 2023, \mnras, 520, 1832

\bibitem[{{Bovy} {et~al}\mbox{.}(2012){Bovy}, {Rix}, {Liu}, {Hogg}, {Beers}, \&
  {Lee}}]{BovyRix2012}
{Bovy} J., {Rix} H.-W., {Liu} C., {Hogg} D.~W., {Beers} T.~C., {Lee} Y.~S.,
  2012, \apj, 753, 148

\bibitem[{{Brook} {et~al}\mbox{.}(2004){Brook}, {Kawata}, {Gibson}, \&
  {Freeman}}]{Brook2004}
{Brook} C.~B., {Kawata} D., {Gibson} B.~K., {Freeman} K.~C., 2004, \apj, 612,
  894

\bibitem[{{Burbidge} {et~al}\mbox{.}(1957){Burbidge}, {Burbidge}, {Fowler}, \&
  {Hoyle}}]{BBFH1957}
{Burbidge} E.~M., {Burbidge} G.~R., {Fowler} W.~A., {Hoyle} F., 1957, Reviews
  of Modern Physics, 29, 547

\bibitem[{{Chen} {et~al}\mbox{.}(2022){Chen}, {Hayden}, {Sharma},
  {Bland-Hawthorn}, {Kobayashi}, \& {Karakas}}]{ChenHayden2022}
{Chen} B., {Hayden} M.~R., {Sharma} S., {Bland-Hawthorn} J., {Kobayashi} C.,
  {Karakas} A.~I., 2022, arXiv e-prints, arXiv:2204.11413

\bibitem[{{Chiappini} {et~al}\mbox{.}(1997){Chiappini}, {Matteucci}, \&
  {Gratton}}]{Chiappini1997}
{Chiappini} C., {Matteucci} F., {Gratton} R., 1997, \apj, 477, 765

\bibitem[{{de Salas} \& {Widmark}(2021)}]{deSalas2021}
{de Salas} P.~F., {Widmark} A., 2021, Reports on Progress in Physics, 84,
  104901

\bibitem[{{Deason} {et~al}\mbox{.}(2019){Deason}, {Belokurov}, \&
  {Sanders}}]{DeasonVasilyJason2019}
{Deason} A.~J., {Belokurov} V., {Sanders} J.~L., 2019, \mnras, 490, 3426

\bibitem[{{Eggen} {et~al}\mbox{.}(1962){Eggen}, {Lynden-Bell}, \&
  {Sandage}}]{ELS1962}
{Eggen} O.~J., {Lynden-Bell} D., {Sandage} A.~R., 1962, \apj, 136, 748

\bibitem[{{Eilers} {et~al}\mbox{.}(2019){Eilers}, {Hogg}, {Rix}, \&
  {Ness}}]{Eilers2019}
{Eilers} A.-C., {Hogg} D.~W., {Rix} H.-W., {Ness} M.~K., 2019, \apj, 871, 120

\bibitem[{{Eilers} {et~al}\mbox{.}(2022){Eilers}, {Hogg}, {Rix}, {Ness},
  {Price-Whelan}, {M{\'e}sz{\'a}ros}, \& {Nitschelm}}]{Eilers2022}
{Eilers} A.-C., {Hogg} D.~W., {Rix} H.-W., {Ness} M.~K., {Price-Whelan} A.~M.,
  {M{\'e}sz{\'a}ros} S., {Nitschelm} C., 2022, \apj, 928, 23

\bibitem[{{Freeman} \& {Bland-Hawthorn}(2002)}]{FrBH02}
{Freeman} K., {Bland-Hawthorn} J., 2002, \araa, 40, 487

\bibitem[{{Fuhrmann}(2011)}]{Fuhrmann2011}
{Fuhrmann} K., 2011, \mnras, 414, 2893

\bibitem[{{Gaia Collaboration} {et~al}\mbox{.}(2021){Gaia Collaboration},
  {Brown}, {Vallenari}, {Prusti}, {de Bruijne}, {Babusiaux}, {Biermann},
  {Creevey}, {Evans}, {Eyer}, {Hutton}, {Jansen}, {Jordi}, {Klioner},
  {Lammers}, {Lindegren}, {Luri}, {Mignard}, {Panem}, {Pourbaix}, {Randich},
  {Sartoretti}, {Soubiran}, {Walton}, {Arenou}, {Bailer-Jones}, {Bastian},
  {Cropper}, {Drimmel}, {Katz}, {Lattanzi}, {van Leeuwen}, {Bakker},
  {Cacciari}, {Casta{\~n}eda}, {De Angeli}, {Ducourant}, {Fabricius},
  {Fouesneau}, {Fr{\'e}mat}, {Guerra}, {Guerrier}, {Guiraud}, {Jean-Antoine
  Piccolo}, {Masana}, {Messineo}, {Mowlavi}, {Nicolas}, {Nienartowicz},
  {Pailler}, {Panuzzo}, {Riclet}, {Roux}, {Seabroke}, {Sordo}, {Tanga},
  {Th{\'e}venin}, {Gracia-Abril}, {Portell}, {Teyssier}, {Altmann}, {Andrae},
  {Bellas-Velidis}, {Benson}, {Berthier}, {Blomme}, {Brugaletta}, {Burgess},
  {Busso}, {Carry}, {Cellino}, {Cheek}, {Clementini}, {Damerdji}, {Davidson},
  {Delchambre}, {Dell'Oro}, {Fern{\'a}ndez-Hern{\'a}ndez}, {Galluccio},
  {Garc{\'\i}a-Lario}, {Garcia-Reinaldos}, {Gonz{\'a}lez-N{\'u}{\~n}ez},
  {Gosset}, {Haigron}, {Halbwachs}, {Hambly}, {Harrison}, {Hatzidimitriou},
  {Heiter}, {Hern{\'a}ndez}, {Hestroffer}, {Hodgkin}, {Holl}, {Jan{\ss}en},
  {Jevardat de Fombelle}, {Jordan}, {Krone-Martins}, {Lanzafame},
  {L{\"o}ffler}, {Lorca}, {Manteiga}, {Marchal}, {Marrese}, {Moitinho}, {Mora},
  {Muinonen}, {Osborne}, {Pancino}, {Pauwels}, {Petit}, {Recio-Blanco},
  {Richards}, {Riello}, {Rimoldini}, {Robin}, {Roegiers}, {Rybizki}, {Sarro},
  {Siopis}, {Smith}, {Sozzetti}, {Ulla}, {Utrilla}, {van Leeuwen}, {van
  Reeven}, {Abbas}, {Abreu Aramburu}, {Accart}, {Aerts}, {Aguado}, {Ajaj},
  {Altavilla}, {{\'A}lvarez}, {{\'A}lvarez Cid-Fuentes}, {Alves}, {Anderson},
  {Anglada Varela}, {Antoja}, {Audard}, {Baines}, {Baker},
  {Balaguer-N{\'u}{\~n}ez}, {Balbinot}, {Balog}, {Barache}, {Barbato},
  {Barros}, {Barstow}, {Bartolom{\'e}}, {Bassilana}, {Bauchet},
  {Baudesson-Stella}, {Becciani}, {Bellazzini}, {Bernet}, {Bertone}, {Bianchi},
  {Blanco-Cuaresma}, {Boch}, {Bombrun}, {Bossini}, {Bouquillon}, {Bragaglia},
  {Bramante}, {Breedt}, {Bressan}, {Brouillet}, {Bucciarelli}, {Burlacu},
  {Busonero}, {Butkevich}, {Buzzi}, {Caffau}, {Cancelliere}, {C{\'a}novas},
  {Cantat-Gaudin}, {Carballo}, {Carlucci}, {Carnerero}, {Carrasco},
  {Casamiquela}, {Castellani}, {Castro-Ginard}, {Castro Sampol}, {Chaoul},
  {Charlot}, {Chemin}, {Chiavassa}, {Cioni}, {Comoretto}, {Cooper}, {Cornez},
  {Cowell}, {Crifo}, {Crosta}, {Crowley}, {Dafonte}, {Dapergolas}, {David},
  {David}, {de Laverny}, {De Luise}, {De March}, {De Ridder}, {de Souza}, {de
  Teodoro}, {de Torres}, {del Peloso}, {del Pozo}, {Delbo}, {Delgado},
  {Delgado}, {Delisle}, {Di Matteo}, {Diakite}, {Diener}, {Distefano},
  {Dolding}, {Eappachen}, {Edvardsson}, {Enke}, {Esquej}, {Fabre}, {Fabrizio},
  {Faigler}, {Fedorets}, {Fernique}, {Fienga}, {Figueras}, {Fouron},
  {Fragkoudi}, {Fraile}, {Franke}, {Gai}, {Garabato}, {Garcia-Gutierrez},
  {Garc{\'\i}a-Torres}, {Garofalo}, {Gavras}, {Gerlach}, {Geyer}, {Giacobbe},
  {Gilmore}, {Girona}, {Giuffrida}, {Gomel}, {Gomez}, {Gonzalez-Santamaria},
  {Gonz{\'a}lez-Vidal}, {Granvik}, {Guti{\'e}rrez-S{\'a}nchez}, {Guy},
  {Hauser}, {Haywood}, {Helmi}, {Hidalgo}, {Hilger}, {H{\l}adczuk}, {Hobbs},
  {Holland}, {Huckle}, {Jasniewicz}, {Jonker}, {Juaristi Campillo}, {Julbe},
  {Karbevska}, {Kervella}, {Khanna}, {Kochoska}, {Kontizas}, {Kordopatis},
  {Korn}, {Kostrzewa-Rutkowska}, {Kruszy{\'n}ska}, {Lambert}, {Lanza}, {Lasne},
  {Le Campion}, {Le Fustec}, {Lebreton}, {Lebzelter}, {Leccia}, {Leclerc},
  {Lecoeur-Taibi}, {Liao}, {Licata}, {Lindstr{\o}m}, {Lister}, {Livanou},
  {Lobel}, {Madrero Pardo}, {Managau}, {Mann}, {Marchant}, {Marconi}, {Marcos
  Santos}, {Marinoni}, {Marocco}, {Marshall}, {Martin Polo},
  {Mart{\'\i}n-Fleitas}, {Masip}, {Massari}, {Mastrobuono-Battisti}, {Mazeh},
  {McMillan}, {Messina}, {Michalik}, {Millar}, {Mints}, {Molina}, {Molinaro},
  {Moln{\'a}r}, {Montegriffo}, {Mor}, {Morbidelli}, {Morel}, {Morris},
  {Mulone}, {Munoz}, {Muraveva}, {Murphy}, {Musella}, {Noval}, {Ord{\'e}novic},
  {Orr{\`u}}, {Osinde}, {Pagani}, {Pagano}, {Palaversa}, {Palicio}, {Panahi},
  {Pawlak}, {Pe{\~n}alosa Esteller}, {Penttil{\"a}}, {Piersimoni}, {Pineau},
  {Plachy}, {Plum}, {Poggio}, {Poretti}, {Poujoulet}, {Pr{\v{s}}a}, {Pulone},
  {Racero}, {Ragaini}, {Rainer}, {Raiteri}, {Rambaux}, {Ramos}, {Ramos-Lerate},
  {Re Fiorentin}, {Regibo}, {Reyl{\'e}}, {Ripepi}, {Riva}, {Rixon}, {Robichon},
  {Robin}, {Roelens}, {Rohrbasser}, {Romero-G{\'o}mez}, {Rowell}, {Royer},
  {Rybicki}, {Sadowski}, {Sagrist{\`a} Sell{\'e}s}, {Sahlmann}, {Salgado},
  {Salguero}, {Samaras}, {Sanchez Gimenez}, {Sanna}, {Santove{\~n}a},
  {Sarasso}, {Schultheis}, {Sciacca}, {Segol}, {Segovia}, {S{\'e}gransan},
  {Semeux}, {Shahaf}, {Siddiqui}, {Siebert}, {Siltala}, {Slezak}, {Smart},
  {Solano}, {Solitro}, {Souami}, {Souchay}, {Spagna}, {Spoto}, {Steele},
  {Steidelm{\"u}ller}, {Stephenson}, {S{\"u}veges}, {Szabados}, {Szegedi-Elek},
  {Taris}, {Tauran}, {Taylor}, {Teixeira}, {Thuillot}, {Tonello}, {Torra},
  {Torra}, {Turon}, {Unger}, {Vaillant}, {van Dillen}, {Vanel}, {Vecchiato},
  {Viala}, {Vicente}, {Voutsinas}, {Weiler}, {Wevers}, {Wyrzykowski}, {Yoldas},
  {Yvard}, {Zhao}, {Zorec}, {Zucker}, {Zurbach}, \&
  {Zwitter}}]{GaiaEDR3general}
{Gaia Collaboration} {et~al.}, 2021, \aap, 650, C3

\bibitem[{{Gaia Collaboration} {et~al}\mbox{.}(2023){Gaia Collaboration},
  {Recio-Blanco}, {Kordopatis}, {de Laverny}, {Palicio}, {Spagna}, {Spina},
  {Katz}, {Re Fiorentin}, {Poggio}, {McMillan}, {Vallenari}, {Lattanzi},
  {Seabroke}, {Casamiquela}, {Bragaglia}, {Antoja}, {Bailer-Jones},
  {Schultheis}, {Andrae}, {Fouesneau}, {Cropper}, {Cantat-Gaudin}, {Bijaoui},
  {Heiter}, {Brown}, {Prusti}, {de Bruijne}, {Arenou}, {Babusiaux}, {Biermann},
  {Creevey}, {Ducourant}, {Evans}, {Eyer}, {Guerra}, {Hutton}, {Jordi},
  {Klioner}, {Lammers}, {Lindegren}, {Luri}, {Mignard}, {Panem}, {Pourbaix},
  {Randich}, {Sartoretti}, {Soubiran}, {Tanga}, {Walton}, {Bastian}, {Drimmel},
  {Jansen}, {van Leeuwen}, {Bakker}, {Cacciari}, {Casta{\~n}eda}, {De Angeli},
  {Fabricius}, {Fr{\'e}mat}, {Galluccio}, {Guerrier}, {Masana}, {Messineo},
  {Mowlavi}, {Nicolas}, {Nienartowicz}, {Pailler}, {Panuzzo}, {Riclet}, {Roux},
  {Sordo}, {Th{\'e}venin}, {Gracia-Abril}, {Portell}, {Teyssier}, {Altmann},
  {Audard}, {Bellas-Velidis}, {Benson}, {Berthier}, {Blomme}, {Burgess},
  {Busonero}, {Busso}, {C{\'a}novas}, {Carry}, {Cellino}, {Cheek},
  {Clementini}, {Damerdji}, {Davidson}, {de Teodoro}, {Nu{\~n}ez Campos},
  {Delchambre}, {Dell'Oro}, {Esquej}, {Fern{\'a}ndez-Hern{\'a}ndez}, {Fraile},
  {Garabato}, {Garc{\'\i}a-Lario}, {Gosset}, {Haigron}, {Halbwachs}, {Hambly},
  {Harrison}, {Hern{\'a}ndez}, {Hestroffer}, {Hodgkin}, {Holl}, {Jan{\ss}en},
  {Jevardat de Fombelle}, {Jordan}, {Krone-Martins}, {Lanzafame},
  {L{\"o}ffler}, {Marchal}, {Marrese}, {Moitinho}, {Muinonen}, {Osborne},
  {Pancino}, {Pauwels}, {Reyl{\'e}}, {Riello}, {Rimoldini}, {Roegiers},
  {Rybizki}, {Sarro}, {Siopis}, {Smith}, {Sozzetti}, {Utrilla}, {van Leeuwen},
  {Abbas}, {{\'A}brah{\'a}m}, {Abreu Aramburu}, {Aerts}, {Aguado}, {Ajaj},
  {Aldea-Montero}, {Altavilla}, {{\'A}lvarez}, {Alves}, {Anders}, {Anderson},
  {Anglada Varela}, {Baines}, {Baker}, {Balaguer-N{\'u}{\~n}ez}, {Balbinot},
  {Balog}, {Barache}, {Barbato}, {Barros}, {Barstow}, {Bartolom{\'e}},
  {Bassilana}, {Bauchet}, {Becciani}, {Bellazzini}, {Berihuete}, {Bernet},
  {Bertone}, {Bianchi}, {Binnenfeld}, {Blanco-Cuaresma}, {Boch}, {Bombrun},
  {Bossini}, {Bouquillon}, {Bramante}, {Breedt}, {Bressan}, {Brouillet},
  {Brugaletta}, {Bucciarelli}, {Burlacu}, {Butkevich}, {Buzzi}, {Caffau},
  {Cancelliere}, {Carballo}, {Carlucci}, {Carnerero}, {Carrasco}, {Castellani},
  {Castro-Ginard}, {Chaoul}, {Charlot}, {Chemin}, {Chiaramida}, {Chiavassa},
  {Chornay}, {Comoretto}, {Contursi}, {Cooper}, {Cornez}, {Cowell}, {Crifo},
  {Crosta}, {Crowley}, {Dafonte}, {Dapergolas}, {David}, {De Luise}, {De
  March}, {De Ridder}, {de Souza}, {de Torres}, {del Peloso}, {del Pozo},
  {Delbo}, {Delgado}, {Delisle}, {Demouchy}, {Dharmawardena}, {Di Matteo},
  {Diakite}, {Diener}, {Distefano}, {Dolding}, {Edvardsson}, {Enke}, {Fabre},
  {Fabrizio}, {Faigler}, {Fedorets}, {Fernique}, {Figueras}, {Fournier},
  {Fouron}, {Fragkoudi}, {Gai}, {Garcia-Gutierrez}, {Garcia-Reinaldos},
  {Garc{\'\i}a-Torres}, {Garofalo}, {Gavel}, {Gavras}, {Gerlach}, {Geyer},
  {Giacobbe}, {Gilmore}, {Girona}, {Giuffrida}, {Gomel}, {Gomez},
  {Gonz{\'a}lez-N{\'u}{\~n}ez}, {Gonz{\'a}lez-Santamar{\'\i}a},
  {Gonz{\'a}lez-Vidal}, {Granvik}, {Guillout}, {Guiraud},
  {Guti{\'e}rrez-S{\'a}nchez}, {Guy}, {Hatzidimitriou}, {Hauser}, {Haywood},
  {Helmer}, {Helmi}, {Sarmiento}, {Hidalgo}, {H{\l}adczuk}, {Hobbs}, {Holland},
  {Huckle}, {Jardine}, {Jasniewicz}, {Jean-Antoine Piccolo},
  {Jim{\'e}nez-Arranz}, {Juaristi Campillo}, {Julbe}, {Karbevska}, {Kervella},
  {Khanna}, {Korn}, {K{\'o}sp{\'a}l}, {Kostrzewa-Rutkowska}, {Kruszy{\'n}ska},
  {Kun}, {Laizeau}, {Lambert}, {Lanza}, {Lasne}, {Le Campion}, {Lebreton},
  {Lebzelter}, {Leccia}, {Leclerc}, {Lecoeur-Taibi}, {Liao}, {Licata},
  {Lindstr{\o}m}, {Lister}, {Livanou}, {Lobel}, {Lorca}, {Loup}, {Madrero
  Pardo}, {Magdaleno Romeo}, {Managau}, {Mann}, {Manteiga}, {Marchant},
  {Marconi}, {Marcos}, {Marcos Santos}, {Mar{\'\i}n Pina}, {Marinoni},
  {Marocco}, {Marshall}, {Martin Polo}, {Mart{\'\i}n-Fleitas}, {Marton},
  {Mary}, {Masip}, {Massari}, {Mastrobuono-Battisti}, {Mazeh}, {Messina},
  {Michalik}, {Millar}, {Mints}, {Molina}, {Molinaro}, {Moln{\'a}r}, {Monari},
  {Mongui{\'o}}, {Montegriffo}, {Montero}, {Mor}, {Mora}, {Morbidelli},
  {Morel}, {Morris}, {Muraveva}, {Murphy}, {Musella}, {Nagy}, {Noval},
  {Oca{\~n}a}, {Ogden}, {Ordenovic}, {Osinde}, {Pagani}, {Pagano}, {Palaversa},
  {Pallas-Quintela}, {Panahi}, {Payne-Wardenaar}, {Pe{\~n}alosa Esteller},
  {Penttil{\"a}}, {Pichon}, {Piersimoni}, {Pineau}, {Plachy}, {Plum},
  {Pr{\v{s}}a}, {Pulone}, {Racero}, {Ragaini}, {Rainer}, {Raiteri}, {Ramos},
  {Ramos-Lerate}, {Regibo}, {Richards}, {Rios Diaz}, {Ripepi}, {Riva}, {Rix},
  {Rixon}, {Robichon}, {Robin}, {Robin}, {Roelens}, {Rogues}, {Rohrbasser},
  {Romero-G{\'o}mez}, {Rowell}, {Royer}, {Ruz Mieres}, {Rybicki}, {Sadowski},
  {S{\'a}ez N{\'u}{\~n}ez}, {Sagrist{\`a} Sell{\'e}s}, {Sahlmann}, {Salguero},
  {Samaras}, {Sanchez Gimenez}, {Sanna}, {Santove{\~n}a}, {Sarasso}, {Sciacca},
  {Segol}, {Segovia}, {S{\'e}gransan}, {Semeux}, {Shahaf}, {Siddiqui},
  {Siebert}, {Siltala}, {Silvelo}, {Slezak}, {Slezak}, {Smart}, {Snaith},
  {Solano}, {Solitro}, {Souami}, {Souchay}, {Spoto}, {Steele},
  {Steidelm{\"u}ller}, {Stephenson}, {S{\"u}veges}, {Surdej}, {Szabados},
  {Szegedi-Elek}, {Taris}, {Taylor}, {Teixeira}, {Tolomei}, {Tonello}, {Torra},
  {Torra}, {Torralba Elipe}, {Trabucchi}, {Tsounis}, {Turon}, {Ulla}, {Unger},
  {Vaillant}, {van Dillen}, {van Reeven}, {Vanel}, {Vecchiato}, {Viala},
  {Vicente}, {Voutsinas}, {Weiler}, {Wevers}, {Wyrzykowski}, {Yoldas}, {Yvard},
  {Zhao}, {Zorec}, {Zucker}, \& {Zwitter}}]{Recio-BlancoGaia2023}
{Gaia Collaboration} {et~al.}, 2023, \aap, 674, A38

\bibitem[{{Gaia Collaboration} {et~al}\mbox{.}(2022){Gaia Collaboration},
  {Vallenari}, {Brown}, {Prusti}, {de Bruijne}, {Arenou}, {Babusiaux},
  {Biermann}, {Creevey}, {Ducourant}, {Evans}, {Eyer}, {Guerra}, {Hutton},
  {Jordi}, {Klioner}, {Lammers}, {Lindegren}, {Luri}, {Mignard}, {Panem},
  {Pourbaix}, {Randich}, {Sartoretti}, {Soubiran}, {Tanga}, {Walton},
  {Bailer-Jones}, {Bastian}, {Drimmel}, {Jansen}, {Katz}, {Lattanzi}, {van
  Leeuwen}, {Bakker}, {Cacciari}, {Casta{\~n}eda}, {De Angeli}, {Fabricius},
  {Fouesneau}, {Fr{\'e}mat}, {Galluccio}, {Guerrier}, {Heiter}, {Masana},
  {Messineo}, {Mowlavi}, {Nicolas}, {Nienartowicz}, {Pailler}, {Panuzzo},
  {Riclet}, {Roux}, {Seabroke}, {Sordo{\o}rcit}, {Th{\'e}venin},
  {Gracia-Abril}, {Portell}, {Teyssier}, {Altmann}, {Andrae}, {Audard},
  {Bellas-Velidis}, {Benson}, {Berthier}, {Blomme}, {Burgess}, {Busonero},
  {Busso}, {C{\'a}novas}, {Carry}, {Cellino}, {Cheek}, {Clementini},
  {Damerdji}, {Davidson}, {de Teodoro}, {Nu{\~n}ez Campos}, {Delchambre},
  {Dell'Oro}, {Esquej}, {Fern{\'a}ndez-Hern{\'a}ndez}, {Fraile}, {Garabato},
  {Garc{\'\i}a-Lario}, {Gosset}, {Haigron}, {Halbwachs}, {Hambly}, {Harrison},
  {Hern{\'a}ndez}, {Hestroffer}, {Hodgkin}, {Holl}, {Jan{\ss}en}, {Jevardat de
  Fombelle}, {Jordan}, {Krone-Martins}, {Lanzafame}, {L{\"o}ffler}, {Marchal},
  {Marrese}, {Moitinho}, {Muinonen}, {Osborne}, {Pancino}, {Pauwels},
  {Recio-Blanco}, {Reyl{\'e}}, {Riello}, {Rimoldini}, {Roegiers}, {Rybizki},
  {Sarro}, {Siopis}, {Smith}, {Sozzetti}, {Utrilla}, {van Leeuwen}, {Abbas},
  {{\'A}brah{\'a}m}, {Abreu Aramburu}, {Aerts}, {Aguado}, {Ajaj},
  {Aldea-Montero}, {Altavilla}, {{\'A}lvarez}, {Alves}, {Anders}, {Anderson},
  {Anglada Varela}, {Antoja}, {Baines}, {Baker}, {Balaguer-N{\'u}{\~n}ez},
  {Balbinot}, {Balog}, {Barache}, {Barbato}, {Barros}, {Barstow},
  {Bartolom{\'e}}, {Bassilana}, {Bauchet}, {Becciani}, {Bellazzini},
  {Berihuete}, {Bernet}, {Bertone}, {Bianchi}, {Binnenfeld}, {Blanco-Cuaresma},
  {Blazere}, {Boch}, {Bombrun}, {Bossini}, {Bouquillon}, {Bragaglia},
  {Bramante}, {Breedt}, {Bressan}, {Brouillet}, {Brugaletta}, {Bucciarelli},
  {Burlacu}, {Butkevich}, {Buzzi}, {Caffau}, {Cancelliere}, {Cantat-Gaudin},
  {Carballo}, {Carlucci}, {Carnerero}, {Carrasco}, {Casamiquela}, {Castellani},
  {Castro-Ginard}, {Chaoul}, {Charlot}, {Chemin}, {Chiaramida}, {Chiavassa},
  {Chornay}, {Comoretto}, {Contursi}, {Cooper}, {Cornez}, {Cowell}, {Crifo},
  {Cropper}, {Crosta}, {Crowley}, {Dafonte}, {Dapergolas}, {David}, {David},
  {de Laverny}, {De Luise}, {De March}, {De Ridder}, {de Souza}, {de Torres},
  {del Peloso}, {del Pozo}, {Delbo}, {Delgado}, {Delisle}, {Demouchy},
  {Dharmawardena}, {Di Matteo}, {Diakite}, {Diener}, {Distefano}, {Dolding},
  {Edvardsson}, {Enke}, {Fabre}, {Fabrizio}, {Faigler}, {Fedorets}, {Fernique},
  {Fienga}, {Figueras}, {Fournier}, {Fouron}, {Fragkoudi}, {Gai},
  {Garcia-Gutierrez}, {Garcia-Reinaldos}, {Garc{\'\i}a-Torres}, {Garofalo},
  {Gavel}, {Gavras}, {Gerlach}, {Geyer}, {Giacobbe}, {Gilmore}, {Girona},
  {Giuffrida}, {Gomel}, {Gomez}, {Gonz{\'a}lez-N{\'u}{\~n}ez},
  {Gonz{\'a}lez-Santamar{\'\i}a}, {Gonz{\'a}lez-Vidal}, {Granvik}, {Guillout},
  {Guiraud}, {Guti{\'e}rrez-S{\'a}nchez}, {Guy}, {Hatzidimitriou}, {Hauser},
  {Haywood}, {Helmer}, {Helmi}, {Sarmiento}, {Hidalgo}, {Hilger},
  {H{\l}adczuk}, {Hobbs}, {Holland}, {Huckle}, {Jardine}, {Jasniewicz},
  {Jean-Antoine Piccolo}, {Jim{\'e}nez-Arranz}, {Jorissen}, {Juaristi
  Campillo}, {Julbe}, {Karbevska}, {Kervella}, {Khanna}, {Kontizas},
  {Kordopatis}, {Korn}, {K{\'o}sp{\'a}l}, {Kostrzewa-Rutkowska},
  {Kruszy{\'n}ska}, {Kun}, {Laizeau}, {Lambert}, {Lanza}, {Lasne}, {Le
  Campion}, {Lebreton}, {Lebzelter}, {Leccia}, {Leclerc}, {Lecoeur-Taibi},
  {Liao}, {Licata}, {Lindstr{\o}m}, {Lister}, {Livanou}, {Lobel}, {Lorca},
  {Loup}, {Madrero Pardo}, {Magdaleno Romeo}, {Managau}, {Mann}, {Manteiga},
  {Marchant}, {Marconi}, {Marcos}, {Marcos Santos}, {Mar{\'\i}n Pina},
  {Marinoni}, {Marocco}, {Marshall}, {Polo}, {Mart{\'\i}n-Fleitas}, {Marton},
  {Mary}, {Masip}, {Massari}, {Mastrobuono-Battisti}, {Mazeh}, {McMillan},
  {Messina}, {Michalik}, {Millar}, {Mints}, {Molina}, {Molinaro}, {Moln{\'a}r},
  {Monari}, {Mongui{\'o}}, {Montegriffo}, {Montero}, {Mor}, {Mora},
  {Morbidelli}, {Morel}, {Morris}, {Muraveva}, {Murphy}, {Musella}, {Nagy},
  {Noval}, {Oca{\~n}a}, {Ogden}, {Ordenovic}, {Osinde}, {Pagani}, {Pagano},
  {Palaversa}, {Palicio}, {Pallas-Quintela}, {Panahi}, {Payne-Wardenaar},
  {Pe{\~n}alosa Esteller}, {Penttil{\"a}}, {Pichon}, {Piersimoni}, {Pineau},
  {Plachy}, {Plum}, {Poggio}, {Pr{\v{s}}a}, {Pulone}, {Racero}, {Ragaini},
  {Rainer}, {Raiteri}, {Rambaux}, {Ramos}, {Ramos-Lerate}, {Re Fiorentin},
  {Regibo}, {Richards}, {Rios Diaz}, {Ripepi}, {Riva}, {Rix}, {Rixon},
  {Robichon}, {Robin}, {Robin}, {Roelens}, {Rogues}, {Rohrbasser},
  {Romero-G{\'o}mez}, {Rowell}, {Royer}, {Ruz Mieres}, {Rybicki}, {Sadowski},
  {S{\'a}ez N{\'u}{\~n}ez}, {Sagrist{\`a} Sell{\'e}s}, {Sahlmann}, {Salguero},
  {Samaras}, {Sanchez Gimenez}, {Sanna}, {Santove{\~n}a}, {Sarasso},
  {Schultheis}, {Sciacca}, {Segol}, {Segovia}, {S{\'e}gransan}, {Semeux},
  {Shahaf}, {Siddiqui}, {Siebert}, {Siltala}, {Silvelo}, {Slezak}, {Slezak},
  {Smart}, {Snaith}, {Solano}, {Solitro}, {Souami}, {Souchay}, {Spagna},
  {Spina}, {Spoto}, {Steele}, {Steidelm{\"u}ller}, {Stephenson}, {S{\"u}veges},
  {Surdej}, {Szabados}, {Szegedi-Elek}, {Taris}, {Taylo}, {Teixeira},
  {Tolomei}, {Tonello}, {Torra}, {Torra}, {Torralba Elipe}, {Trabucchi},
  {Tsounis}, {Turon}, {Ulla}, {Unger}, {Vaillant}, {van Dillen}, {van Reeven},
  {Vanel}, {Vecchiato}, {Viala}, {Vicente}, {Voutsinas}, {Weiler}, {Wevers},
  {Wyrzykowski}, {Yoldas}, {Yvard}, {Zhao}, {Zorec}, {Zucker}, \&
  {Zwitter}}]{GaiaDR3general}
{Gaia Collaboration} {et~al.}, 2022, arXiv e-prints, arXiv:2208.00211

\bibitem[{{Gilmore} \& {Reid}(1983)}]{GiRe83}
{Gilmore} G., {Reid} N., 1983, \mnras, 202, 1025

\bibitem[{{Gilmore} \& {Wyse}(1998)}]{GilmoreWyse1998}
{Gilmore} G., {Wyse} R. F.~G., 1998, \aj, 116, 748

\bibitem[{{Grand} {et~al}\mbox{.}(2017){Grand}, {G{\'o}mez}, {Marinacci},
  {Pakmor}, {Springel}, {Campbell}, {Frenk}, {Jenkins}, \& {White}}]{Grand2017}
{Grand} R. J.~J. {et~al.}, 2017, \mnras, 467, 179

\bibitem[{{GRAVITY Collaboration} {et~al}\mbox{.}(2022){GRAVITY Collaboration},
  {Abuter}, {Aimar}, {Amorim}, {Ball}, {Baub{\"o}ck}, {Berger}, {Bonnet},
  {Bourdarot}, {Brandner}, {Cardoso}, {Cl{\'e}net}, {Dallilar}, {Davies}, {de
  Zeeuw}, {Dexter}, {Drescher}, {Eisenhauer}, {F{\"o}rster Schreiber},
  {Foschi}, {Garcia}, {Gao}, {Gendron}, {Genzel}, {Gillessen}, {Habibi},
  {Haubois}, {Hei{\ss}el}, {Henning}, {Hippler}, {Horrobin}, {Jochum}, {Jocou},
  {Kaufer}, {Kervella}, {Lacour}, {Lapeyr{\`e}re}, {Le Bouquin}, {L{\'e}na},
  {Lutz}, {Ott}, {Paumard}, {Perraut}, {Perrin}, {Pfuhl}, {Rabien},
  {Shangguan}, {Shimizu}, {Scheithauer}, {Stadler}, {Stephens}, {Straub},
  {Straubmeier}, {Sturm}, {Tacconi}, {Tristram}, {Vincent}, {von Fellenberg},
  {Widmann}, {Wieprecht}, {Wiezorrek}, {Woillez}, {Yazici}, \&
  {Young}}]{Gravity2022}
{GRAVITY Collaboration} {et~al.}, 2022, \aap, 657, L12

\bibitem[{{Hayden} {et~al}\mbox{.}(2015){Hayden}, {Bovy}, {Holtzman},
  {Nidever}, {Bird}, {Weinberg}, {Andrews}, {Majewski}, {Allende Prieto},
  {Anders}, {Beers}, {Bizyaev}, {Chiappini}, {Cunha}, {Frinchaboy}, \&
  {Garc{\'\i}a-Her{\'n}and ez}}]{Hayden2015}
{Hayden} M.~R. {et~al.}, 2015, \apj, 808, 132

\bibitem[{{Helmi} {et~al}\mbox{.}(2018){Helmi}, {Babusiaux}, {Koppelman},
  {Massari}, {Veljanoski}, \& {Brown}}]{Helmi2018}
{Helmi} A., {Babusiaux} C., {Koppelman} H.~H., {Massari} D., {Veljanoski} J.,
  {Brown} A. G.~A., 2018, \nat, 563, 85

\bibitem[{{Hunt} {et~al}\mbox{.}(2019){Hunt}, {Bub}, {Bovy}, {Mackereth},
  {Trick}, \& {Kawata}}]{HuntBubBovy2019}
{Hunt} J. A.~S., {Bub} M.~W., {Bovy} J., {Mackereth} J.~T., {Trick} W.~H.,
  {Kawata} D., 2019, \mnras, 490, 1026

\bibitem[{{Juri{\'c}} {et~al}\mbox{.}(2008){Juri{\'c}}, {Ivezi{\'c}}, {Brooks},
  {Lupton}, {Schlegel}, {Finkbeiner}, {Padmanabhan}, {Bond}, {Sesar},
  {Rockosi}, {Knapp}, {Gunn}, {Sumi}, {Schneider}, {Barentine}, {Brewington},
  \& {Brinkmann}}]{Juea08}
{Juri{\'c}} M. {et~al.}, 2008, \apj, 673, 864

\bibitem[{{Khanna} {et~al}\mbox{.}(2023){Khanna}, {Sharma}, {Bland-Hawthorn},
  \& {Hayden}}]{Khanna2023}
{Khanna} S., {Sharma} S., {Bland-Hawthorn} J., {Hayden} M., 2023, \mnras, 520,
  5002

\bibitem[{{Li} \& {Binney}(2022)}]{LiBinneyYoungD}
{Li} C., {Binney} J., 2022, \mnras, 516, 3454

\bibitem[{{Lian} {et~al}\mbox{.}(2022){Lian}, {Zasowski}, {Mackereth}, {Imig},
  {Holtzman}, {Beaton}, {Bird}, {Cunha}, {Fern{\'a}ndez-Trincado}, {Horta},
  {Lane}, {Masters}, {Nitschelm}, \& {Roman-Lopes}}]{Lian2022}
{Lian} J. {et~al.}, 2022, \mnras, 513, 4130

\bibitem[{{Majewski} {et~al}\mbox{.}(2017){Majewski}, {Schiavon}, {Frinchaboy},
  {Allende Prieto}, {Barkhouser}, {Bizyaev}, {Blank}, {Brunner}, {Burton},
  {Carrera}, {Chojnowski}, {Cunha}, {Epstein}, {Fitzgerald}, \& {Garc{\'\i}a
  P{\'e}rez}}]{Majewski2017}
{Majewski} S.~R. {et~al.}, 2017, \aj, 154, 94

\bibitem[{{Matteucci} \& {Francois}(1989)}]{Matteucci1989}
{Matteucci} F., {Francois} P., 1989, \mnras, 239, 885

\bibitem[{{McMillan} {et~al}\mbox{.}(2022){McMillan}, {Petersson},
  {Tepper-Garcia}, {Bland-Hawthorn}, {Antoja}, {Chemin}, {Figueras}, {Khanna},
  {Kordopatis}, {Ramos}, {Romero-G{\'o}mez}, \& {Seabroke}}]{McMillan2022}
{McMillan} P.~J. {et~al.}, 2022, \mnras, 516, 4988

\bibitem[{{M{\'e}ndez-Delgado} {et~al}\mbox{.}(2022){M{\'e}ndez-Delgado},
  {Amayo}, {Arellano-C{\'o}rdova}, {Esteban}, {Garc{\'\i}a-Rojas}, {Carigi}, \&
  {Delgado-Inglada}}]{MendezDelgado2022}
{M{\'e}ndez-Delgado} J.~E., {Amayo} A., {Arellano-C{\'o}rdova} K.~Z., {Esteban}
  C., {Garc{\'\i}a-Rojas} J., {Carigi} L., {Delgado-Inglada} G., 2022, \mnras,
  510, 4436

\bibitem[{{Mr{\'o}z} {et~al}\mbox{.}(2019){Mr{\'o}z}, {Udalski}, {Skowron},
  {Skowron}, {Soszy{\'n}ski}, {Pietrukowicz}, {Szyma{\'n}ski}, {Poleski},
  {Koz{\l}owski}, \& {Ulaczyk}}]{Mroz2019}
{Mr{\'o}z} P. {et~al.}, 2019, \apjl, 870, L10

\bibitem[{{Myeong} {et~al}\mbox{.}(2018){Myeong}, {Evans}, {Belokurov},
  {Sanders}, \& {Koposov}}]{MyeongEvans2018}
{Myeong} G.~C., {Evans} N.~W., {Belokurov} V., {Sanders} J.~L., {Koposov}
  S.~E., 2018, \apjl, 856, L26

\bibitem[{{Pagel}(1997)}]{Pagel1997}
{Pagel} B. E.~J., 1997, {Nucleosynthesis and Chemical Evolution of Galaxies}.
  Cambridge University Press

\bibitem[{{Portail} {et~al}\mbox{.}(2015){Portail}, {Wegg}, {Gerhard}, \&
  {Martinez-Valpuesta}}]{Portail2015}
{Portail} M., {Wegg} C., {Gerhard} O., {Martinez-Valpuesta} I., 2015, \mnras,
  448, 713

\bibitem[{{Queiroz} {et~al}\mbox{.}(2023){Queiroz}, {Anders}, {Chiappini},
  {Khalatyan}, {Santiago}, {Nepal}, {Steinmetz}, {Gallart}, {Valentini}, {Dal
  Ponte}, {Barbuy}, {P{\'e}rez-Villegas}, {Masseron}, {Fern{\'a}ndez-Trincado},
  {Khoperskov}, {Minchev}, {Fern{\'a}ndez-Alvar}, {Lane}, \&
  {Nitschelm}}]{StarHorse3}
{Queiroz} A.~B.~A. {et~al.}, 2023, \aap, 673, A155

\bibitem[{{Queiroz} {et~al}\mbox{.}(2018){Queiroz}, {Anders}, {Santiago},
  {Chiappini}, {Steinmetz}, {Dal Ponte}, {Stassun}, {da Costa}, {Maia},
  {Crestani}, {Beers}, {Fern{\'a}ndez-Trincado}, {Garc{\'\i}a-Hern{\'a}ndez},
  {Roman-Lopes}, \& {Zamora}}]{StarHorse}
{Queiroz} A.~B.~A. {et~al.}, 2018, \mnras, 476, 2556

\bibitem[{{Read} \& {Steger}(2017)}]{Read2017}
{Read} J.~I., {Steger} P., 2017, \mnras, 471, 4541

\bibitem[{{Reid} \& {Brunthaler}(2020)}]{ReidBrunthaler2020}
{Reid} M.~J., {Brunthaler} A., 2020, \apj, 892, 39

\bibitem[{{Robin} {et~al}\mbox{.}(2022){Robin}, {Bienaym{\'e}}, {Salomon},
  {Reyl{\'e}}, {Lagarde}, {Figueras}, {Mor}, {Fern{\'a}ndez-Trincado}, \&
  {Montillaud}}]{Robin2022}
{Robin} A.~C. {et~al.}, 2022, \aap, 667, A98

\bibitem[{{Robin} {et~al}\mbox{.}(2003){Robin}, {Reyl{\'e}}, {Derri{\`e}re}, \&
  {Picaud}}]{Roea03}
{Robin} A.~C., {Reyl{\'e}} C., {Derri{\`e}re} S., {Picaud} S., 2003, \aap, 409,
  523

\bibitem[{{Roman}(1950)}]{Roman1950}
{Roman} N.~G., 1950, \apj, 112, 554

\bibitem[{{Roman}(1999)}]{Roman1999}
{Roman} N.~G., 1999, \apss, 267, 37

\bibitem[{{Rybizki} {et~al}\mbox{.}(2022){Rybizki}, {Green}, {Rix}, {El-Badry},
  {Demleitner}, {Zari}, {Udalski}, {Smart}, \& {Gould}}]{Rybizki2022}
{Rybizki} J. {et~al.}, 2022, \mnras, 510, 2597

\bibitem[{{Sanders} \& {Binney}(2015)}]{SaJJB15:EDF}
{Sanders} J.~L., {Binney} J., 2015, \mnras, 449, 3479

\bibitem[{{Sch{\"o}nrich}(2012)}]{Schoenrich2012}
{Sch{\"o}nrich} R., 2012, \mnras, 427, 274

\bibitem[{{Sch{\"o}nrich} \& {Binney}(2009{\natexlab{a}})}]{SchoenrichB2009a}
{Sch{\"o}nrich} R., {Binney} J., 2009{\natexlab{a}}, \mnras, 396, 203

\bibitem[{{Sch{\"o}nrich} \& {Binney}(2009{\natexlab{b}})}]{SchoenrichB2009b}
{Sch{\"o}nrich} R., {Binney} J., 2009{\natexlab{b}}, \mnras, 399, 1145

\bibitem[{{Sch{\"o}nrich} \& {McMillan}(2017)}]{SchoenPJM}
{Sch{\"o}nrich} R., {McMillan} P.~J., 2017, \mnras, 467, 1154

\bibitem[{{Sellwood} \& {Binney}(2002)}]{SellwoodB2002}
{Sellwood} J.~A., {Binney} J.~J., 2002, \mnras, 336, 785

\bibitem[{{Sharma} {et~al}\mbox{.}(2021){Sharma}, {Hayden}, \&
  {Bland-Hawthorn}}]{Sharma2021}
{Sharma} S., {Hayden} M.~R., {Bland-Hawthorn} J., 2021, \mnras, 507, 5882

\bibitem[{{Spina} {et~al}\mbox{.}(2022){Spina}, {Magrini}, \&
  {Cunha}}]{Spina2022}
{Spina} L., {Magrini} L., {Cunha} K., 2022, Universe, 8, 87

\bibitem[{{Tinsley}(1980)}]{Tinsley1980}
{Tinsley} B.~M., 1980, \fcp, 5, 287

\bibitem[{{Trick} {et~al}\mbox{.}(2019){Trick}, {Coronado}, \&
  {Rix}}]{TrickRix2019}
{Trick} W.~H., {Coronado} J., {Rix} H.-W., 2019, \mnras, 484, 3291

\bibitem[{{van den Bosch} {et~al}\mbox{.}(2008){van den Bosch}, {van de Ven},
  {Verolme}, {Cappellari}, \& {de Zeeuw}}]{vdBea08}
{van den Bosch} R.~C.~E., {van de Ven} G., {Verolme} E.~K., {Cappellari} M.,
  {de Zeeuw} P.~T., 2008, \mnras, 385, 647

\bibitem[{{Vasiliev}(2019)}]{AGAMA}
{Vasiliev} E., 2019, \mnras, 482, 1525

\bibitem[{{Vasiliev} \& {Baumgardt}(2021)}]{VasilievBaumgardt2021}
{Vasiliev} E., {Baumgardt} H., 2021, \mnras, 505, 5978

\bibitem[{{Wegg} \& {Gerhard}(2013)}]{Wegg2013}
{Wegg} C., {Gerhard} O., 2013, \mnras, 435, 1874

\bibitem[{{Zhu} {et~al}\mbox{.}(2018){Zhu}, {van de Ven}, {M{\'e}ndez-Abreu},
  \& {Obreja}}]{Zhu2018}
{Zhu} L., {van de Ven} G., {M{\'e}ndez-Abreu} J., {Obreja} A., 2018, \mnras,
  479, 945

\end{thebibliography}

\appendix

\section{Validation of our likelihood}\label{app:validate}

We show that our likelihood is maximised when the model perfectly reproduces
the data.

By the cloud-in-cell algorithm we distribute stars over a grid in some space
(three-dimensional velocity space or four dimensional chemical-action space).
Then we use the same algorithm to interpolate a density at the location of
each real star and add to $\cL$ the logarithm of that density. To analyse
this algorithm we cover the grid actually used by a virtual grid so fine that
the interpolated mock-star density can be treated as constant in each of its
cells. Then $\cL=N^{-1}\sum_in_i\ln m_i$ is the sum over the fine grid's
cells of the number $n_i$ of real stars in that cell times the logarithm of
the number $m_i$ of mock stars in the cell, and $N\equiv\sum_in_i$. We use a
Lagrange multiplier $\lambda$ to maximise $\cL$ with respect to $m_i$ subject
to the constraint that $M=\sum m_i$ is constant.
\[
0={\p\over\p m_i}(\cL-\lambda M)={n_i\over Nm_i}-\lambda.
\]
Hence $\cL$ is maximum when $m_i\propto n_i$ as was to be shown.

\end{document}